\documentclass[twocolumn,superscriptaddress,aps,longbibliography]{revtex4-1}
\usepackage{amsxtra,amssymb,bm,amsbsy,amsmath,amsthm,latexsym,dsfont,array,layout,graphicx,physics,mathrsfs,color,soul,subfigure,enumerate,footmisc,scrextend}
\usepackage[colorlinks=true,citecolor=blue,linkcolor=blue,anchorcolor=blue,urlcolor=blue]{hyperref}
\hypersetup{linktoc=all}

\begin{document}

\title{Quantum transport theory of hybrid superconducting  systems}
\author{Chuan-Zhe Yao}
\thanks{These two authors contributed equally to this work.}
\affiliation{Department of Physics and Center for Quantum Information Science,
National Cheng Kung University, Tainan 70101, Taiwan}
\author{Hon-Lam Lai}
\thanks{These two authors contributed equally to this work.}
\affiliation{Department of Physics and Center for Quantum Information Science,
National Cheng Kung University, Tainan 70101, Taiwan}
\author{Wei-Min Zhang}
\email{wzhang@mail.ncku.edu.tw}
\affiliation{Department of Physics and Center for Quantum Information Science,
National Cheng Kung University, Tainan 70101, Taiwan}
%\affiliation{Physics Division, National Center for Theoretical Sciences, Taipei 10617, Taiwan}

\begin{abstract}
We present a quantum transport theory for hybrid superconducting systems based on our exact master equation approach. 
The system-terminal transport current dynamics are fully captured by the extended non-equilibrium Green's function incorporating pair correlations via spectral density matrices. The total transient transport current is decomposed into components that describe coherent transports through different paths of particle and hole channels. We show that these coherent transports are resultant interferences of numerous repeated tunneling processes and cannot be rendered as a simple normal transmission or Andreev reflection as usually described in the steady quantum transport involving superconductivity. As a practical application, we apply the theory to a two-terminal superconductor-semiconductor nanowire to study the transport dynamics through a pair of Majorana zero modes. We find that the coherent transport currents passing through a pair of well-separated Majorana zero modes vanish due to the totally destructive interference between the particle and hole channels. This provides a new understanding to the scenario of ``teleportation"  via a pair of delocalized Majorana zero modes, namely a pair of delocalized Majorana zero modes  does not possess the nonlocality of an entangled pair for quantum teleportation.
\end{abstract}

\maketitle
\section{Introduction}
Topological quantum computation has been considered as the promising candidate because it has been thought to be robust against decoherence~\cite{K01,K03,FKL03,NSS08,AOR11,VHF15,PLS16,AHM16,LPS16,LBK18,PSD20,OV20}. The building blocks of topological quantum computation, i.e. topological qubits, are proposed to be made by anyons which obey the non-Abelian statistics \cite{K03,FKL03,NSS08}. One of the realizations for topological qubits is the spatially well-separated Majorana zero modes (MZMs) which forms a highly degenerate ground state space \cite{AOR11,VHF15,PLS16,AHM16,LPS16,LBK18,PSD20,OV20}. The existence of MZMs is theoretically predicted to manifest at the boundaries of $p$-wave topological superconductors~\cite{K01,K03}. Both theoretical and experimental investigations aimed at confirming the presence of MZMs have heavily relied on their transport properties. For instance, tunneling conductance measurements have been conducted in two-terminal~\cite{LLN09,F10,DNS16,MST18,ABA19} or three-terminal experimental setups~\cite{DHH20,MAM20}. However, the origin of the observed zero-bias conductance peaks (ZBCPs) in these experiments, whether they arise from MZMs or other non-topological bound states, remains a subject of ongoing debate~\cite{CPS15,SCP16,PAS17,LSS17,DVP18,APP19,ACB19}. Furthermore, current power spectroscopy has been proposed as another approach of detecting evidence for MZMs, also utilizing the transport dynamics~\cite{CWX12,ZR13,LCL15,QFL22,S22}.

In the experimental studies of transport properties, particularly those concerning MZMs, a significant portion of research has centered with hybrid superconducting structures. With proximity-induced conventional $s$-wave superconductivity in various systems, it becomes possible to generate effective $p$-wave superconductivity. A prime example is the utilization of the surface states of a topological insulator combined with induced $s$-wave superconductivity to create effective $p$-wave superconductivity. Consequently, MZMs can manifest as vortices at the interface between a well-engineered topological insulator and a conventional superconductor~\cite{FK08,FK09}. Such achievements can also be accomplished by inducing $s$-wave superconductivity in a semiconductor with strong spin-orbit coupling when the time-reversal symmetry is intentionally broken. Agents capable of breaking time-reversal symmetry include external magnetic fields~\cite{A10,LSD10,ORv10}, ferromagnetic insulators~\cite{LSD10} or layers of half-metals~\cite{CZQ11,DB11}, or alternatively, positioning a chain of magnetic adatoms on a conventional superconductor substrate~\cite{NDB13,PGV13}.

In the literature, the investigation of tunneling currents and conductances in hybrid superconducting  systems primarily relies on the scattering matrix formalism applied to a normal metal (or semiconductor)-superconductor junction~\cite{ADH11,LPL12,SBS15,RVK2018}. The transport processes described by the scattering matrix are classified as normal transmissions and Andreev reflections. However, the scattering theory may be valid only in the steady-state limit with a semiclassical particle picture but cannot capture the transient transport dynamics adequately. In fact, the simple picture of the normal transmission and Andreev reflection for transient quantum transport fails quantum mechanically because the actual transport processes are the resultant interferences of numerous repeated tunneling processes between the system and the terminals that mixes the normal transmissions and Andreev reflections in a very complicated dynamical way, as we will show in details in this paper. 

It is therefore desired to develop a nonequilibrium transport theory that specifies the transient dynamics from the system-terminal interactions and clarifies the connection between transport current and the quantum states of the system, including both topological and non-topological states. Over the past decade, we have derived an exact master equation that incorporates transport current, applicable to the study of non-Markovian dissipation, decoherence dynamics, and nonequilibrium transport physics in various bosonic and fermionic open systems~\cite{TZ08,JTZ10,LZ12,ZLX12,Z19,TZJ12,TAZ14}. Recently, our exact master equation has been extended to topological systems~\cite{LYH18,LZ20,YZ20,HYZ20,XLZ21,YZ22}, including MZM systems~\cite{LYH18,LZ20,YZ22} and Majorana Aharonov-Bohm interferometers~\cite{XLZ21}. The exact master equation reveals an important consequence that the dissipative and the transient transport dynamics of a quantum system is fully captured by the extended nonequilibrium Green's functions incorporating pair correlations via spectral density matrices. The spectral density matrices are given by the product of the terminal densities of state (which characterizes the spectral structures of the terminals) and the conjugate products of the system-terminal coupling amplitudes (which involves wave-function overlaps between the Majorana bound states of the system and the terminal states). Since the topological properties of MZMs in the system are characterized by both their energy spectra and energy eigenfunctions, topological structures of the system are manifested in the dissipative and transport dynamics through these spectral density matrices~\cite{YZ20}. 

In this paper, our aim is to extend our transport theory based on the exact master equation approach to explore the general transient transport dynamics of hybrid superconducting  systems. A hybrid superconducting system comprises a central system that is coupled to multiple leads with adjustable tunneling couplings, given by the Hamiltonian $H_{\rm tot}=H_S+H_{\rm lead}+H_T$. Notably, both Hamiltonians $H_S$ and $H_{\rm lead}$ may incorporate superconducting pairing terms, enabling the analysis of a broad range of systems, including both topological and non-topological ones. By performing Bogoliubov transformations, the central system and terminals can be diagonalized and the pairing terms can always be incorporated into the tunneling Hamiltonian between the central system and terminals \cite{HYZ20}. 

It is important to notice that after Bogoliubov transformations, the coupling strengths in the tunneling Hamiltonian, see the detailed expressions given in Sec.~II~A, are proportional to the wave-function overlap between the topological (or non-topological) states and the terminals states, and are therefore crucial in showing the transport  
dynamics involving different topological structures.  However, in most of researches  ~\cite{DRB11,VRM12,CLD12,RVM13,BBK13,MRG14,GL14}, to simplify the calculations, people use a Majorana tunneling Hamiltonian with the Lindblad-type master equation plus the wide-band limit to study dissipation, decoherence and transport dynamics of topological systems. In these studies, the spectral density matrices are treated as constant decay rates. Such a simplification ignores the wave-function structures of the topological states for transport and thereby removes the significance coming from topological contribution in transport dynamics. As a result, it fails to describe correctly the topological transport dynamics, namely fails to account for the topological structures embedded in the system wave-functions and is thereby unable to distinguish the difference of transport dynamics between topological states and non-topological states, as we have pointed out in our previous work \cite{HYZ20}. 
A very recent work \cite{JL22} follows our derivation of the exact master 
equation for MZMs \cite{LYH18,HYZ20} to study the transport  through MZMs. Although the approach can apply to 
arbitrary spectral density matrices,  they also focus their study only in the  
wide-band limit. Once the wide-band limit is used, it faces the same problem of being unable to capture the topological properties of MZMs (which is characterized by the MZMs wave-functions through the spectral density matrices) in transport dynamics. 

With a general Hamiltonian of the hybrid superconducting systems, we have extended and generalized our exact master equation to include superconducting pairing between the system and the terminals~\cite{HYZ20}. In this paper, we further derive the transient transport dynamics of hybrid superconducting systems or MZM topological systems in both the partitioned and partition-free schemes. In the partitioned scheme, the central system is initially decoupled to the biased terminals before the system-terminal couplings are turned on. In the partition-free scheme, the central system is initially in equilibrium with the terminals before the bias is turned on. The resulting particle/hole transport dynamics between the terminals and the system can be fully described with the extended nonequilibrium Green's functions, which incorporate pairing correlations through the spectral density matrices. Consequently, the transient transport dynamics effectively reveal the topological properties of the system and terminals. This allows our transport theory to unambiguously identify the influence of the non-local MZM wave-function on the transient transport current through the spectral density matrices, which is a crucial aspect in the experimental search for MZM signatures. Moreover, the total transient transport current can be decomposed into components that coherently transfer between different terminal-channels. As a result, the simple picture of the normal transmission and Andreev reflection cannot capture the transient transport dynamics adequately.

As an application of the theory, we study the transport dynamics of a MZM topological system modeled by a superconductor-semiconductor nanowire in the partitioned and partition-free schemes. 
%In the partitioned scheme, the central system is prepared to host MZMs and initially decoupled to the two biased terminals 
%before the system-terminal couplings are turned on. In the partition-free scheme, MZMs are firstly prepared to exist in the 
%central system, which is then coupled to the unbiased leads, and the bias will be turned on after the total system reaches 
%equilibrium. 
We also provide a new understanding to the scenario of ``teleportation" via a pair of well-separated MZMs. Here, ``teleportation" introduced by Fu~\cite{Fu10} refers to an incident electron tunneling into one MZM and coming out from its spatially separated partner while preserving phase coherence. We found that, in both partitioned and partition-free schemes, if the MZMs are perfectly delocalized (or spatially well-separated, namely, their wave-functions do not overlap with each other), the current components tunneling through the particle channel and the hole channel cancel each other. In other words, no quantum teleportation occurs between a pair of perfectly delocalized MZMs because of the totally destructive interference between the particle and hole channels. In other words, a pair of perfectly delocalized MZMs cannot be thought of being non-locally entangled. Note that the key issue of the teleportation in quantum technology is the utilizing of the non-locality of a spatially separated entangled 
pair of states. In \cite{Fu10}, for the "teleportation" to occur a finite charging energy term coupling the two MZMs must explicitly 
appear in the Hamiltonian.  Thus, electron transport caused by the direct coupling energy between the two MZMs, rather than the non-locality of the two MZM entanglement,  is not the effect of teleportation refereed commonly in quantum technology. In fact, in a previous work \cite{LYH18}, we already proved that in the topological phase, when one of the two delocalized MZMs in a nanowire is disturbed, only the disturbed MZM decoheres, leaving the other MZM unchanged. This indicates that the two delocalized MZMs do not entangled together. Therefore, no teleportation occurs via two delocalized MZMs.

The rest of the paper is organized as follows. In section \ref{Sec1}, we discuss the general Hamiltonian of the hybrid superconducting systems, derive their exact master equation and develop the transient quantum transport theory in both partitioned and partition-free schemes. In section \ref{Sec2}, we identify the current contributions of coherent tunnelings through different paths. We then study the transport dynamics of a MZM topological system.  Also, we analyze the behavior of the cross differential conductance and the differential conductance that is experimentally measurable, where the simple description of the normal transmission and Andreev reflection becomes insufficient in the transient regime.  We discuss in details the "teleportation" scenario of MZMs by analyzing interferences between coherent transportations through the particle/hole channels of different leads, and show that the coherent transport currents through a pair of perfectly delocalized MZMs vanishes due to totally destructive interference. Finally, a conclusion is given in section \ref{Sec3}.

\section{Quantum transport theory for hybrid superconducting systems}
\label{Sec1}
\subsection{The general Hamiltonian for hybrid superconducting systems}
A hybrid superconducting system comprises a central system that is coupled to multiple leads with adjustable tunneling couplings, given by the Hamiltonian 
\begin{align}
H_{\rm tot} = H_S + H_{\rm lead} + H_T . 
\end{align}
Notably, both Hamiltonians $H_S$ and $H_{\rm lead}$ may incorporate superconducting pairing terms, enabling the analysis of a broad range of systems, including both topological and non-topological ones. By performing Bogoliubov transformations, the central system and terminals can be diagonalized and the pairing terms can always be incorporated into the tunneling Hamiltonian between the central system and terminals, namely,
\begin{align}
H_{\rm tot}&=\sum_{i}\epsilon_{i}a^{\dag}_{i}a_{i}+\sum_{\alpha k}[\epsilon_{\alpha k}+U_{\alpha}(t)]b^{\dag}_{\alpha k}b_{\alpha k} \notag\\
&+\sum_{j\alpha k}\big[\eta_{\alpha k}b^{\dag}_{\alpha k}(\kappa_{\alpha j}a_{j}+\kappa'_{\alpha j}a^{\dag}_{j})+H.c.\big],
\label{H_tot}
\end{align}
where $a_{i}$ ($b_{\alpha k}$) is the annihilation operator of the $i$th energy level (spectrum $k$ mode) of the central system (terminal $\alpha$), $U_{\alpha}(t)$ is the applied bias to lead $\alpha$, $\eta_{\alpha k}$ is the tunneling strength between the central system and the spectrum $k$ mode of lead $\alpha$. In this context, the topological properties of the central system and the leads are manifested through their respective wave-functions, which are given by the Bogoliubov transformation coefficients $\kappa_{\alpha j}$ and $\kappa'_{\alpha j}$. 

As a specific example, let us consider a hybrid superconducting system modeled by a tight-binding N-site p-wave superconducting wire with its left/right ends coupled respectively to the left/right leads ($\alpha = L, R$). More explicitly, the left and right leads are coupled to the leftmost and the rightmost cites of the wire. Thus, the Hamiltonians of the 
superconducting wire and the two leads plus the tunneling Hamiltonian is given by \cite{HYZ20, YZ22},
\begin{subequations}
\label{kcth}
\begin{align}
H_{\rm tot}  = & \sum^{N-1}_{j=1}(\mu_w c^{\dag}_{j}c_{j}+wc^{\dag}_{j+1}c_{j}+\Delta c_{j+1}c_{j}+{\rm H.c.}), \notag \\
& +  \sum_{\alpha k} [\epsilon_{\alpha k}+U_{\alpha}(t)] b^{\dag}_{\alpha k} b_{\alpha k} \notag \\
& +   \sum_{k}(\eta_{L k}b^{\dag}_{\alpha k}c_{1}+\eta_{R k}b^{\dag}_{R k}c_{N}+{\rm H.c.}). 
\label{K_chain}
\end{align}
where $c_j$ ($c^{\dag}_j$) is the electron annihilation (creation) operator of the system chain cite $j$.
Consequently, by diagonalizing the system Hamiltonian with a Bogoliubov transformation,  the above 
Hamiltonian is reduced to the form of Eq. (\ref{H_tot}) in terms of bogoliubon operators,
\begin{align}
H_{\rm tot}=& \sum_{j}\epsilon_ja^{\dag}_ja_j  +  \sum_{\alpha k} [\epsilon_{\alpha k}+U_{\alpha}(t)] b^{\dag}_{\alpha k} b_{\alpha k}\notag\\
 & + \sum_{kj}\big[\eta_{Lk}(\kappa_{Lj}b^{\dag}_{Lk}a_j+\kappa'_{Lj}b^{\dag}_{Lk}a^{\dag}_j) \notag\\
 & \qquad \quad + \eta_{Rk}(\kappa_{Rj}b^{\dag}_{Rk}a_j+\kappa'_{Rj}b^{\dag}_{Rk}a^{\dag}_j)+H.c.\big], 
\label{H_chain_bov}
\end{align}
\end{subequations}
where $a_j$ are bogoliubon operators of the wire. 
An analytical diagonalization of such a system Hamiltonian with asymmetrical distributed chemical 
potentials $\mu_i$ has been presented in our previous work \cite{YZ22}. The Bogoliubov transformation and the 
corresponding coefficients 
$\kappa_{\alpha j}$ and $\kappa'_{\alpha j}$ representing the wave-functions structure of the Majorana zero modes and 
non-zero modes are given explicitly by 
\begin{subequations}
\label{btkc}
 \begin{align}
 c_1&=\sum_{j}(\kappa_{Lj}a_j+\kappa'_{Lj}a^{\dag}_j) \\
 c_N&=\sum_{j}(\kappa_{Rj}a_j+\kappa'_{Rj}a^{\dag}_j). 
 \end{align}
 \end{subequations}
A numerical calculation of these Bogoliubov transformation coefficients are shown in Fig.~\ref{kappa} in Sec.~III.  
For a large-number chain of the 
p-wave superconducting wire with small chemical potentials, it is well-known that two Majorana zero modes are localized at the ends of 
the wire with exponentially decaying wave-functions along the wire, also see our previous works \cite{YZ20,YZ22}. It is the Majorana zero modes and non-zero 
modes wave-functions embedded in the tunneling Hamiltonian that can distinguish the difference in the transport phenomena 
through the topological and non-topological states in such a two-terminal device.

In a similar way, we can also apply the Hamiltonian (\ref{H_tot}) to the two-dimensional superconductor-semiconductor heterostructure, where the system can be modeled, for example, as a two-dimensional topological Haldane model \cite{H1988}. 
By diagonalizing the system Hamiltonian of the Haldane model with a Bogoliubov transformation, one obtains again 
the general form of the Hamiltonian (\ref{H_tot}), and the wave-functions of topological (or non-topological) states will 
enter explicitly into the coupling strength 
between the topological (or non-topological) states and the terminals states \cite{YLCZ}. The resulting transport theory, 
if one formulates it correctly, can manifest the difference in transport through the topological and non-topological states.

In the literature, one usually starts with a general 
topological system containing, for example, $2N$ MZMs, and models the system Hamiltonian and the tunneling Hamiltonian
by
\begin{subequations}
\begin{align}
H_S=\frac{i}{2}\sum^{2N}_{ij=1}\epsilon_{ij}\gamma_{i}\gamma_{j} , ~~
H_T=\sum_{i\alpha k}(V_{i\alpha k}b^{\dag}_{\alpha k}\gamma_{i}+H.c.),  \label{mhh}
\end{align}
where $\gamma_i$ represents the $i$-th Majorana zero mode.  This Majorana Hamiltonian  
can be easily rewritten in terms of $N$ bogoliubon operators $a_{n}=\frac{1}{2}(\gamma_{2n-1}+i\gamma_{2n})$, namely, 
\begin{align}
H_S&=\sum^{N}_{mn}\tilde{\epsilon}_{mn}(a^{\dag}_{m}a_{n}-a_{m}a^{\dag}_{n}) \\
H_T&=\sum_{n\alpha k}(\tilde{V}_{n\alpha k}b^{\dag}_{\alpha k}a_{n}+\tilde{V}'_{n\alpha k}b^{\dag}_{\alpha k}a^{\dag}_{n}+H.c.).  
\end{align}
\end{subequations}
By diagonalizing $H_S$, one again obtains a total Hamiltonian with the form of Eq.~(\ref{H_tot}). Note that the coupling strength $V_{i\alpha k}$ in Eq.~(\ref{mhh}) is proportional to the wave-function overlap between the topological (or non-topological) states and the terminals states, and is therefore crucial in showing the transport  
dynamics involving different topological structures. However, in most of researches  ~\cite{DRB11,VRM12,CLD12,RVM13,BBK13,MRG14,GL14}, one usually takes the wide-band limit to simplify the calculations.
Such a simplification ignores the wave-function structures of the topological states for transport and thereby removes the significance coming from topological contribution in transport dynamics. As a result, it fails to describe correctly the topological transport dynamics, as we have pointed out in our previous work \cite{HYZ20}, also see the detailed formulation of the transport theory given below.

\subsection{The generalized quantum transport theory for partition-free and partitioned initial states}
From the total Hamiltonian (\ref{H_tot}), the total density matrix of the system and the leads (terminals) is determined by the von Neumann equation: $\frac{d}{dt}\rho_{tot}(t)=\frac{1}{i\hbar}[H_{tot},\rho_{tot}(t)]$, the reduced density matrix of the system is then obtained by partially tracing out the degrees of freedom of the leads, i.e. $\rho_S={\rm Tr}_{E}[\rho_{tot}(t)]$. We consider firstly the partition-free scheme, in which the total system is initially in an equilibrium entangled state between the system and terminals, namely $\rho_{tot}(t_0)=\frac{1}{Z}e^{-\beta H_{tot}}$. A bias $U_{\alpha}(t)=U_{\alpha}\Theta(t-t_0)$ is then applied to the leads at time $t_0$. Since the system and the environment are highly entangled initially, the Feynman-Vernon influence functional method is no longer applicable. We have found \cite{HYZ20} that the exact master equation is given by 
\begin{align}
\frac{d}{dt}\rho_S(t)=\frac{1}{i\hbar}[H_S(t),\rho_S(t)]+[\mathcal{L}^{+}(t)+\mathcal{L}^{-}(t)]\rho_S(t).
\label{ME}
\end{align}
The total current flow out of the system to the leads $I_T(t)=\sum_{\alpha}I_{\alpha}(t)$ can be obtained by tracing the current superoperators $\mathcal{L}^{+}(t)\rho_S(t)$ and $\mathcal{L}^{-}(t)\rho_S(t)$, 
\begin{align}
I_T(t)=e{\rm Tr}[\mathcal{L}^{+}(t)\rho_S(t)]=-e{\rm Tr}[\mathcal{L}^{-}(t)\rho_S(t)]
\label{I_superoperator}
\end{align}
(see Appendix \ref{App2} for a detailed derivation). The current $I_{\alpha}$ passing through the lead $\alpha$ can be expressed as
\begin{align}
I_{\alpha}(t)=&-\frac{e}{\hbar^2}{\rm Tr}\Big[\int^{t}_{t_0}d\tau\bm{g}^{-}_{\alpha}(t,\tau)\tilde{\bm{\rho}}(\tau,t) \notag\\
&-\int^{t}_{t_0}d\tau\tilde{\bm{g}}^{-}_{\alpha}(t,\tau)\bm{U}^{\dag}(t,\tau)+H.c.\Big],
\label{I_alpha0}
\end{align}
where $\tilde{\bm{\rho}}_{ij}(\tau,t)=\begin{pmatrix}
\langle a_{i}(\tau)a^{\dag}_{j}(t)\rangle && \langle a_{i}(\tau)a_{j}(t)\rangle \\
\langle a^{\dag}_{i}(\tau)a^{\dag}_{j}(t)\rangle &&\langle a^{\dag}_{i}(\tau)a_{j}(t)\rangle
\end{pmatrix}$ is the extended particle correlation functional matrix and can be explicitly written as 
\begin{align}
\tilde{\bm{\rho}}&(\tau,t)=\bm{U}(\tau,t_0)\tilde{\bm{\rho}}(t_0,t_0)\bm{U}^{\dag}(t,t_0) \notag\\
&+\frac{1}{\hbar^2}\!\!\int^{\tau}_{t_0}\!\!d\tau_1\!\!\int^{t}_{t_0}\!\!d\tau_2
\bm{U}(\tau,\tau_1)\tilde{\bm{g}}^{+}(\tau_1,\tau_2)\bm{U}^{\dag}(t,\tau_2) \notag\\
&+\frac{1}{\hbar^2}\!\!\int^{\tau}_{t_0}\!\!d\tau_1\!\!\int^{t}_{t_0}\!\!d\tau_2
\bm{U}(\tau,\tau_1)[\bar{\bm{g}}(\tau_1,\tau_2)\!+\!\bar{\bm{g}}^\dag(\tau_2,\tau_1)]\bm{U}^{\dag}(t,\tau_2),
\label{rho_tilde}
\end{align}
while $\bm{U}_{ij}(t,t_0)=
\begin{pmatrix}
\langle \{a_{i}(t),a^{\dag}_{j}(t_0)\}\rangle && \langle \{a_{i}(t),a_{j}(t_0)\}\rangle \\
\langle \{a^{\dag}_{i}(t),a^{\dag}_{j}(t_0)\}\rangle && \langle \{a^{\dag}_{i}(t),a_{j}(t_0)\}\rangle
\end{pmatrix}$ is the extended retarded Green function matrix that 
obeys the following differential-integral equation,
\begin{align}
\frac{d}{dt}\bm{U}(t,t_0)+\!\frac{i}{\hbar}\!
&\begin{pmatrix}
\bm{\epsilon} && \bm{0} \\
\bm{0} && -\bm{\epsilon}
\end{pmatrix}
\bm{U}(t,t_0) \notag\\
&+\!\frac{1}{\hbar^2}\!\int^{t}_{t_0}\!d\tau\bm{g}^{+}(t,\tau)\bm{U}(\tau,t_0)\!=\!0.
\label{U_eq}
\end{align}
The time-correlation functions  $\bar{\bm{g}}(\tau_1,\tau_2)=\sum_{\alpha}\bar{\bm{g}}_{\alpha}(\tau_1,\tau_2)$, $\bm{g}^{\pm}(t,\tau)=\sum_{\alpha}\bm{g}^{\pm}_{\alpha}(t,\tau)$ and $\tilde{\bm{g}}^{\pm}(\tau_1,\tau_2)=\sum_{\alpha}\tilde{\bm{g}}^{\pm}_{\alpha}(\tau_1,\tau_2)$ describe the system-leads correlations that encompass all the initial correlations of the system and the leads and between them. 
They can be expressed in terms of spectral density matrices 
\begin{subequations}
\label{integral kernel}
\begin{align}
\bm{g}^{\pm}_{\alpha}(t,\tau)\!\!=&\!\!\int\!\!\frac{d\omega}{2\pi}e^{-i(\omega+\frac{U_\alpha}{\hbar})(t-\tau)}\bm{\mathcal{J}}^{\pm}_{\alpha}(\omega) \\
\tilde{\bm{g}}^{\pm}_{\alpha}(t,\tau)\!\!=&\!\!\int\!\!\frac{d\omega}{2\pi}e^{-i(\omega+\frac{U_\alpha}{\hbar})(t-\tau)}
\tilde{\bm{\mathcal{J}}}^{\pm}_{\alpha}(\omega) \\
\bar{\bm{g}}_{\alpha}(t,\tau)\!\!=&-2i\!\!\int\!\!\frac{d\omega}{2\pi}\begin{pmatrix}
e^{-i(\omega+\frac{U_\alpha}{\hbar})(t-t_0)}&0\\
0&e^{i(\omega+\frac{U_\alpha}{\hbar})(t-t_0)}
\end{pmatrix}\notag\\
&\times\tilde{\bm{\mathcal{J}}}^{+}_{\alpha}(\omega)\bm{\chi}(\omega)\delta(\tau-t_0),
\end{align}
\end{subequations}
where $\bm{\chi}(\omega)$ is given by Eq.~(\ref{chi}). The spectral density matrices are given by $\bm{\mathcal{J}}^{\pm}_{\alpha}(\omega)=\bm{\mathcal{J}}_{\alpha}(\omega)\pm\bm{\mathcal{J}}'_{\alpha}(-\omega)$ and $\tilde{\bm{\mathcal{J}}}^{\pm}_{\alpha}(\omega)=n_{\alpha}(\omega)\bm{\mathcal{J}}_{\alpha}(\omega)\pm[1-n_{\alpha}(-\omega)]\bm{\mathcal{J}}'_{\alpha}(-\omega)$. In the partition-free scheme, $n_{\alpha}(\omega)$ is given by Eq.~(\ref{PFN}). The spectral density matrices are given by 
\begin{subequations}
\begin{align}
\bm{\mathcal{J}}_{\alpha}(\omega)&=
\begin{pmatrix}
\bm{J}_{\alpha}(\omega) && \bar{\bm{J}}_{\alpha}(\omega) \\
\bar{\bm{J}}^{\dag}_{\alpha}(\omega) && \hat{\bm{J}}_{\alpha}(\omega)
\end{pmatrix}  \\
\bm{\mathcal{J}}'_{\alpha}(\omega)&=
\begin{pmatrix}
\hat{\bm{J}}_{\alpha}(\omega) && \bar{\bm{J}}^{\dag}_{\alpha}(\omega) \\
\bar{\bm{J}}_{\alpha}(\omega) && \bm{J}_{\alpha}(\omega)
\end{pmatrix},
\end{align} 
\end{subequations}
with the matrix elements 
\begin{subequations}
\label{spectral density matrices}
\begin{align}
J&_{\alpha ij}(\omega)=\kappa^*_{\alpha i}\kappa_{\alpha j}J_{0\alpha}(\omega)  \\
\hat{J}&_{\alpha ij}(\omega)=\kappa'^*_{\alpha i}\kappa'_{\alpha j}J_{0\alpha}(\omega) \\
\bar{J}&_{\alpha ij}(\omega)=\kappa^*_{\alpha i}\kappa'_{\alpha j}J_{0\alpha}(\omega),
\end{align}
\end{subequations}
and $J_{0\alpha}(\omega)=2\pi\sum_{k}|\eta_{\alpha k}|^2\delta(\omega-\epsilon_{\alpha k}/\hbar)$, which closely lies on the wave-function structures of the system and the terminal states.

On the other hand, in the partitioned scheme, the central system is initially decoupled from the leads, namely $\rho_{tot}(t_0)=\rho_{S}(t_0)\otimes\rho_{E}(t_0)$. We have derived the exact master equation in this case by using the Feynman-Vernon influence functional in the coherent state representation (see Appendix \ref{App1} for the detailed derivation). The exact master equation and the current $I_{\alpha}$ can be written in the same form as Eq.~(\ref{ME}) and Eq.~(\ref{I_alpha0}) respectively. In this case, the extended particle reduced density matrix is given by
\begin{align}
\tilde{\bm{\rho}}(\tau,t)=& \bm{U}(\tau,t_0)\tilde{\bm{\rho}}(t_0,t_0)\bm{U}^{\dag}(t,t_0) \notag\\
&+\frac{1}{\hbar^2}\!\!\int^{\tau}_{t_0}\!\!d\tau_1\!\!\int^{t}_{t_0}\!\!d\tau_2
\bm{U}(\tau,\tau_1)\tilde{\bm{g}}^{+}(\tau_1,\tau_2)\bm{U}^{\dag}(t,\tau_2) .
\end{align}
In the partitioned scheme, the lead $\alpha$ are initially biased with $n_{\alpha}(\omega)=\frac{1}{e^{(\omega-\mu_{\alpha})/k_BT_{\alpha}}+1}$ for $U_{\alpha}=0$ [see Eq. (\ref{integral kernel})].

\section{Transport dynamics of hybrid superconducting systems}
\label{Sec2}
\subsection{General discussion}
The transient transport current $I_{\alpha}(t)$ [see Eq. (\ref{I_alpha0})] can be expressed as the combination of contributions flowing into the particle and hole channel of the lead $\alpha$, namely
\begin{align}
I_{\alpha}(t)=I_{\alpha p}(t)-I_{\alpha h}(t). 
\label{I_alpha}
\end{align}
Notice that the minus sign comes from the correlation functions $\bm{g}^{-}_{\alpha}(t,\tau)$ and $\tilde{\bm{g}}^{-}_{\alpha}(t,\tau)$ in Eq. (\ref{I_alpha0}). By utilizing the following identity,
\begin{align}
\bm{U}^{\dag}(t,\tau)&=\bm{U}(\tau,t_0)\bm{U}^{\dag}(t,t_0) \notag\\
&+\!\frac{1}{\hbar^2}\!\int^{\tau}_{t_0}\!d\tau_1\!\int^{t}_{t_0}\!d\tau_2\bm{U}(\tau,\tau_1)\bm{g}^{+}(\tau_1,\tau_2)\bm{U}^{\dag}(t,\tau_2), 
\end{align}
the current contributions can further be decomposed into components that describe coherent transports through different paths, 
\begin{align}
I_{\alpha\sigma}(t)&=I_{\alpha\sigma0}(t)+\sum_{\beta\sigma'}I_{\alpha\sigma,\beta\sigma'}(t)+\sum_{\beta\sigma'}\bar{I}_{\alpha\sigma,\beta\sigma'}(t),
\label{I_as_PF}
\end{align}
where $\sigma,\sigma'=p,h$ indicate the particle or hole channels, and
\begin{subequations}\label{I_ph_LR:0}
\begin{align}
I&_{\alpha\sigma0}(t)
=\frac{-e}{\hbar^2}\!\!\int^{t}_{t_0}\!\!d\tau{\rm Tr}\Big[\bm{g}^{\sigma}_{\alpha}(t,\tau)\bm{U}(\tau,t_0)\tilde{\bm{\rho}}(t_0,t_0)\bm{U}^{\dag}(t,t_0) \notag\\
&\;\;\;\;\;\;\;\;-\tilde{\bm{g}}^{\sigma}_{\alpha}(t,\tau)\bm{U}(\tau,t_0)\bm{U}^{\dag}(t,t_0)+\!\!H.c.\Big] \label{I_ph_LR:1}\\
I&_{\alpha\sigma,\beta\sigma'}(t) \notag\\
&\!\!\!\!=\!\!\frac{-e}{\hbar^4}\!\!\int^{t}_{t_0}\!\!d\tau\!\!\int^{\tau}_{t_0}\!\!d\tau_1\!\!\int^{t}_{t_0}\!\!d\tau_2
{\rm Tr}\Big[\bm{g}^{\sigma}_{\alpha}(t,\!\tau)\bm{U}(\tau,\!\tau_1)\tilde{\bm{g}}^{\sigma'}_{\beta}(\tau_1,\!\tau_2)\bm{U}^{\dag}(t,\!\tau_2) \notag\\
&-\tilde{\bm{g}}^{\sigma}_{\alpha}(t,\tau)\bm{U}(\tau,\tau_1)\bm{g}^{\sigma'}_{\beta}(\tau_1,\tau_2)\bm{U}^{\dag}(t,\tau_2)\!\!+\!\!H.c.\Big] \label{I_ph_LR:2}\\ 
\bar{I}&_{\alpha\sigma,\beta\sigma'}(t)=\frac{-e}{\hbar^4}\int^{t}_{t_0}d\tau\int^{\tau}_{t_0}d\tau_1\int^{t}_{t_0}d\tau_2{\rm Tr}\Big[\bm{g}^{\sigma}_{\alpha}(t,\tau)\bm{U}(\tau,\tau_1) \notag\\
&\times[\bar{\bm{g}}^{\sigma'}_{\beta}(\tau_1,\tau_2)+\bar{\bm{g}}^{\dag\sigma'}_{\beta}(\tau_2,\tau_1)]\bm{U}^{\dag}(t,\tau_2)+H.c.\Big], \label{I_ph_LR:3}
\end{align}
\end{subequations}
with the system-lead time-correlations through the particle and hole channels,
\begin{subequations}
\begin{align}
\bm{g}^{p}_{\alpha}(t,\tau)\!\!=&\!\!\int\!\!\frac{d\omega}{2\pi}e^{-i(\omega+\frac{U_\alpha}{\hbar})(t-\tau)}\bm{\mathcal{J}}_{\alpha}(\omega) \\
\bm{g}^{h}_{\alpha}(t,\tau)\!\!=&\!\!\int\!\!\frac{d\omega}{2\pi}e^{-i(\omega+\frac{U_\alpha}{\hbar})(t-\tau)}\bm{\mathcal{J}}'_{\alpha}(-\omega) \\
\tilde{\bm{g}}^{p}_{\alpha}(t,\tau)\!\!=&\!\!\int\!\!\frac{d\omega}{2\pi}e^{-i(\omega+\frac{U_\alpha}{\hbar})(t-\tau)}n_{\alpha}(\omega)
\bm{\mathcal{J}}_{\alpha}(\omega) \\
\tilde{\bm{g}}^{h}_{\alpha}(t,\tau)\!\!=&\!\!\int\!\!\frac{d\omega}{2\pi}e^{-i(\omega+\frac{U_\alpha}{\hbar})(t-\tau)}[1\!-\!n_{\alpha}(-\omega)]
\bm{\mathcal{J}}'_{\alpha}(-\omega) \\
\bar{\bm{g}}^p_{\alpha}(t,\tau)\!\!=&\!-\!2i\delta(\tau\!-\!t_0)\!\!\int\!\!\dfrac{d\omega}{2\pi}e^{-i(\omega+\frac{U_\alpha}{\hbar})(t-t_0)}\bm{\mathcal{J}}_\alpha(\omega)\bm{\chi}(\omega)\\
\bar{\bm{g}}^h_{\alpha}(t,\tau)\!\!=&\!-\!2i\delta(\tau\!-\!t_0)\!\!\int\!\!\dfrac{d\omega}{2\pi}e^{i(\omega+\frac{U_\alpha}{\hbar})(t-t_0)}\bm{\mathcal{J}'}_\alpha(-\omega)\bm{\chi}(\omega).
\end{align}
\end{subequations}
The current $I_{\alpha\sigma0}(t)$ describes the coherent transport between the system and the channel $\sigma$ of lead $\alpha$. The contribution of these processes will eventually decay to zero in the steady-state limit if there is no localized bound state \cite{ZLX12,Z19}. This is because $U(t\rightarrow\infty,t_0)=0$ if no localized bound state exists. On the other hand, both currents $I_{\alpha\sigma,\beta\sigma'}(t)$ and $\bar{I}_{\alpha\sigma,\beta\sigma'}(t)$ describe the coherent transport between the channel $\sigma'$ of lead $\beta$ and the channel $\sigma$ of lead $\alpha$, where $\bar{I}_{\alpha\sigma,\beta\sigma'}(t)$ is caused by the initial system-lead correlations through the correlation function $\bar{\bm{g}}^{\sigma'}_{\beta}(\tau_1,\tau_2)$. Note that in the partitioned scheme, $\bar{I}_{\alpha\sigma,\beta\sigma'}(t)=0$.

In the steady-state limit, the coherent transport current $I_{\alpha\sigma,\beta\sigma'}(t)$ can be explicitly expressed as follows,
\begin{subequations}
\label{I_infty}
\begin{align}
I_{\alpha p,\beta p}(t\rightarrow\infty)
&=\frac{-e}{\hbar^4}\int d\omega\bm{\mathcal{T}}^{pp}_{\alpha\beta}(\omega)[n_{\beta}(\omega)-n_{\alpha}(\omega)] \\
I_{\alpha h,\beta h}(t\rightarrow\infty)
&=\frac{-e}{\hbar^4}\int d\omega\bm{\mathcal{T}}^{hh}_{\alpha\beta}(\omega)[n_{\alpha}(-\omega)-n_{\beta}(-\omega)] \\
I_{\alpha p,\beta h}(t\rightarrow\infty)
&=\frac{-e}{\hbar^4}\int d\omega\bm{\mathcal{T}}^{ph}_{\alpha\beta}(\omega)[1\!\!-\!\!n_{\beta}(-\omega)\!\!-\!\!n_{\alpha}(\omega)] \notag \\
&=-I_{\beta h,\alpha p}(t\rightarrow\infty),
\end{align}
\end{subequations}
where $\tilde{\bm{U}}(\omega)=\int^{\infty}_{t_0}d\tau\bm{U}(\tau,t_0)e^{-i\omega(\tau-t_0)}$ is the modified Laplace transform of $\bm{U}(\tau,t_0)$ and the transmission matrices are given by
\begin{subequations}
\begin{align}
\bm{\mathcal{T}}^{pp}_{\alpha\beta}(\omega)&\!=\!{\rm Tr}\Big[\bm{\mathcal{J}}_{\alpha}(\omega)\tilde{\bm{U}}(\omega\!+\!\frac{U_\alpha}{\hbar})
\bm{\mathcal{J}}_{\beta}(\omega)\tilde{\bm{U}}^{\dag}(\omega\!+\!\frac{U_\beta}{\hbar})\!+\!H.c.\Big] \\
\bm{\mathcal{T}}^{hh}_{\alpha\beta}(\omega)&\!=\!{\rm Tr}\Big[\bm{\mathcal{J}}'_{\alpha}(\!-\!\omega)\tilde{\bm{U}}(\omega\!\!+\!\!\frac{U_\alpha}{\hbar})
\bm{\mathcal{J}}'_{\beta}(\!-\!\omega)\tilde{\bm{U}}^{\dag}(\omega\!\!+\!\!\frac{U_\beta}{\hbar})\!+\!H.c.\Big] \\
\bm{\mathcal{T}}^{ph}_{\alpha\beta}(\omega)&\!=\!{\rm Tr}\Big[\bm{\mathcal{J}}_{\alpha}(\omega)\tilde{\bm{U}}(\omega\!+\!\frac{U_\alpha}{\hbar})
\bm{\mathcal{J}}'_{\beta}(\!-\!\omega)\tilde{\bm{U}}^{\dag}(\omega\!+\!\frac{U_\beta}{\hbar})\!+\!H.c.\Big] \notag\\
&=\bm{\mathcal{T}}^{hp}_{\beta\alpha}(\omega). \\ \notag
\end{align}
\end{subequations}
It is obvious that $I_{\alpha p,\alpha p}(t\rightarrow\infty)=I_{\alpha h,\alpha h}(t\rightarrow\infty)=0$. 

The current components $I_{\alpha p,\beta p}$ ($I_{\alpha h,\beta h}$) and $I_{\alpha p,\beta h}$ ($I_{\alpha h,\beta p}$) are commonly called respectively the normal transmission and the Andreev reflection in the scattering matrix theory. However, this picture may be misleading in the transient regime. Notice that the Green function matrix $\bm{U}(t,t_0)$ has taken into account all the system-lead tunnelings through the spectral density matrices [see Eq. (\ref{U_eq})], which is not a free propagator of the system. In order to clarify the transient physical processes, we expand the Green function matrix $\bm{U}(t,t_0)$ with respect to the system-lead time correlation functions $\bm{g}$ ($\bm{g}\propto|\eta|^2$) order by order,
\begin{align}
\bm{U}&(t,t_0)=\bm{U}_0(t,t_0) \notag\\
&\!\!-\!\!\frac{1}{\hbar^2}\!\!\int^{t}_{t_0}\!\!d\tau_{1}\!\!\int^{\tau_1}_{t_0}\!\!\!\!d\tau_2
\bm{U}_0(t,\tau_1)\bm{g}^{+}(\tau_1,\tau_2)\bm{U}_0(\tau_2,t_0)\!+\cdots, 
\end{align}
where $\bm{U}_0(t,t_0)=\exp{-\frac{i}{\hbar}H_{S}(t-t_0)}$ is the free propagator of the system. Likewise, we can expand $I_{\alpha\sigma,\beta\sigma'}(t)$ in the same way,
\begin{align}
I&_{\alpha\sigma,\beta\sigma'}(t)=I^{(2)}_{\alpha\sigma,\beta\sigma'}(t)+I^{(3)}_{\alpha\sigma,\beta\sigma'}(t)+\cdots,
\label{I_per}
\end{align}
where 
\begin{widetext}
\begin{subequations}
\label{Ipe}
\begin{align}
I^{(2)}_{\alpha\sigma,\beta\sigma'}(t)&=\frac{-e}{\hbar^4}\int^{t}_{t_0}\!\!d\tau\int^{\tau}_{t_0}\!\!d\tau_1\int^{t}_{t_0}\!\!d\tau_2
{\rm Tr}\Big[\bm{g}^{\sigma}_{\alpha}(t,\!\tau)\bm{U}_0(\tau,\!\tau_1)\tilde{\bm{g}}^{\sigma'}_{\beta}(\tau_1,\!\tau_2)\bm{U}^{\dag}_0(t,\!\tau_2)
-\tilde{\bm{g}}^{\sigma}_{\alpha}(t,\tau)\bm{U}_0(\tau,\tau_1)\bm{g}^{\sigma'}_{\beta}(\tau_1,\tau_2)\bm{U}^{\dag}_0(t,\tau_2)+H.c.\Big] \label{I2}\\
I^{(3)}_{\alpha\sigma,\beta\sigma'}(t)&=\sum_{\sigma''\alpha'}
\frac{e}{\hbar^6}\int^{t}_{t_0}\!\!\!d\tau\int^{\tau}_{t_0}\!\!\!d\tau_1\int^{t}_{t_0}\!\!\!d\tau_2\int^{\tau}_{\tau_1}\!\!\!d\tau_3\int^{\tau_3}_{\tau_1}\!\!\!d\tau_4
{\rm Tr}\Big[\bm{g}^{\sigma}_{\alpha}(t,\!\tau)\bm{U}_0(\tau,\tau_3)\bm{g}^{\sigma''}_{\alpha'}(\tau_3,\tau_4)\bm{U}_0(\tau_4,\tau_1)
\tilde{\bm{g}}^{\sigma'}_{\beta}(\tau_1,\!\tau_2)\bm{U}^{\dag}_0(t,\!\tau_2) \notag\\
&\qquad\qquad\qquad\qquad\qquad\qquad\qquad\qquad
-\tilde{\bm{g}}^{\sigma}_{\alpha}(t,\tau)\bm{U}_0(\tau,\tau_3)\bm{g}^{\sigma''}_{\alpha'}(\tau_3,\tau_4)\bm{U}_0(\tau_4,\tau_1)
\bm{g}^{\sigma'}_{\beta}(\tau_1,\tau_2)\bm{U}^{\dag}_0(t,\tau_2)+H.c.\Big] \notag\\
&+\sum_{\sigma''\alpha'}
\frac{e}{\hbar^6}\int^{t}_{t_0}\!\!\!d\tau\int^{\tau}_{t_0}\!\!\!d\tau_1\int^{t}_{t_0}\!\!\!d\tau_2\int^{t}_{\tau_2}\!\!\!d\tau_3\int^{\tau_3}_{\tau_2}\!\!\!d\tau_4
{\rm Tr}\Big[\bm{g}^{\sigma}_{\alpha}(t,\!\tau)\bm{U}_0(\tau,\!\tau_1)\tilde{\bm{g}}^{\sigma'}_{\beta}(\tau_1,\!\tau_2)
\bm{U}^{\dag}_0(\tau_4,\tau_2)[\bm{g}^{\sigma''}_{\alpha'}]^{\dag}(\tau_3,\tau_4)\bm{U}^{\dag}_0(t,\tau_3) \notag\\
&\qquad\qquad\qquad\qquad\qquad\qquad\qquad\qquad
-\tilde{\bm{g}}^{\sigma}_{\alpha}(t,\tau)\bm{U}_0(\tau,\tau_1)\bm{g}^{\sigma'}_{\beta}(\tau_1,\tau_2)
\bm{U}^{\dag}_0(\tau_4,\tau_2)[\bm{g}^{\sigma''}_{\alpha'}]^{\dag}(\tau_3,\tau_4)\bm{U}^{\dag}_0(t,\!\tau_3)+H.c.\Big]. \label{I3}
\end{align}
\end{subequations}
\end{widetext}
Here, only the first two leading-order contributions are shown. 

\begin{figure}
\begin{center}
\includegraphics[width=0.48\textwidth]{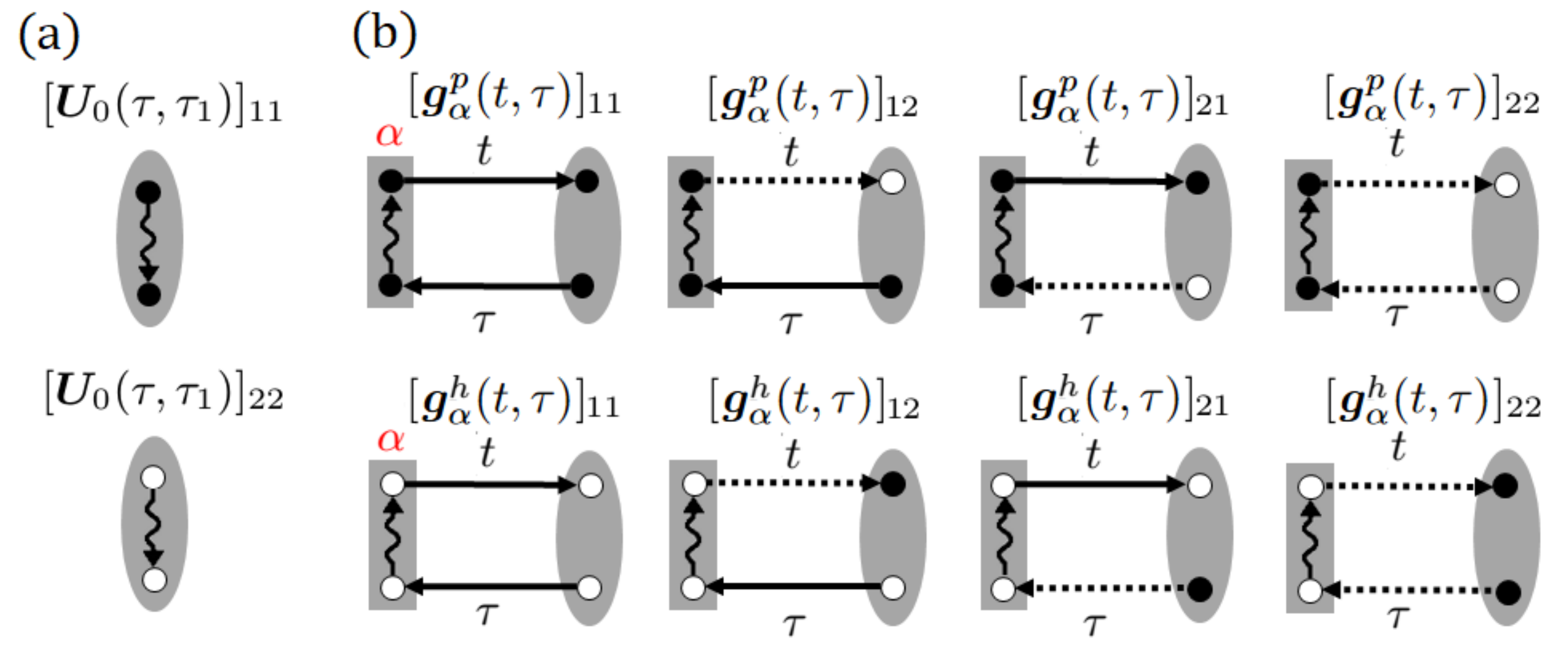}
\caption{(Color online) Feynman diagrams of  (a) the free propagator of the system $\bm{U}_0(\tau,\tau_1)$ for particles and holes, respectively, (b) the system-lead time-correlation functions $\bm{g}^{p}_{\alpha}(t,\tau)$ and $\bm{g}^{h}_{\alpha}(t,\tau)$.}
\label{example}
\end{center}
\end{figure}

In the following discussions, we focus on a central system coupled to two leads (left lead $L$ and right lead $R$). 
Figure \ref{example} is the basic Feynman diagrams for particle and hole free propagators $\bm{U}_0(\tau,\tau_1)$, respectively, of the system and the system-lead time-correlation functions $\bm{g}^{p}_{\alpha}(t,\tau)$ and $\bm{g}^{h}_{\alpha}(t,\tau)$ 
in the expansions of the transient transport currents of Eq.~(\ref{I_as_PF}) given by Eq.~(\ref{I_per}) and Eq.~(\ref{Ipe}).  Note that in this transport current obtained from the
exact master equation or from the nonequilibrium Green function technique, the lead (reservoir) degrees of freedoms have been
completely integrated out. The current is determined only by the particle/hole propagating functions of the system plus the system-lead 
correlations (also called as the self-energy correlation functions to the system). It is the system-lead correlations show how the particle and hole channels are opened between the system and leads. 
If one does not look at the physical processes happened in the system and only pay attension on the measured current at time $t$, then the resulting current $I_\alpha(t)$ only shows apparently the electrons (holes) transferring from lead $\alpha$ into the system at time $t$. But the real physical processes are much more complicated, as shown by Fig.~\ref{example}b.

More specifically, the correlation function $\bm{g}^{p}_{\alpha}(t,\tau)$ contains 
the processes of a quasiparticle transferring from the system to lead $\alpha$ at time $\tau$, propagating freely in the lead from time $\tau$ to time $t$ and then transferring back to the system at time $t$, as shown by the first diagram in Fig.~\ref{example}b.  But it also contains other three processes (corresponding to the other three diagrams in top of Fig.~\ref{example}b): a quasiparticle transfers into the lead at time $\tau$, propagates freely in the lead, and then is annihilated with another quasiparticle (or a quasihole is created) in the system at time $t$; a quasihole is annihilated (or a quasiparticle is created) in the system accompanied with a quasiparticle creating in the lead at the same time $\tau$, the quasiparticle propagates freely in the lead and then is transferred into the system at time $t$; a quasihole is annihilated (or a quasiparticle created) in the system with another quasiparticle created in the lead at the same time $\tau$, then the quasiparticle propagates freely in the lead and is annihilated with a quasiparticle (or a quasihole is created) in the system at time $t$. The later three processes are induced by the superconductivity of the system, as shown in the tunneling Hamiltonian of Eq.~(\ref{H_tot}) after the Bogoliubov transformation is performed. But if one pays attention only on the current at time $t$, then only the part of electron transferring from lead $\alpha$ to the system at time $t$ is observed, the electron dynamics before time $t$ is hardly manifested in the scattering matrix theory, for example.

\begin{figure*}
\begin{center}
\includegraphics[width=0.8\textwidth]{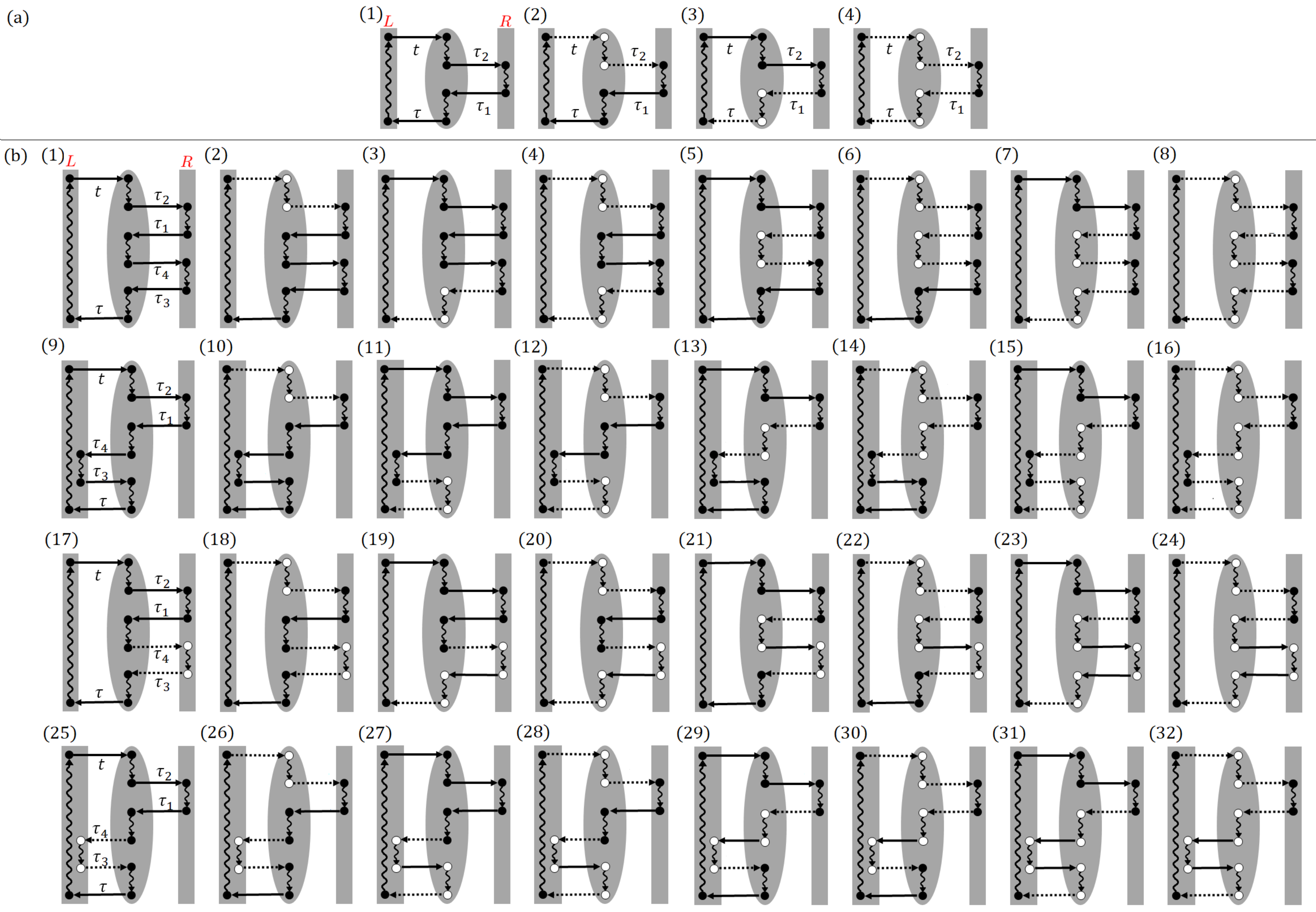}
\caption{(Color online) Representative diagrams of (a) the second-order processes in $I^{(2)}_{Lp,Rp}$ and (b) the third-order processes in $I^{(3)}_{Lp,Rp}$. The black circle represents a particle and the white circle represents a hole. Only diagrams of the first two lines of Eq. (\ref{I3}) are shown, other diagrams can be drawn similarly.}
\label{Ipp}
\end{center}
\end{figure*}

\begin{figure*}
\centering
\includegraphics[width=0.8\textwidth]{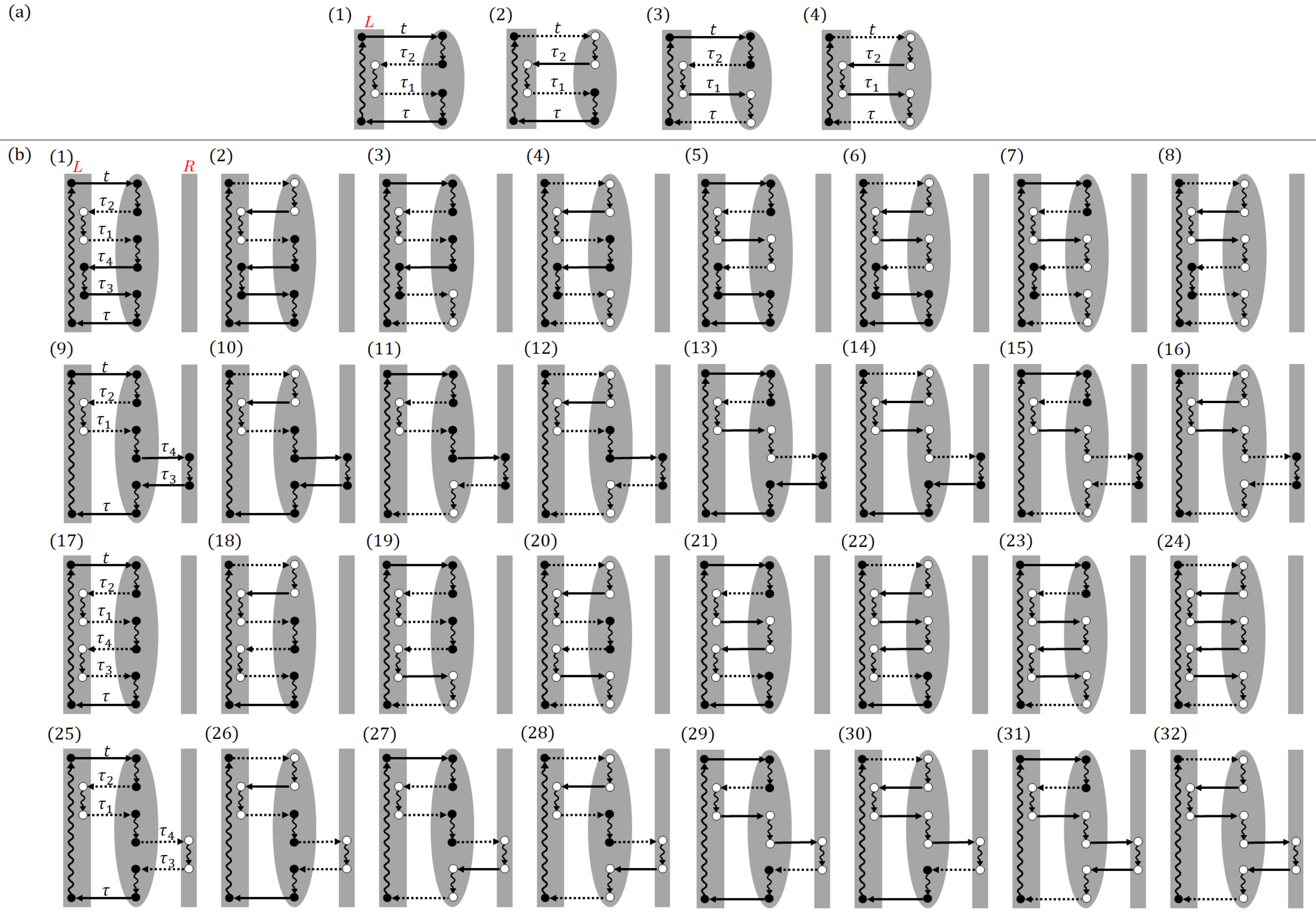}
\caption{(Color online) Representative diagrams of (a) the second-order processes in $I^{(2)}_{Lp,Lh}$ and (b) the third-order processes in $I^{(3)}_{Lp,Lh}$. The black circle represents a particle and the white circle represents a hole. Only diagrams of the first two lines of Eq. (\ref{I3}) are shown, other diagrams can be drawn similarly.}
\label{Iph}
\end{figure*}

To see the detailed processes of electrons transfer through the superconductor in the transient regime, we show diagrammatically 
in Figs.~\ref{Ipp} and \ref{Iph} a part of the first two leading-order contributions to the transport current of Eq.~(\ref{I_as_PF}).  The lowest-order contributions shown in Fig. \ref{Ipp}a and Fig. \ref{Iph}a are the second-order processes (with respect to $\bm{g}$) to the current, given by $I^{(2)}_{Lp,Rp}(t)$ of Eq.~(\ref{Ipe}). It involves four processes of two particle/hole exchanges between the two leads through the system [characterized by two $\bm{g}$-functions, see Eq.~(\ref{I2})]. Explicitly, Fig.~\ref{Ipp}(a1) depicts the process of particle transmissions between lead $L$ and lead $R$ through the two particle channels of the system, while Fig.~\ref{Ipp}(a4) depicts the process of particle transmissions between the leads through two hole channels of the system. On the other hand, Fig.~\ref{Ipp}(a2) and Fig.~\ref{Ipp}(a3) depict the processes of particle transmissions between the leads through a combination of both the particle channel and hole channel of the system. However, with respect to the current $I_{Lp}(t)$ which measures the current passing the lead $L$ at time $t$, these four processes depicted by Fig.~\ref{Ipp}a correspond to two different physical processes. One is the normal particle transmission from lead $L$ to the system, 
contributed with Fig.~\ref{Ipp}(a1) and Fig.~\ref{Ipp}(a3), as a resultant particle transmission between lead $L$ and lead $R$ through a combination of the particle and hole channels of the system. The other one is the particle (hole) pair production/annihilation  
of the lead and the system, respectively, contributed with Fig.~\ref{Ipp}(a2) and Fig.~\ref{Ipp}(a4), which is also  a resultant particle transmission between lead $L$ and lead $R$ through a combination of the particle and hole channels of the system (but the particle channels and hole channels are exchanged), as shown in Fig.~\ref{Ipp}a. Thus, it is difficult both theoretically and experimentally to distinguish the contributions from the particle transmission or the pair production/annihilation. 
Likewise, Fig.~\ref{Iph}a depicts processes that a particle in lead $L$ is transmitted to the system and back to in the same lead through different channels of the system. All the four processes plotted in Fig. \ref{Iph}a involve the resultant combinations of
particle transport through the mixture of the particle and hole channels of the system. This cannot be depicted by the scattering matrix in terms of the simple picture of normal particle transmission and Andreev reflection.

The next-order contributions are the third-order processes which corresponds to three particle/hole exchanges between the leads and the system [see Eq.~(\ref{I3})], as shown in Fig.~\ref{Ipp}b and Fig.~\ref{Iph}b. These processes are much more complicated and certainly cannot be rendered as a simple normal transmission or Andreev reflection in the scattering matrix theory. For example, Fig.~\ref{Ipp}b depict processes involved by the third-order contribution $I^{(3)}_{Lp,Rp}(t)$. Figure \ref{Ipp}(b1)-(b8) depict processes resulting from the mixing of two normal transmissions between the two leads. Furthermore, Fig.~\ref{Ipp}(b17)-(b24) depict processes resulting from the mixing of a normal transmission and a cross Andreev reflection between the two leads. Similarly, Fig.~\ref{Iph}b depicts processes involved by $I^{(3)}_{Lp,Rh}(t)$, which also result from the mixing of the normal transmissions and Andreev reflections (or cross Andreev reflections) between the leads. 

Obviously, the higher-order contributions in the expansion Eq.~(\ref{I_per}) consist of more complicated mixtures of numerous normal transmissions and Andreev reflections between the leads and the system. As a result, the resultant coherent transport $I_{L\sigma,R\sigma'}(t)$ between lead $L$ and lead $R$ is the total sum of all order contributions, and consists of numerous but more complicated processes that cannot be rendered simply as normal transmissions and Andreev reflections. Therefore, the simple picture of normal transmission and Andreev reflection based on the scattering matrix theory is not applicable to the transient quantum transport processes. In fact, the coherent transport currents Eq.~(\ref{I_ph_LR:0}) are the resultant interferences of all the tunneling processes between the particle/hole channels of the leads and the system. Of course, if there is no superconductivity involved, all the hole channels do not occurs, and our theory reproduces the Meir-Wingreen formula based on the nonequilibrium Green function technique, as we have shown in our previous work \cite{JTZ10,Yang2017}. In the steady-state limit, it also reproduce the Landauer-B\"{u}ttiker formula based on the scattering matrix theory \cite{JTZ10,Yang2017} 

\subsection{Coherent transports through Majorana zero modes}
\begin{figure}
\begin{center}
\includegraphics[width=0.42\textwidth]{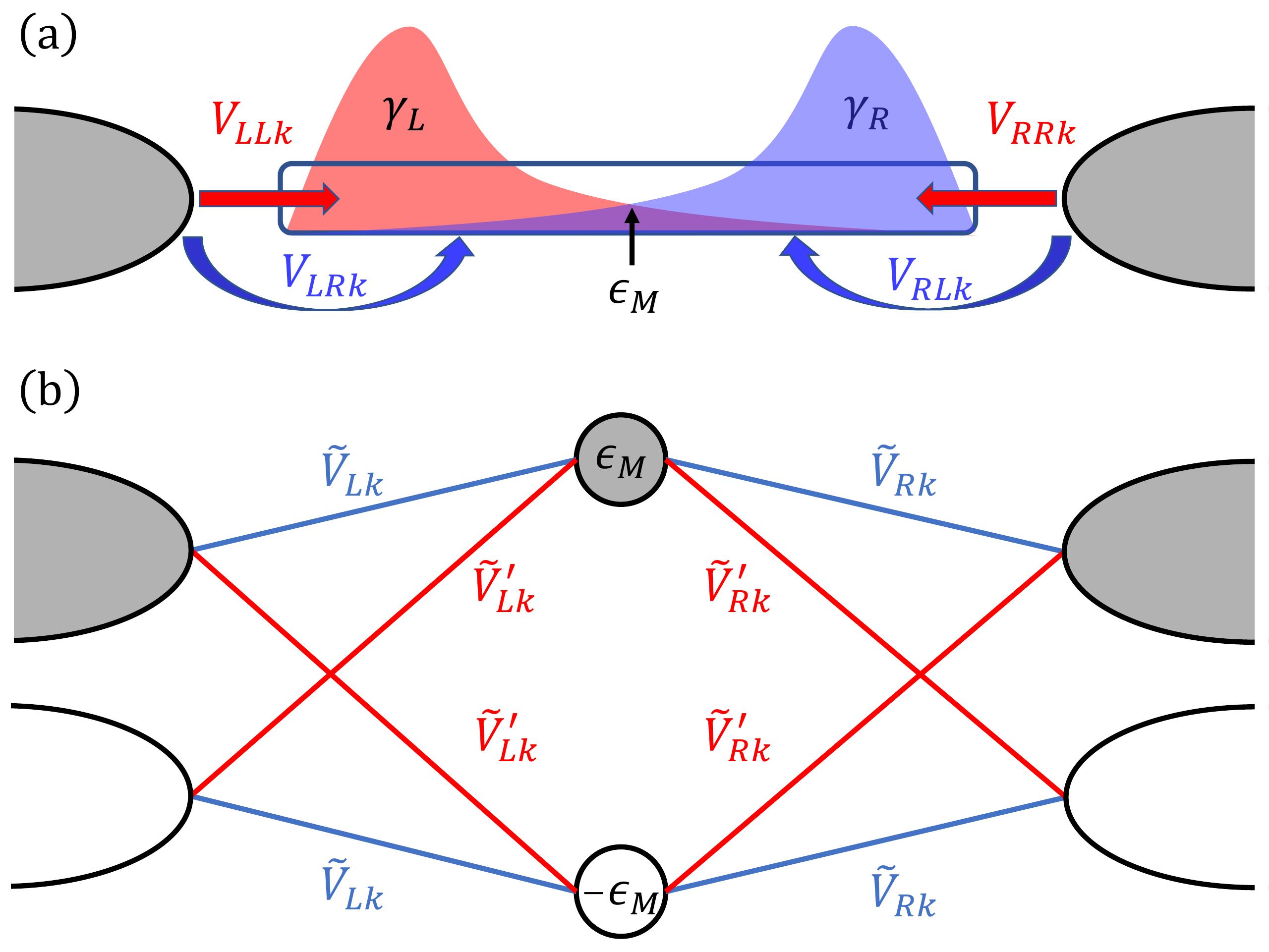}
\caption{(Color online) A schematic plot of the two-terminal MZM system. (a) The colored profiles labeled with $\gamma_{L,R}$ represent the wave-function distributions of the left and the right MZMs. In the MZM basis,  $\epsilon_M$ depicts the wave-function overlap of two MZMs, $V_{LLk}$ ($V_{RRk}$) is the tunneling coupling between the left (right) lead and the left (right) MZM, and $V_{LRk}$ ($V_{RLk}$) is the cross coupling between the left (right) lead and the right (left) MZM. (b) In the bogoliubon basis,  $\tilde{V}_{\alpha k}$ is the particle-particle (or hole-hole) tunneling coupling between the system and lead $\alpha$, and $\tilde{V}'_{\alpha k}$ is the particle-hole tunneling coupling between the system and lead $\alpha$.}
\label{H_int}
\end{center}
\end{figure}

In the following, we will apply our transport theory to the system consisting of a pair of MZMs at the ends of the wire and discuss the coherent transport dynamics through the left and the right leads coupled to the system. With only the pair of MZMs being considered, the total Hamiltonian can be written as,
\begin{align} 
H_{tot}=&\epsilon_M(a^{\dag}_0a_0-\frac{1}{2})+\sum_{\alpha k}[\epsilon_{\alpha k}+U_{\alpha}(t)]b^{\dag}_{\alpha k}b_{\alpha k} \notag\\
&+\sum_{\alpha k}\eta_{\alpha k}(\kappa_{\alpha 0}b^{\dag}_{\alpha k}a_0+\kappa'_{\alpha 0}b^{\dag}_{\alpha k}a^{\dag}_0+H.c.), 
\label{mzmH}
\end{align}
where $a_0$ ($a^{\dag}_0$) is the zero-mode bogoliubon operator, the non-zero $\epsilon_M$ is caused by the wave-function overlap between the two separated MZMs and $b_{\alpha k}$ ($b^{\dag}_{\alpha k}$) is the annihilation (creation) operator of the lead $\alpha=L, R$. Again, we apply the bias voltage $U_{\alpha}(t)=U_{\alpha}\Theta(t-t_0)$ to the two leads for the partition-free scheme and let $U_{\alpha}=0$ for the partitioned scheme. The system Hamiltonian and the tunneling Hamiltonian can be rewritten in terms of Majorana operators $\gamma_L=a_0+a^{\dag}_0$ and $\gamma_R=-i(a_0+a^{\dag}_0)$, 
\begin{subequations}
\begin{align}
H_{S}=& \frac{i}{2}\epsilon_M\gamma_L\gamma_R  \\
H_{T}=& \!\sum_{k}\frac{\eta_{Lk}}{2}\big[(\kappa_{L0}\!+\!\kappa'_{L0})b^{\dag}_{Lk}\gamma_L
\!+\!i(\kappa_{L0}\!-\!\kappa'_{L0})b^{\dag}_{Lk}\gamma_R\big] \notag\\
&+\!\sum_{k}\frac{\eta_{Rk}}{2}\big[(\kappa_{R0}\!+\!\kappa'_{R0})b^{\dag}_{Rk}\gamma_L
\!+\!i(\kappa_{R0}\!-\!\kappa'_{R0})b^{\dag}_{Rk}\gamma_R\big]\!\! \notag \\
& + \! H.c. 
\end{align}
\end{subequations}
In the literature, this tunneling Hamiltonian is usually given in the following form,
\begin{align}
H_{T}&=\sum_{\alpha k}(\tilde{V}_{\alpha k}b^{\dag}_{\alpha k}a_0+\tilde{V}'_{\alpha k}b^{\dag}_{\alpha k}a^{\dag}_0+H.c.) \notag\\
&=\sum_{k}(V_{LLk}b^{\dag}_{Lk}\gamma_L+V_{LRk}b^{\dag}_{Lk}\gamma_R \notag\\
& \qquad\quad+V_{RLk}b^{\dag}_{Rk}\gamma_L+V_{RRk}b^{\dag}_{Rk}\gamma_R+H.c.) 
\label{H_bov}
\end{align}
The coefficient $V_{\alpha\alpha k}$ is the coupling between the MZM $\gamma_\alpha$ and the lead $\alpha$, and the coefficient $V_{LRk}$ ($V_{RLk}$) depicts the cross coupling between the left (right) lead and the right (left) MZM. In our formalism, it is clear that these couplings and cross couplings are determined by the wave-function structures of the MZMs, which are characterized by $\kappa_{\alpha 0}$ and $\kappa'_{\alpha 0}$ through the following relations, 
\begin{align}
\tilde{V}_{\alpha k}&=\eta_{\alpha k}\kappa_{\alpha 0}, ~~\tilde{V}'_{\alpha k}=\eta_{\alpha k}\kappa'_{\alpha 0}, 
\label{VV'0}
\end{align}
and
\begin{subequations}
\label{VV'}
\begin{align}
V_{LLk}&=\frac{\eta_{Lk}}{2}(\kappa_{L0}+\kappa'_{L0}), ~~V_{LRk}=\frac{i\eta_{Lk}}{2}(\kappa_{L0}-\kappa'_{L0}) \\
V_{RLk}&=\frac{\eta_{Rk}}{2}(\kappa_{R0}+\kappa'_{R0}), ~~V_{RRk}=\frac{i\eta_{Rk}}{2}(\kappa_{R0}-\kappa'_{R0}). 
\end{align}
\end{subequations}
The coupling $V_{LLk}$ ($V_{RRk}$) is determined by the coefficient $(\kappa_{L0}+\kappa'_{L0})$ [$i(\kappa_{R0}-\kappa'_{R0})$], which describes the MZM amplitude of $\gamma_L$ ($\gamma_R$) coupling to the leftmost (rightmost) cite. On the other hand, the cross coupling $V_{LRk}$ ($V_{RLk}$) is determined by the coefficient $i(\kappa_{L0}-\kappa'_{L0})$ [$(\kappa_{R0}+\kappa'_{R0})$], which describes the MZM amplitude of $\gamma_R$ ($\gamma_L$) coupling to its opposite-end cite. In other words, if a MZM is not perfectly localized, i.e. $i(\kappa_{L0}-\kappa'_{L0})\neq0$ and $\kappa_{R0}+\kappa'_{R0}\neq0$, it can be directly coupled to its opposite-end lead (see Fig. \ref{H_int}a). 

Moreover, as shown in Fig. \ref{H_int}b, this tunneling Hamiltonian involves couplings between different channels of quantum states of the leads and the system, respectively. The hopping term $\tilde{V}_{\alpha k}$ describes the tunnelings from the particle (hole) channel of lead $\alpha$ to the quasi-particle (quasi-hole) state of the bogoliubon in the system (see the blue lines in Fig.~{\ref{H_int}}b). The pairing term $\tilde{V}'_{\alpha k}$ describes the tunnelings from the particle (hole) channel of lead $\alpha$ to the quasi-hole (quasi-particle) state of the bogoliubon (see the red lines in Fig.~{\ref{H_int}}b). Therefore, the coherent transport current $I_{L\sigma,R\sigma'}(t)$ is induced from an equivalent interferometer formed by the $\sigma$ channel of the left lead and the $\sigma'$ channel of the right lead via the quasi-particle and quasi-hole states of the bogoliubon in the system. Eq.~(\ref{VV'}) shows that when the MZMs are perfectly delocalized, i.e. $i(\kappa_{L0}-\kappa'_{L0})=0$ and $\kappa_{R0}+\kappa'_{R0}=0$, the cross couplings between the MZMs and the leads vanish, i.e. $V_{LRk}=V_{RLk}=0$. In this case, the coupling coefficients become $\tilde{V}_{Lk}=\tilde{V}'_{Lk}$ and $\tilde{V}_{Rk}=-\tilde{V}'_{Rk}=\tilde{V}'_{Rk}e^{i\pi}$ [see Eq. (\ref{VV'0})], of which the interference dynamics is given equivalently by that of the double-dot Aharonov-Bohm interferometer with a $\pi$ phase difference and symmetric couplings to the two system states, which produces a totally destructive interference~\cite{TZJ12}.

We are now going to show that the coherent transport currents $I_{L\sigma,R\sigma'}(t)$ and $\bar{I}_{L\sigma,R\sigma'}(t)$ between the left and right lead through the zero-energy bogoliubon (i.e. a pair of MZMs) vanishes when the MZMs are perfectly delocalized, which implies that the left and right MZM wave-function overlap is zero and there is no cross coupling, i.e. $\epsilon_M=0$ and $V_{LRk}=V_{RLk}=0$. Under such conditions, the elements of the correlation function matrices satisfy the following relations,
%\begin{subequations} 
\begin{align}
[\bm{g}^{p(h)}_{L}]_{11}&=[\bm{g}^{p(h)}_{L}]_{22}=[\bm{g}^{p(h)}_{L}]_{12}=[\bm{g}^{p(h)}_{L}]_{21}, \notag \\
[\bm{g}^{p(h)}_{R}]_{11}&=[\bm{g}^{p(h)}_{R}]_{22}=-[\bm{g}^{p(h)}_{R}]_{12}=-[\bm{g}^{p(h)}_{R}]_{21}, \notag \\
[\tilde{\bm{g}}^{p(h)}_{L}]_{11}&=[\tilde{\bm{g}}^{p(h)}_{L}]_{22}=[\tilde{\bm{g}}^{p(h)}_{L}]_{12}=[\tilde{\bm{g}}^{p(h)}_{L}]_{21},  \notag  \\
[\tilde{\bm{g}}^{p(h)}_{R}]_{11}&=[\tilde{\bm{g}}^{p(h)}_{R}]_{22}=-[\tilde{\bm{g}}^{p(h)}_{R}]_{12}=-[\tilde{\bm{g}}^{p(h)}_{R}]_{21}  \label{g_MZM} \\
[\bar{\bm{g}}^{p(h)}_{L}]_{11}&=[\bar{\bm{g}}^{p(h)}_{L}]_{22}=[\bar{\bm{g}}^{p(h)}_{L}]_{12}=[\bar{\bm{g}}^{p(h)}_{L}]_{21}, \notag  \\
[\bar{\bm{g}}^{p(h)}_{R}]_{11}&=[\bar{\bm{g}}^{p(h)}_{R}]_{22}=-[\bar{\bm{g}}^{p(h)}_{R}]_{12}=-[\bar{\bm{g}}^{p(h)}_{R}]_{21} \notag .
\end{align}
%\end{subequations}
Furthermore, the elements of $\bm{g}^{+}_{\alpha}(t,\tau)$ becomes real numbers
\begin{align}
\bm{g}&^{+}_{\alpha}(t,\tau)\notag\\
&=2
\begin{pmatrix}
{\rm Re}[g_L(t,\tau)\!+\!g_R(t,\tau)] && {\rm Re}[g_L(t,\tau)\!-\!g_R(t,\tau)] \\
{\rm Re}[g_L(t,\tau)\!-\!g_R(t,\tau)] && {\rm Re}[g_L(t,\tau)\!+\!g_R(t,\tau)]
\end{pmatrix}, 
\end{align}
with $g_{\alpha}(t,\tau)=\int d\omega/(2\pi)e^{-i(\omega+\frac{U_{\alpha}}{\hbar})(t-\tau)} J_{0\alpha}(\omega)$. Therefore, from Eq. (\ref{U_eq}), the extended retarded Green function matrix $\bm{U}(t,t_0)$ become real and satisfy the following relations,
\begin{subequations}
 \label{U_MZM}
\begin{align}
[\bm{U}(\tau_1,\tau_2)]_{11}& =[\bm{U}(\tau_1,\tau_2)]_{22}, \\
[\bm{U}(\tau_1,\tau_2)]_{12}& =[\bm{U}(\tau_1,\tau_2)]_{21}.
\end{align}
\end{subequations}

The coherent transport currents $I_{L\sigma,R\sigma'}(t)$ and $\bar{I}_{L\sigma,R\sigma'}(t)$ between the left and right lead through the zero-energy bogoliubons (a pair of MZMs) come from the interference of all the contributions of various paths. From Eq.~(\ref{I_ph_LR:2}) and (\ref{I_ph_LR:3}), one can write $I_{L\sigma,R\sigma'}(t)=\sum^{2}_{a,b,c,d=1}I^{(abcd)}_{L\sigma,R\sigma'}(t)$ and $\bar{I}_{L\sigma,R\sigma'}(t)=\sum^{2}_{a,b,c,d=1}\bar{I}^{(abcd)}_{L\sigma,R\sigma'}(t)$, where $a,b,c,d$ are the matrix indices that indicate the path. Explicitly,
\begin{widetext}
\begin{subequations}
\begin{align}
I^{(abcd)}_{L\sigma,R\sigma'}(t)
&=\frac{-e}{\hbar^4}\!\!\int^{t}_{t_0}\!\!d\tau\!\!\int^{\tau}_{t_0}\!\!d\tau_1\!\!\int^{t}_{t_0}\!\!d\tau_2
[\bm{g}^{\sigma}_{L}(t,\!\tau)]_{ab}[\bm{U}(\tau,\!\tau_1)]_{bc}[\tilde{\bm{g}}^{\sigma'}_{R}(\tau_1,\!\tau_2)]_{cd}
[\bm{U}^{\dag}(t,\!\tau_2)]_{da} \notag\\
&+\frac{e}{\hbar^4}\!\!\int^{t}_{t_0}\!\!d\tau\!\!\int^{\tau}_{t_0}\!\!d\tau_1\!\!\int^{t}_{t_0}\!\!d\tau_2
[\tilde{\bm{g}}^{\sigma}_{L}(t,\tau)]_{ab}[\bm{U}(\tau,\tau_1)]_{bc}[\bm{g}^{\sigma'}_{R}(\tau_1,\tau_2)]_{cd}[\bm{U}^{\dag}(t,\tau_2)]_{da}+H.c.  \\
\bar{I}^{(abcd)}_{L\sigma,R\sigma'}(t)
&=\frac{-e}{\hbar^4}\!\!\int^{t}_{t_0}\!\!d\tau\!\!\int^{\tau}_{t_0}\!\!d\tau_1\!\!\int^{t}_{t_0}\!\!d\tau_2
[\bm{g}^{\sigma}_{L}(t,\!\tau)]_{ab}[\bm{U}(\tau,\!\tau_1)]_{bc}[\bar{\bm{g}}^{\sigma'}_{R}(\tau_1,\!\tau_2)]_{cd}
[\bm{U}^{\dag}(t,\!\tau_2)]_{da}+H.c. 
\label{I_abcd}
\end{align}
\end{subequations}
\end{widetext}
By utilizing the relations Eq.~(\ref{g_MZM}) and Eq.~(\ref{U_MZM}) and also the fact that all the matrix elements of $\bm{U}$ are real, it can be shown that the contributions from all paths in $I_{L\sigma,R\sigma'}(t)$ and $\bar{I}_{L\sigma,R\sigma'}(t)$ cancel each other, respectively,
\begin{subequations}
\begin{align}
&I^{(1111)}_{L\sigma,R\sigma'}(t)+I^{(1221)}_{L\sigma,R\sigma'}(t)=0,~ 
I^{(1121)}_{L\sigma,R\sigma'}(t)+I^{(1211)}_{L\sigma,R\sigma'}(t)=0,~ \notag \\
&I^{(1112)}_{L\sigma,R\sigma'}(t)+I^{(1222)}_{L\sigma,R\sigma'}(t)=0,~
I^{(1212)}_{L\sigma,R\sigma'}(t)+I^{(1122)}_{L\sigma,R\sigma'}(t)=0, \notag \\
&I^{(2121)}_{L\sigma,R\sigma'}(t)+I^{(2211)}_{L\sigma,R\sigma'}(t)=0,~
I^{(2111)}_{L\sigma,R\sigma'}(t)+I^{(2221)}_{L\sigma,R\sigma'}(t)=0,~ \notag \\
&I^{(2122)}_{L\sigma,R\sigma'}(t)+I^{(2212)}_{L\sigma,R\sigma'}(t)=0,~
I^{(2112)}_{L\sigma,R\sigma'}(t)+I^{(2222)}_{L\sigma,R\sigma'}(t)=0, \notag\\
\end{align}
\begin{align}
&\bar{I}^{(1111)}_{L\sigma,R\sigma'}(t)+\bar{I}^{(1221)}_{L\sigma,R\sigma'}(t)=0,~
\bar{I}^{(1121)}_{L\sigma,R\sigma'}(t)+\bar{I}^{(1211)}_{L\sigma,R\sigma'}(t)=0,~ \notag \\
&\bar{I}^{(1112)}_{L\sigma,R\sigma'}(t)+\bar{I}^{(1222)}_{L\sigma,R\sigma'}(t)=0,~
\bar{I}^{(1212)}_{L\sigma,R\sigma'}(t)+\bar{I}^{(1122)}_{L\sigma,R\sigma'}(t)=0, \notag\\
&\bar{I}^{(2121)}_{L\sigma,R\sigma'}(t)+\bar{I}^{(2211)}_{L\sigma,R\sigma'}(t)=0,~
\bar{I}^{(2111)}_{L\sigma,R\sigma'}(t)+\bar{I}^{(2221)}_{L\sigma,R\sigma'}(t)=0,~ \notag \\
&\bar{I}^{(2122)}_{L\sigma,R\sigma'}(t)+\bar{I}^{(2212)}_{L\sigma,R\sigma'}(t)=0,~
\bar{I}^{(2112)}_{L\sigma,R\sigma'}(t)+\bar{I}^{(2222)}_{L\sigma,R\sigma'}(t)=0. 
\end{align}
\end{subequations}
As a result, the coherent transport currents $I_{L\sigma,R\sigma'}(t)$ and $\bar{I}_{L\sigma,R\sigma'}(t)$ vanish because of the totally destructive interferences between various path. Therefore, in both partitioned and partition-free schemes, a particle or hole from one lead cannot coherently transport to the other lead when the two MZMs are perfectly delocalized (well-separated).
In other words, a delocalized MZM pair does not have the nonlocal properties of an entangled pair.

When the MZMs are not perfectly delocalized, the cross couplings or the MZM wave-function overlap become finite, particles (holes) can transport coherently between different leads through MZMs. As an illustration, we calculate the cross current $I_{RL}(t)=\sum_{\sigma,\sigma'}I_{R\sigma,L\sigma'}(t)$ through a pair of MZMs generated from a semiconductor-superconductor nanowire.
%\begin{align}
%H_S=\sum^{N-1}_{j=1}(\mu_w c^{\dag}_{j}c_{j}+wa^{\dag}_{j+1}c_{j}+\Delta c_{j+1}c_{j}+H.c.), \notag
%\end{align}
%where $c_j$ ($c^{\dag}_j$) is the annihilation (creation) operator of the system chain cite $j$. 
Also, the left and right leads are coupled to the leftmost and the rightmost cites of the wire.
% namely 
%\begin{align}
%H_{int}=\sum_{k}(\eta_{Lk}b^{\dag}_{Lk}c_{1}+\eta_{Rk}b^{\dag}_{Rk}c_{N}+H.c.). \notag
%\end{align}
Thus, the total Hamiltonian is given by Eq.~(\ref{kcth}).
The topological structures of the wire are manifested in the transport dynamics through the spectral density matrices given by Eq. (\ref{spectral density matrices}), with the coefficients $\kappa_{\alpha i}$ and $\kappa'_{\alpha i}$  given by the bogoliubov transformation of Eq.~(\ref{btkc}),
If we only consider the coherent transport through the MZMs, namely focus on the system ground state and neglect all other system excited states, then
\begin{subequations}
\begin{align}
H_{S}=&\epsilon_Ma^{\dag}_0a_0 \\
H_{T}=&\sum_{k}\big[\eta_{Lk}(\kappa_{L0}b^{\dag}_{Lk}a_0+\kappa'_{L0}b^{\dag}_{Lk}a^{\dag}_0) \notag\\
&+\eta_{Rk}(\kappa_{R0}b^{\dag}_{Rk}a_0+\kappa'_{R0}b^{\dag}_{Rk}a^{\dag}_0)+H.c.\big]. 
\end{align}
\end{subequations}
which is just a realization of Eq.~(\ref{mzmH}).
We will compute the cross current $I_{RL}(t)$ in the partitioned scheme, with Lorentzian spectral densities,
\begin{align}
J_{0\alpha}(\omega)=2\pi\sum_{k}|\eta_{\alpha k}|^2\delta(\omega-\epsilon_{\alpha k}/\hbar)=\frac{\Gamma_{\alpha}d^2}{\omega^2+d^2}, 
\label{lorentz}
\end{align}
where $\Gamma_{\alpha}$ is the coupling strength to the lead $\alpha$ and $d$ is the width of the spectrum. 
Note that the full spectral density matrices containing all the topological properties of MZMs are given by Eq.~(\ref{spectral density matrices}) which is the above spectral densities multiplying the Bogoliubov transformation coefficients 
$\kappa_{\alpha i}$ and $\kappa'_{\alpha i}$.

\begin{figure}
\begin{center}
\includegraphics[width=0.49\textwidth]{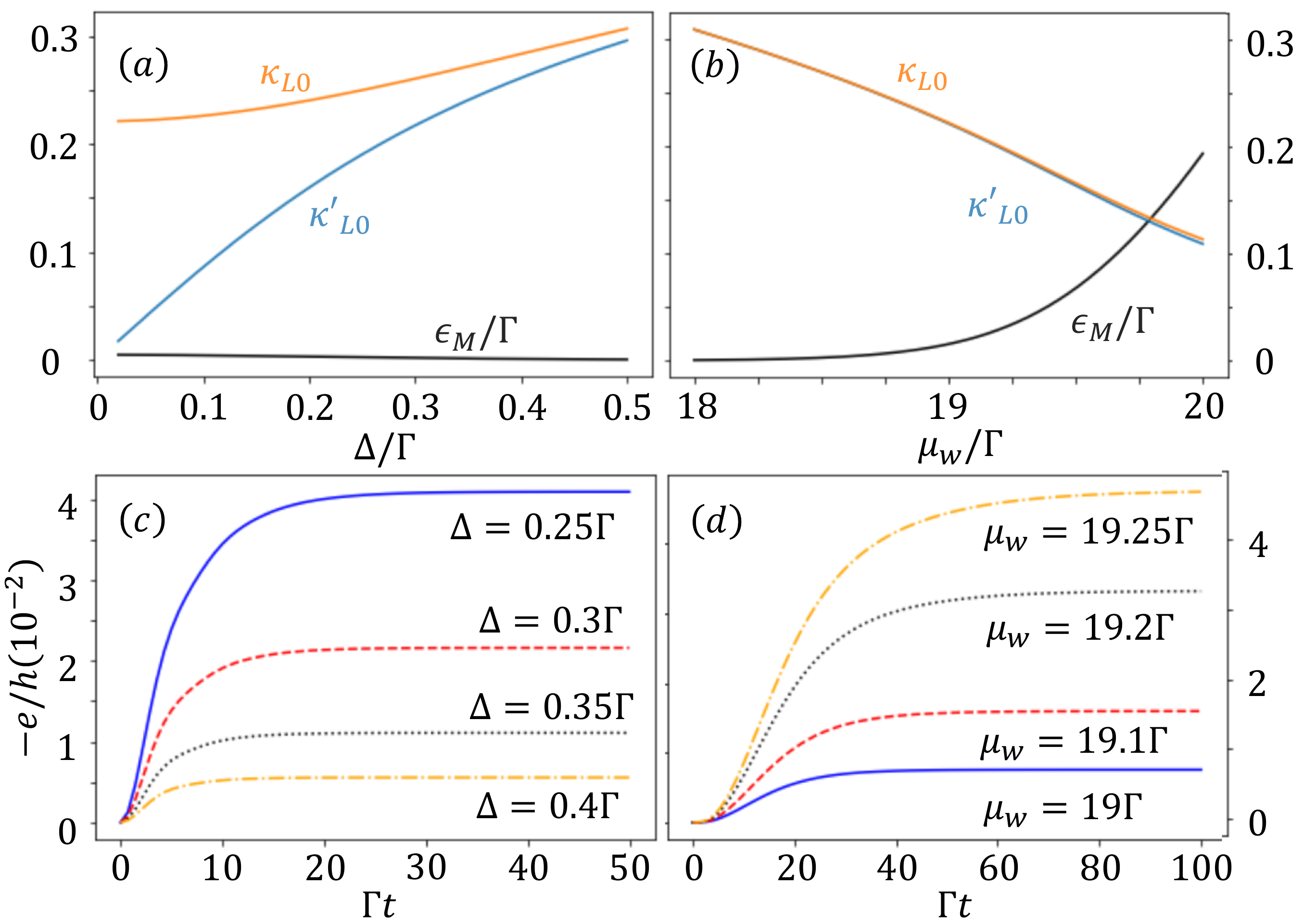}
\caption{(Color online) The coherent transport passing through a nanowire, with the ground state energy and wave-function structure of the chain. The coefficients $\kappa_{L0}$, $\kappa'_{L0}$ and the MZM wave-function overlap $\epsilon_M/\Gamma$ (a) at various $\Delta$, with $\mu_w=0.01\Gamma$, and (b) at various $\mu_w$, with $\Delta=w$. The time evolution of the cross current $I_{RL}(t)$ for (c) different $\Delta$ and (d) different $\mu_w$. These calculations are performed with Lorentzian spectral densities $J_{0\alpha}(\omega)=\frac{\Gamma_{\alpha}d^2}{\omega^2+d^2}$ with the tunneling strengths to the left and right leads are given by $\Gamma_L=\Gamma_R=\Gamma$ and the spectrum width $d=50\Gamma$. Other parameters are set to be $w=10\Gamma$, $\mu_R=-\mu_L=\Gamma$, $k_BT_L=k_BT_R=0.1\Gamma$, and the total number of chain cites $N=81$.}
\label{I_cross}
\end{center}
\end{figure}

In Fig. \ref{I_cross}, the cross current is studied in two scenarios: (1) with a fixed $\mu_w$ and different $\Delta$ (see Fig. \ref{I_cross}a, c), and (2) with a fixed $\Delta$ and different $\mu_w$ (see Fig. \ref{I_cross}b, d). In the first scenario, the cross coupling between MZM $\gamma_L$ and lead $R$, which is determined by the difference $i(\kappa_{\alpha 0}-\kappa'_{\alpha 0})$ [see Eq. (\ref{VV'})], increases when $\Delta$ decreases, as shown in Fig. \ref{I_cross}a. As a result, the particles (holes) can transport coherently to the other lead because the MZMs are directly coupled to the opposite-end leads (see Fig. \ref{I_cross}c). In this scenario, the Majorana energy $\epsilon_M/\Gamma$ (MZM wave-function overlap) is negligible and the cross current is mainly caused by the explicit cross coupling. In the second scenario, the Majorana energy $\epsilon_M$ emerges from zero when the chemical potential of the chain is increased near the topological phase transition point $\mu_w\sim 2\Delta$ (see Fig. \ref{I_cross}b), where the cross-coherent transports can happen through the wave-function overlap between the two MZMs (see Fig. \ref{I_cross}d). In this scenario, the cross coupling [$i(\kappa_{\alpha 0}-\kappa'_{\alpha 0})$] is negligible and the cross current is mainly caused by the MZM wave-function overlap. In both scenarios, there is no ``quantum teleportation" between a delocalized pair of MZMs in this topological system, because the above electron transports are not caused by the nonlocality of an entangled pair.  As we have shown in our previous work \cite{LYH18}, in the topological phase, when one of the two delocalized MZMs in a nanowire is disturbed, only the disturbed MZM decoheres, leaving the other MZM unchanged. That is, the two delocalized MZMs do not entangled together. In other words, no teleportation can occur via delocalized MZM pairs.

%It is known that an electron can transport coherently through a pair of MZMs in the presence of charging energy.  
%In this case, an additional term $U$ is added to the system Hamiltonian. The charging energy should be small comparing 
%to the topologically protected energy gap, while it should also be large enough so that the total number of charges 
%in the system is confined to be either $n_0$ or $n_0+1$ during all transport processes. 
In Ref. \cite{Fu10}, by constraining the system Hilbert space to a subspace with allowed charge number, 
Fu obtained the following effective Hamiltonian (i.e. Eq.~(10) of Ref. \cite{Fu10}), 
\begin{align}
\tilde{H}=&H_L+\delta(f^{\dag}f-\frac{1}{2})+(\lambda_1b^{\dag}_1f+H.c.) \notag\\
&+(-1)^{n_0}(-i\lambda_2b^{\dag}_2f+H.c.). 
\end{align}
Here, $H_L$ is the Hamiltonian of the leads, $b_{\alpha}$ is the fermion operator of the lead $\alpha$ and $f$ is a single-level fermion operator. Taking lead $1$ as the left lead and lead $2$ as the right lead, the above effective Hamiltonian is of the same form as our Eq. (\ref{H_bov}) with 
\begin{align}
\epsilon_M&\rightarrow\delta \notag\\
\tilde{V}_{Lk}&\rightarrow\lambda_1,~\tilde{V}'_{Lk}\rightarrow0  \\
\tilde{V}_{Rk}&\rightarrow\pm i\lambda_2,~\tilde{V}'_{Rk}\rightarrow0. \notag
\end{align}
The $+$/$-$ sign corresponds to an odd/even $n_0$. Utilizing the relation Eq.~(\ref{VV'}), for both $\tilde{V}'_{Lk}$ and $\tilde{V}'_{Rk}$ are zero, the cross coupling coefficients $V_{LRk}=V_{LLk}\rightarrow\lambda_1$ and $V_{RLk}=V_{RRk}\rightarrow\pm i\lambda_2$. In this case, electrons can be coherently transported through a pair of delocalized MZMs because the charging energy induces direct cross-coupling between the MZMs and the opposite-side leads, i.e., both electrons and holes are directly coupled to the two leads.  These processes should not be called as "teleportation" because teleportation is defined as a realization through the non-locality of an entangled pair rather than a direct coupling between two objects \cite{QT1993}. A delocalized (well-separated) MZM pair does not have the non-local property of entanglement pairs that can be used as a resource for quantum teleportation. Calling the coherent transport via a direct coupling of delocalized MZMs induced by the charging energy between them as "teleportation" is conceptually misleading.

The coherent transport between the left and right leads through a pair of MZMs can be explored in experiments by measuring the cross differential conductance (CDC) in a superconductor-semiconductor nanowire. The CDC is defined by the differentiation of the left-lead (right-lead) current with respect to the right (left) bias, namely $dI_{L}/d\mu_{R}$ ($dI_{R}/d\mu_{L}$), with $\mu_{\alpha}$ being the bias voltage of lead $\alpha$. As we have already shown, a particle or hole cannot transport coherently through a pair of perfectly delocalized MZMs, therefore, one may expect that the measured CDC vanish within the topologically protected energy gap when MZMs exist. This behavior of CDC is contrary to the usual differential conductance (simply called as DC) $dI_L/d\mu_L$ ($dI_R/d\mu_R$). The DC shows a peak value at zero bias when MZMs exist, which is the well-known zero-bias conductance peak (ZBCP). This ZBCP is caused by local coherent transport processes, i.e. particles and holes flow coherently in and out of the same lead, no cross coherent transport contribution takes place through a pair of MZMs in DC. On the other hand, if the transport processes are contributed by non-topological system states other than MZMs, one may expect that both CDC and the DC behave similarly.

\begin{figure}
\begin{center}
\includegraphics[width=0.47\textwidth]{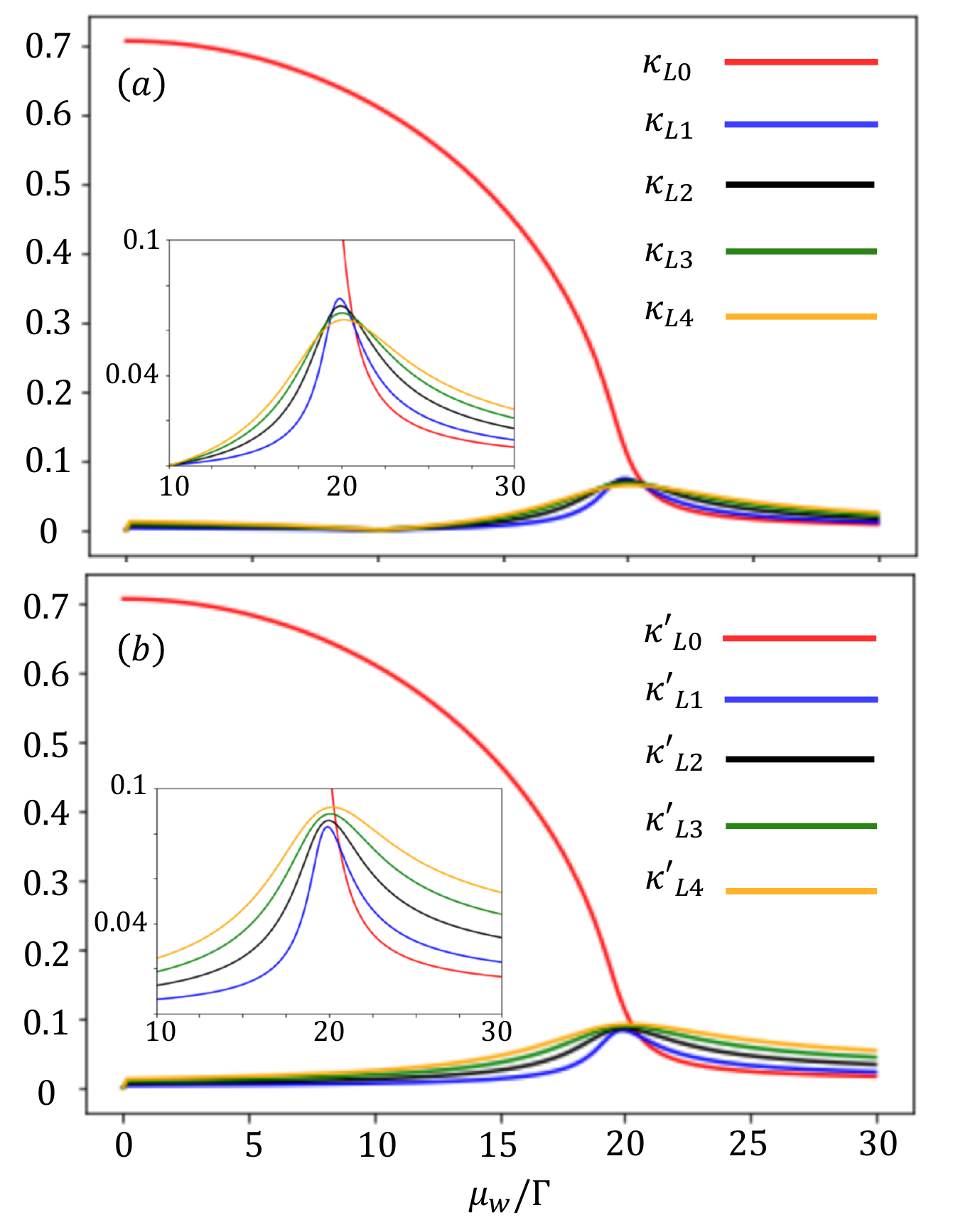}
\caption{(Color online) The coefficients (a) $\kappa_{Li}$ and (b) $\kappa'_{Li}$ that characterize the wave-function structures of the quantum states of the system wire. Here, the coefficients of the ground state and first four excited states are shown. The parameters are set to be $w=\Delta=10\Gamma$ and the total number of system cites $N=81$.}
\label{kappa}
\end{center}
\end{figure}

As a further illustration, we extend our calculations of the CDC and DC for the semiconductor-superconductor nanowire from the Hamiltonian (\ref{H_chain_bov}) that includes all the non-topological excited states of the wire. The wave-function structures of both topological and non-topological states are captured by the coefficients $\kappa_{\alpha i}$ and $\kappa'_{\alpha i}$, see Eq.~(\ref{btkc}). As shown in Fig.~\ref{kappa}, in the topological regime of the wire ($\mu_w<2\Delta$), the coefficient $\kappa_{\alpha 0}$ ($\kappa'_{\alpha 0}$), which characterizes the ground state wave-function of the wire, are much larger than other coefficients $\kappa_{\alpha i}$ ($\kappa'_{\alpha i}$), which characterize the non-topological excited states of the wire. This indicates that the wave-function of the system ground state is localized at the end of the wire and the transport dynamics is mainly contributed by the electrons tunneling through the topological ground state of the wire. On the other hand, as the system chemical potential $\mu_w$ increases, $\kappa_{\alpha 0}$ ($\kappa'_{\alpha 0}$) decreases because the wave-function of the ground state begins to spread along the wire. Eventually, in the non-topological regime of the wire ($\mu_w>2\Delta$), the coefficient $\kappa_{\alpha 0}$ ($\kappa'_{\alpha 0}$) becomes smaller than other coefficients $\kappa_{\alpha i}$ ($\kappa'_{\alpha i}$), which indicates that transport dynamics is mainly contributed by electrons tunneling through the non-topological states of the wire. 

\begin{figure*}
\begin{center}
\includegraphics[width=1.04\textwidth]{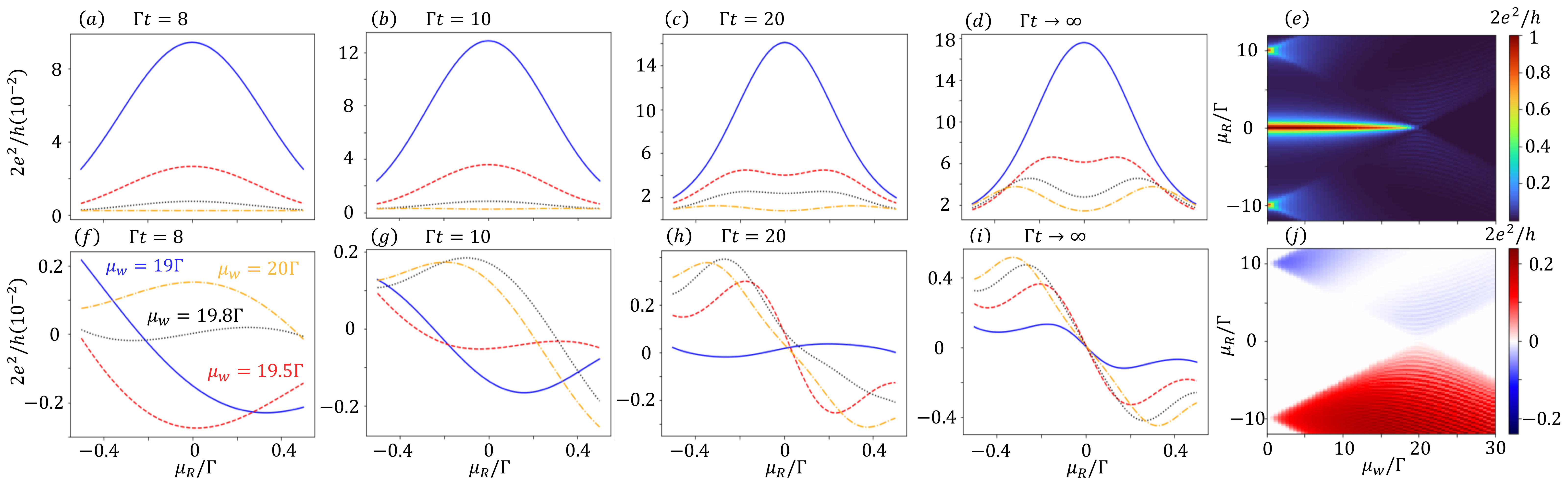}
\caption{(Color online) (a)-(d) The transient differential conductance $dI_{R}/d\mu_{R}$, and (f)-(i) the transient cross differential conductance $dI_{L}/d\mu_{R}$ with various wire chemical potential $\mu_w$ at different times. (e) The usual differential conductance and (j) the cross differential conductance with different bias and wire chemical potential in the steady-state limit. These calculations are performed with lorenzian spectral densities $J_{0\alpha}(\omega)=\frac{\Gamma_{\alpha}d^2}{\omega^2+d^2}$, and the coupling strengths to the left and right leads are set to be $\Gamma_L=\Gamma_R=\Gamma$, the spectrum width $d=50\Gamma$. Other parameters $w=\Delta=10\Gamma$, $k_BT_L=k_BT_R=0.1\Gamma$. The total number of system cites $N=51$.}
\label{CDC}
\end{center}
\end{figure*}

The following calculations are performed in the partitioned scheme. 
%with the Lorentzian spectral densities of the leads [see Eq. (\ref{lorentz})]. 
In the partitioned scheme, the wire and the leads are assumed to be initially decoupled. Then one can turn on the tunneling couplings $\eta_{Lk}$ and $\eta_{Rk}$ so that the system starts to be driven by the biased leads. Figure \ref{CDC} shows the CDC ($dI_{L}/d\mu_{R}$), and the DC ($dI_{R}/d\mu_{R}$), respectively. The DC gradually forms peaks from plateaus as time evolves (see Fig.~\ref{CDC}a-d), while the CDC shows very different behaviors in the transient regime than that in the steady-state limit (see Fig.~\ref{CDC}f-i). In the transient regime, the CDC does not vanish at zero bias in the topological phase of the wire (see Fig.~\ref{CDC}f-h). This is because particles/holes can transport to the opposite leads through the overlap of the MZM wave-function, which can be measured by Majorana energy (see Fig.~\ref{I_cross}b), as we have discussed above. In the steady-state limit, near the topological phase transition point of the wire ($\mu_w=2\Delta$), the CDC becomes approximately an odd function and a linear function of the bias in the low bias limit (see Fig. \ref{CDC}i). However, these behaviors cannot be observed in the transient regime, in which the CDC is neither an odd function nor a linear function even in the low bias limit (see Fig. \ref{CDC}f-h). Furthermore,  the CDC vanishes at zero bias in the steady-state limit (see Fig. \ref{CDC}j), while the DC shows a ZBCP (see Fig. \ref{CDC}e). On the other hand, near the topological phase transition point of the wire ($\mu_w=2\Delta$), the MZM wave-functions spread along the wire and the two MZMs overlap so that coherent transport processes occur between the two leads. Therefore, one can find that the CDC begins to emerge while the ZBCP of the DC begins to split (see Fig. \ref{CDC}d and i). Away from zero bias, particles can transport coherently between two leads through non-topological finite-energy bogoliubon states so that the CDC and the DC behave similarly (see Fig. \ref{CDC}e and j). A similar steady-state behavior of the CDC has also be demonstrated in a three-terminal device with the scattering matrix theory \cite{RVK2018}, but the scattering matrix theory cannot describe the above transient transport dynamics.

\section{Conclusion and perspective}
\label{Sec3}
We have presented a quantum transport theory for hybrid superconducting  systems in both partition-free and partitioned schemes. The transient transport dynamics is fully captured in the extended nonequilibrium Green's functions which incorporate pair correlations via the spectral density matrices. Especially, the spectral density matrices are proportional to the wave-function overlaps of the system and terminal states and therefore can characterize the topological structures of the system and terminals if they are in the topological phases. Our transport theory shows that all coherent transport dynamics of particles/holes between different terminals are resultant interferences of various tunneling processes between the system and the terminals, and cannot be rendered simply as the picture of normal transmission and Andreev reflection that one usually used. We then applied our theory to study the transport dynamics via a pair of MZMs in a two-terminal nanowire. We showed that, when the MZMs are well-delocalized so that their wavefunctions do not overlap with each other, the transport process corresponds to interferences in double-dot AB interferometers with a $\pi$ phase difference so that totally destructive interferences occur. Consequently, there is no coherent current flowing through the pair of delocalized MZMs and the so-called ``teleportation" cannot happen between a pair of delocalized MZMs.  Electron transport induced by a finite charging energy between the pair of MZMs is a consequence of the direct coupling between the two MZMs, it is not the teleportation utilizing the nonlocality of an entangled pair. The pair of delocalized MZMs generated in topological systems does not form an entangled pair, as we have shown in our previous work \cite{LYH18}. For the application to a superconductor-semiconductor nanowire, which could be experimentally measured, it also shows signatures of these destructive interferences. The cross differential conductance vanishes in the topological regime of the nanowire, while the direct differential conductance shows the well known ZBCP, indicates that local coherent transports can happen only locally in a MZM system and there is no coherent transport through a pair of perfectly delocalized MZMs. 

In addition to the quantum transport theory, it is natural to apply our exact master equation to the study of thermoelectric properties, for example the thermopower, of Majorana systems, which are recently highly discussed as signatures of MZMs~\cite{LLS14,HSZ21,ZWZ22,S23}. It is worth noting that all the physical observables of the system can be computed from the reduced density matrix at any time, including all the thermodynamics quantities such as the internal energy, particle number and also entropy $S(t)=-k_B{\rm Tr}_{S}[\rho_S(t){\rm ln}\rho_S(t)]$. Based on our exact master equation, from which the reduced density matrix can be solved, we have developed the quantum thermodynamics far from equilibrium~\cite{AHZ20,HZ22,HZ22a}. We have pointed out that, when the system-bath couplings become strong, the thermodynamic quantities of the system must be renormalized~\cite{HZ22,HZ22a}. By considering the renormalization of the central-system energy due to the system-lead couplings, the transient heat current as well as the transient electric current through the central system can be changed significantly and the thermopower far from equilibrium can be explored. Thus, it is straightforward to extend our theory to the theory of thermoelectric transport for hybrid superconducting systems, which goes far beyond the steady-state limit and linear response theory. All the properties of the topological state wave-functions are fully captured by the extended nonequilibrium Green function through the spectral density matrices incorporating pair correlations, so that topological features manifested in the heat currents and other thermodynamics quantities can be investigated. 

\section*{Acknowledgement}
This work is supported by National Science and Technology Council of Taiwan, Republic of China, under Contract No. MOST-111-2811-M-006-014-MY3.

\appendix
\section{Derivation of the exact master equation}
\label{App}
\begin{widetext}
In this appendix, we will derive the transient transport current incorporating all the dissipation and fluctuation processes through the particle and hole channels in the partition-free scheme and partitioned scheme, respectively. The derivation is based on our master equation approach proposed in Ref.~\cite{HYZ20} for partition-free scheme and Ref.~\cite{HZ22b} for partitioned scheme, in which the exact master equation is given by
\begin{align}
\frac{d}{dt}\rho_S(t)=\frac{1}{i\hbar}[H_S(t),\rho_S(t)]+[\mathcal{L}^{+}(t)+\mathcal{L}^{-}(t)]\rho_S(t),
\end{align}
where
\begin{subequations} 
\begin{align}
\label{A2}
\mathcal{L}^{+}(t)\rho_S(t)&=\bm{a}^{\dag}\cdot\bm{A}(t)+\bm{A}^{\dag}(t)\cdot\bm{a}(t)\\
\mathcal{L}^{-}(t)\rho_S(t)&=-\bm{a}\cdot\bm{A}^{\dag}(t)-\bm{A}(t)\cdot\bm{a}^{\dag}(t).
\end{align}
\end{subequations}
The collective operator $\bm{A}(t)$ is defined as
\begin{align}
\label{A3}
A[\rho_S(t)]\equiv\dfrac{i}{\hbar}\Tr_E[(\boldsymbol{\eta}\boldsymbol{\kappa^\prime}\boldsymbol{b^\dagger}-\boldsymbol{\eta^*}\boldsymbol{\kappa^*}\boldsymbol{b})\rho_{tot}(t)]
\end{align}

\subsection{Partition-free scheme}
\label{App2}
We first consider the case that the system and the environment (leads) are initially correlated. More specifically, we suppose the total system is initially in a partition-free state (a Gaussian-type state including initial system-environment correlations), the Gaussian-type total density matrix in the coherent state representation is given by
\begin{align}
&\langle\boldsymbol{\xi}|\rho_{tot}(t)|\boldsymbol{\xi^\prime}\rangle=\dfrac{1}{Z_{tot}(t)}\exp\left[\dfrac{1}{2}\begin{pmatrix}
\boldsymbol{\xi^\dagger} & \boldsymbol{\xi^{\prime T}}\end{pmatrix}
\begin{pmatrix}
\boldsymbol{\Omega}(t) & \boldsymbol{\Pi}(t)\\
-\boldsymbol{\Pi^*}(t) &-\boldsymbol{\Omega^*}(t)
\end{pmatrix}
\begin{pmatrix}
\boldsymbol{\xi^\prime}\\
\boldsymbol{\xi^*}
\end{pmatrix}
\right]
\end{align}
where $\boldsymbol{\Omega}(t)$ is a Hermitian matrix and $\boldsymbol{\Pi}(t)$ is an antisymmetric matrix.
The collective operator $A[\rho_S(t)]$ can be solved with the generalized Gaussian integral
\begin{align}
&\left\langle\boldsymbol{\xi_S}\left\vert\begin{pmatrix}
\boldsymbol{A}(t)\\
-\boldsymbol{A^\dagger}(t)^T
\end{pmatrix}
\right\vert\boldsymbol{\xi^\prime_S}\right\rangle=
\dfrac{1}{i\hbar}\sum\limits_{\alpha}\begin{pmatrix}
\boldsymbol{\eta_{\alpha}^*}\boldsymbol{\kappa_{\alpha}^*} & -\boldsymbol{\eta_{\alpha}}\boldsymbol{\kappa_{\alpha}^\prime}\\
\boldsymbol{\eta_{\alpha}^*}\boldsymbol{\kappa_{\alpha}^{\prime*}} & -\boldsymbol{\eta_{\alpha}}\boldsymbol{\kappa_{\alpha}}
\end{pmatrix}\int d\mu(\boldsymbol{\xi_{E_{\alpha}}})\left\langle\boldsymbol{\xi_S},\boldsymbol{\xi_{E_{\alpha}}}\left\vert
\begin{pmatrix}
\boldsymbol{b_{\alpha}}\\
(\boldsymbol{b_{\alpha}}^\dagger)^T
\end{pmatrix}
\rho_{tot}(t)
\right\vert\boldsymbol{\xi_S^\prime},-\boldsymbol{\xi_{E_{\alpha}}}\right\rangle\notag\\
&\hspace{3mm}=\dfrac{1}{i\hbar}\sum\limits_{\alpha}\begin{pmatrix}
\boldsymbol{\eta_{\alpha}^*}\boldsymbol{\kappa_{\alpha}^*} & -\boldsymbol{\eta_{\alpha}}\boldsymbol{\kappa_{\alpha}^\prime}\\
\boldsymbol{\eta_{\alpha}^*}\boldsymbol{\kappa_{\alpha}^{\prime*}} & -\boldsymbol{\eta_{\alpha}}\boldsymbol{\kappa_{\alpha}}
\end{pmatrix}\dfrac{1}{Z_{tot}(t)}\int d\mu(\boldsymbol{\xi_{E_{\alpha}}})\begin{pmatrix}
\frac{\partial}{\partial\boldsymbol{\xi_{E_{\alpha}}^\dagger}}\\
\boldsymbol{\xi_{E_{\alpha}}^\dagger}
\end{pmatrix}
\exp\left[\dfrac{1}{2}\begin{pmatrix}
\boldsymbol{\xi_{\alpha}^\dagger} & \boldsymbol{\xi_{\alpha}^{\prime T}}\end{pmatrix}
\begin{pmatrix}
\boldsymbol{\Omega_{\alpha}}(t) & \boldsymbol{\Pi_{\alpha}}(t)\\
-\boldsymbol{\Pi_{\alpha}^*}(t) &-\boldsymbol{\Omega_{\alpha}^*}(t)
\end{pmatrix}
\begin{pmatrix}
\boldsymbol{\xi_{\alpha}^\prime}\\
\boldsymbol{\xi_{\alpha}^*}
\end{pmatrix}
\right]
\end{align}
where the D algebra of fermion creation and annihilation operators $\langle\boldsymbol{\xi_{E_{\alpha}}}|\boldsymbol{b}_{\alpha}=\frac{\partial}{\partial\boldsymbol{\xi_{E_{\alpha}}^\dagger}}\langle\boldsymbol{\xi_{E_{\alpha}}}|$ and $\langle\boldsymbol{\xi_{E_{\alpha}}}|\boldsymbol{b}_{\alpha}^\dagger=\boldsymbol{\xi_{E_{\alpha}}^\dagger}\langle\boldsymbol{\xi_{E_{\alpha}}}|$ are used.\\

The Gaussian integral $\int d\mu(\boldsymbol{\xi})e^{\boldsymbol{\xi^\dagger}\cdot\boldsymbol{\Theta}\cdot\boldsymbol{\xi}+\boldsymbol{\eta^\dagger}\cdot\boldsymbol{\xi}+\boldsymbol{\xi^\dagger}\cdot\boldsymbol{\eta^\prime}}=||\boldsymbol{1}-\boldsymbol{\Theta}||e^{\boldsymbol{\eta^\dagger}\cdot(\boldsymbol{1}-\boldsymbol{\Theta})^{-1}\cdot\boldsymbol{\eta^\prime}}$ can be generalized to the system with pairing terms
\begin{align}
&\int d\mu(\boldsymbol{\xi})\exp\left[\frac{1}{2}\begin{pmatrix}
\boldsymbol{\xi^*} & \boldsymbol{\xi^T}
\end{pmatrix}
\begin{pmatrix}
\boldsymbol{\Theta}&\boldsymbol{\mathcal{P}^\prime}\\
\boldsymbol{\mathcal{P}^*}&-\boldsymbol{\Theta^\dagger}
\end{pmatrix}
\begin{pmatrix}
\boldsymbol{\xi}\\
\boldsymbol{\xi^\dagger}
\end{pmatrix}
+\begin{pmatrix}
\boldsymbol{\xi^*}&\boldsymbol{\xi^T}
\end{pmatrix}
\begin{pmatrix}
\boldsymbol{\eta^\prime}\\
-\boldsymbol{\eta^\dagger}
\end{pmatrix}
\right]\notag\\
&=\left\Vert\boldsymbol{1}-\begin{pmatrix}
\boldsymbol{\Theta}&\boldsymbol{\mathcal{P}^\prime}\\
-\boldsymbol{\mathcal{P}^*}&\boldsymbol{\Theta^\dagger}
\end{pmatrix}\right\Vert^{1/2}\exp\left\{\frac{1}{2}\begin{pmatrix}
\boldsymbol{\eta^*} & -\boldsymbol{\eta^{\prime T}}\end{pmatrix}
\left[\boldsymbol{1}-\begin{pmatrix}
\boldsymbol{\Theta}&\boldsymbol{\mathcal{P}^\prime}\\
-\boldsymbol{\mathcal{P}^*}&\boldsymbol{\Theta^\dagger}
\end{pmatrix}
\right]^{-1}\begin{pmatrix}
1&0\\
0&-1
\end{pmatrix}
\begin{pmatrix}
\boldsymbol{\eta^\prime}\\
-\boldsymbol{\eta^\dagger}
\end{pmatrix}
\right\}
\end{align}
Using this generalized Gaussian integral, we have
\begin{align}
&\left\langle\boldsymbol{\xi_S}\left\vert\begin{pmatrix}
\boldsymbol{A}(t)\\
-\boldsymbol{A^\dagger}(t)^T
\end{pmatrix}
\right\vert\boldsymbol{\xi^\prime_S}\right\rangle\notag\\
&=\dfrac{1}{i\hbar}\sum\limits_{\alpha}\begin{pmatrix}
\boldsymbol{\eta_{\alpha}^*}\boldsymbol{\kappa_{\alpha}^*} & -\boldsymbol{\eta_{\alpha}}\boldsymbol{\kappa_{\alpha}^\prime}\\
\boldsymbol{\eta_{\alpha}^*}\boldsymbol{\kappa_{\alpha}^{\prime*}} & -\boldsymbol{\eta_{\alpha}}\boldsymbol{\kappa_{\alpha}}
\end{pmatrix}
\begin{pmatrix}
\boldsymbol{1}+\boldsymbol{\Omega_{E_{\alpha}E_{\alpha}}}(t) & -\boldsymbol{\Pi_{E_{\alpha}E_{\alpha}}}(t)\\
-\boldsymbol{\Pi_{E_{\alpha}E_{\alpha}}^*}(t) & \boldsymbol{1}+\boldsymbol{\Omega^*_{E_{\alpha}E_{\alpha}}}(t)
\end{pmatrix}^{-1}\begin{pmatrix}
\boldsymbol{\Omega_{E_{\alpha}S}}(t) & \boldsymbol{\Pi_{E_{\alpha}S}}(t)\\
-\boldsymbol{\Pi^*_{E_{\alpha}S}}(t) & -\boldsymbol{\Omega^*_{E_{\alpha}S}}(t)
\end{pmatrix}\begin{pmatrix}
\boldsymbol{\xi^\prime_S}\\
\boldsymbol{\xi_S^*}
\end{pmatrix}
\langle\boldsymbol{\xi_S}|\rho_S(t)|\boldsymbol{\xi_S^\prime}\rangle
\end{align}
where 
\begin{subequations}
\begin{align}
&\langle\boldsymbol{\xi_S}|\rho_S(t)|\boldsymbol{\xi_S^\prime}\rangle=\dfrac{1}{Z_S(t)}\exp\left[\begin{pmatrix}
\boldsymbol{\xi_S^\dagger} & \boldsymbol{\xi_S^{\prime T}}
\end{pmatrix}
\begin{pmatrix}
\boldsymbol{\Omega_{S}}(t) & \boldsymbol{\Pi_{S}}(t)\\
-\boldsymbol{\Pi_{S}^*}(t) & -\boldsymbol{\Omega^*_{S}}(t)
\end{pmatrix}
\begin{pmatrix}
\boldsymbol{\xi_{S}^\prime}\\
\boldsymbol{\xi_S^*}
\end{pmatrix}
\right]\\ 
&Z_S(t)=Z_{tot}(t)\left\Vert\boldsymbol{1}-\begin{pmatrix}
\boldsymbol{\Omega_{E_{\alpha}E_{\alpha}}}(t) & \boldsymbol{\Pi_{E_{\alpha}E_{\alpha}}}(t)\\
\boldsymbol{\Pi_{E_{\alpha}E_{\alpha}}^*}(t) & \boldsymbol{\Omega^*_{E_{\alpha}E_{\alpha}}}(t)
\end{pmatrix}\right\Vert^{-1/2}
\end{align}
\end{subequations}
and
\begin{align}
\begin{pmatrix}
\boldsymbol{\Omega_{S}}(t) & \boldsymbol{\Pi_{S}}(t)\\
-\boldsymbol{\Pi_{S}^*}(t) & -\boldsymbol{\Omega^*_{S}}(t)
\end{pmatrix}=&\begin{pmatrix}
\boldsymbol{\Omega_{SS}}(t) & \boldsymbol{\Pi_{SS}}(t)\\
-\boldsymbol{\Pi_{SS}^*}(t) & -\boldsymbol{\Omega^*_{SS}}(t)
\end{pmatrix}\notag\\
&+\begin{pmatrix}
\boldsymbol{\Omega_{SE_{\alpha}}}(t) & \boldsymbol{\Pi_{SE_{\alpha}}}(t)\\
-\boldsymbol{\Pi_{SE_{\alpha}}^*}(t) & -\boldsymbol{\Omega^*_{SE_{\alpha}}}(t)
\end{pmatrix}
\left[\boldsymbol{1}-\begin{pmatrix}
\boldsymbol{\Omega_{E_{\alpha}E_{\alpha}}}(t) & \boldsymbol{\Pi_{E_{\alpha}E_{\alpha}}}(t)\\
\boldsymbol{\Pi_{E_{\alpha}E_{\alpha}}^*}(t) & \boldsymbol{\Omega^*_{E_{\alpha}E_{\alpha}}}(t)
\end{pmatrix}\right]^{-1}\begin{pmatrix}
\boldsymbol{\Omega_{E_{\alpha}S}}(t) & \boldsymbol{\Pi_{E_{\alpha}S}}(t)\\
\boldsymbol{\Pi_{E_{\alpha}S}^*}(t) & \boldsymbol{\Omega^*_{E_{\alpha}S}}(t)
\end{pmatrix}
\end{align}

Then using the D-algebra of fermion creation and annihilation operators again, the collective operators can be expressed as
\begin{align}
\begin{pmatrix}
\boldsymbol{A}(t)\\
-\boldsymbol{A^\dagger}(t)^T
\end{pmatrix}=\dfrac{i}{\hbar}\sum\limits_{\alpha}\begin{pmatrix}
\boldsymbol{\eta_{\alpha}^*}\boldsymbol{\kappa_{\alpha}^*} & -\boldsymbol{\eta_{\alpha}}\boldsymbol{\kappa_{\alpha}^\prime}\\
\boldsymbol{\eta_{\alpha}^*}\boldsymbol{\kappa_{\alpha}^{\prime*}} & -\boldsymbol{\eta_{\alpha}}\boldsymbol{\kappa_{\alpha}}
\end{pmatrix}
\begin{pmatrix}
\boldsymbol{1}+\boldsymbol{\Omega_{E_{\alpha}E_{\alpha}}}(t) & -\boldsymbol{\Pi_{E_{\alpha}E_{\alpha}}}(t)\\
-\boldsymbol{\Pi_{E_{\alpha}E_{\alpha}}^*}(t) & \boldsymbol{1}+\boldsymbol{\Omega^*_{E_{\alpha}E_{\alpha}}}(t)
\end{pmatrix}^{-1}\begin{pmatrix}
\boldsymbol{\Omega_{E_{\alpha}S}}(t) & \boldsymbol{\Pi_{E_{\alpha}S}}(t)\\
-\boldsymbol{\Pi^*_{E_{\alpha}S}}(t) & -\boldsymbol{\Omega^*_{E_{\alpha}S}}(t)
\end{pmatrix}\begin{pmatrix}
\rho_S(t)\boldsymbol{a}\\
\boldsymbol{a^\dagger}\rho_S(t)
\end{pmatrix}
\end{align}
Similarly, we can use the same technique to calculate the correlation functions in terms of the
Gaussian kernel elements
\begin{subequations}
\begin{align}
\boldsymbol{N_S}(t)&\equiv\begin{pmatrix}
\langle \boldsymbol{a^\dagger}(t)\boldsymbol{a}(t)\rangle & \langle \boldsymbol{a}(t)\boldsymbol{a}(t)\rangle\\
\langle \boldsymbol{a^\dagger}(t)\boldsymbol{a^\dagger}(t)\rangle & \langle \boldsymbol{a}(t)\boldsymbol{a^\dagger}(t)\rangle
\end{pmatrix}\notag\\
&=\begin{pmatrix}
\boldsymbol{\Omega_{S}}(t) & \boldsymbol{\Pi_{S}}(t)\\
-\boldsymbol{\Pi_{S}^*}(t) & -\boldsymbol{\Omega^*_{S}}(t)
\end{pmatrix}\left[\boldsymbol{1}-\begin{pmatrix}
\boldsymbol{\Omega_{S}}(t) & \boldsymbol{\Pi_{S}}(t)\\
\boldsymbol{\Pi_{S}^*}(t) & \boldsymbol{\Omega^*_{S}}(t)
\end{pmatrix}\right]^{-1}\\
\boldsymbol{N_{E_{\alpha}S}}(t)&\equiv\begin{pmatrix}
\langle \boldsymbol{a^\dagger}(t)\boldsymbol{b_{\alpha}}(t)\rangle & \langle \boldsymbol{a}(t)\boldsymbol{b_{\alpha}}(t)\rangle\\
\langle \boldsymbol{a^\dagger}(t)\boldsymbol{b_{\alpha}^\dagger}(t)\rangle & \langle \boldsymbol{a}(t)\boldsymbol{b_{\alpha}^\dagger}(t)\rangle
\end{pmatrix}\notag\\
&=\begin{pmatrix}
\boldsymbol{1}+\boldsymbol{\Omega_{E_{\alpha}E_{\alpha}}}(t) & -\boldsymbol{\Pi_{E_{\alpha}E_{\alpha}}}(t)\\
-\boldsymbol{\Pi_{E_{\alpha}E_{\alpha}}^*}(t) & \boldsymbol{1}+\boldsymbol{\Omega^*_{E_{\alpha}E_{\alpha}}}(t)
\end{pmatrix}^{-1}\begin{pmatrix}
\boldsymbol{\Omega_{E_{\alpha}S}}(t) & \boldsymbol{\Pi_{E_{\alpha}S}}(t)\\
-\boldsymbol{\Pi^*_{E_{\alpha}S}}(t) & -\boldsymbol{\Omega^*_{E_{\alpha}S}}(t)
\end{pmatrix}\left[\boldsymbol{1}-\begin{pmatrix}
\boldsymbol{\Omega_{S}}(t) & \boldsymbol{\Pi_{S}}(t)\\
\boldsymbol{\Pi_{S}^*}(t) & \boldsymbol{\Omega^*_{S}}(t)
\end{pmatrix}\right]^{-1}
\end{align}
\end{subequations}
On the other hand, the time evolution of these correlation functions can be solved by the Heisenberg equation
of motion
\begin{align}
\label{A27}
\boldsymbol{\dot{N}_S}(t,\tau)=\frac{1}{i\hbar}\left[\begin{pmatrix}
\boldsymbol{\varepsilon}&\boldsymbol{0}\\
\boldsymbol{0}&-\boldsymbol{\varepsilon}
\end{pmatrix}\boldsymbol{N_S}(t)+
\sum\limits_{\alpha}\begin{pmatrix}
\boldsymbol{\eta_{\alpha}^*}\boldsymbol{\kappa_{\alpha}^*} & -\boldsymbol{\eta_{\alpha}}\boldsymbol{\kappa_{\alpha}^\prime}\\
\boldsymbol{\eta_{\alpha}^*}\boldsymbol{\kappa_{\alpha}^{\prime*}} & -\boldsymbol{\eta_{\alpha}}\boldsymbol{\kappa_{\alpha}}
\end{pmatrix}
\boldsymbol{N_{E_{\alpha}S}}(t)\right],
\end{align}
and it solution is
\begin{align}
\label{A28}
\boldsymbol{N_S}(t,\tau)=\boldsymbol{U}(t,t_0)\boldsymbol{N_S}(t_0)\boldsymbol{U}^{\dagger}(\tau,t_0)+\boldsymbol{V_C}(t,\tau)
\end{align}
where the retarded Green function $\boldsymbol{U}(t,t_0)$ satisfies the differential-integral equation Eq.~(\ref{U_eq}), while the noise-induced correlation Green function of the initially correlated state $\boldsymbol{V_C}(t,\tau)$ is diven by
\begin{align}
\label{VC}
\boldsymbol{V_C}(t,\tau)=\int_{t_0}^t d\tau_1\int_{t_0}^\tau d\tau_2\boldsymbol{U}(t,\tau_1)\left[\tilde{\bm{g}}^{+}(\tau_1,\tau_2)+\bar{\bm{g}}(\tau_1,\tau_2)+\bar{\bm{g}}^\dag(\tau_2,\tau_1)\right]\boldsymbol{U}^{\dagger}(\tau,\tau_2).
\end{align}
The effects of the initial correlations between the system and the environment $\langle \boldsymbol{a^\dagger}(t_0)\boldsymbol{b_{\alpha}}(t_0)\rangle$ and $\langle \boldsymbol{a}(t_0)\boldsymbol{b_{\alpha}}(t_0)\rangle$ are manifested in the additional terms of $\bar{\bm{g}}(\tau_1,\tau_2)=\sum_{\alpha}\bar{\bm{g}}_{\alpha}(\tau_1,\tau_2)$ in the integral kernel, which is given by
\begin{align}
\label{gbar_NES}
\bar{\bm{g}}_{\alpha}(\tau_1,\tau_2)=-2i\sum\limits_{j\alpha k}\begin{pmatrix}
\eta_{\alpha k}\kappa_{\alpha j} e^{-i(\epsilon_{\alpha k}+U_\alpha)(\tau_1-t_0)/\hbar}&
-\eta_{\alpha k}\kappa'_{\alpha j} e^{i(\epsilon_{\alpha k}+U_\alpha)(\tau_1-t_0)/\hbar}\\
\eta^*_{\alpha k}\kappa^{\prime*}_{\alpha j} e^{-i(\epsilon_{\alpha k}+U_\alpha)(\tau_1-t_0)/\hbar}&
-\eta^*_{\alpha k}\kappa^*_{\alpha j} e^{i(\epsilon_{\alpha k}+U_\alpha)(\tau_1-t_0)/\hbar}
\end{pmatrix}[\bm{N_{E_\alpha S}}(t_0)]_{jk}\delta(\tau_2-t_0).
\end{align}
The initial correlation $\bm{N_{E_\alpha S}}(t_0)$ can be exactly solved by diagonalizing the total system $H_{\rm{tot}}=\sum\limits_{\alpha k}\epsilon_{\alpha k}c_{\alpha k}^\dag c_{\alpha k}$ with the following transformation \cite{YLZ15}
\begin{subequations}
\label{tot_diagonal}
\begin{align}
\begin{pmatrix}
a_j^\dag\\ a_j
\end{pmatrix}=&\sum_{i\alpha k}\begin{pmatrix}
\eta_{\alpha k}^*\kappa^*_{\alpha i}& -\eta_{\alpha k}\kappa'_{\alpha i}\\
-\eta_{\alpha k}^*\kappa^{\prime*}_{\alpha i}& \eta_{\alpha k}\kappa_{\alpha i}
\end{pmatrix}\bm{G}_{ij}(\epsilon_{\alpha k})\begin{pmatrix}
c_{\alpha k}^\dag\\ c_{\alpha k}
\end{pmatrix}\\
\begin{pmatrix}
b_{\alpha k}^\dag\\ b_{\alpha k}
\end{pmatrix}=&\sum_{ij}\begin{pmatrix}
\eta_{\alpha k}\kappa_{\alpha i}& \eta^*_{\alpha k}\kappa'^*_{\alpha i}\\
\eta_{\alpha k}\kappa_{\alpha i}^{\prime}& \eta^*_{\alpha k}\kappa_{\alpha i}^*
\end{pmatrix}\bm{Z}_{ij\alpha k}^{-1}\begin{pmatrix}
(\eta^*_{\alpha k}\kappa^*_{\alpha j}-\eta_{\alpha k}\kappa'_{\alpha j})c_{\alpha k}^\dag\\
(\eta_{\alpha k}\kappa_{\alpha j}-\eta^*_{\alpha k}\kappa'^*_{\alpha j})c_{\alpha k}
\end{pmatrix}
+\sum_{ij\alpha' k'}\bm{Z}_{ij\alpha' k'}\dfrac{\bm{G}_{ij}(\epsilon_{\alpha' k'})}{\epsilon_{\alpha' k'}-\epsilon_{\alpha k}+i\delta}\begin{pmatrix}
c^\dag_{\alpha' k'}\\ c_{\alpha' k'}
\end{pmatrix},
\end{align}
\end{subequations}
where $\delta\rightarrow0^+$ and
\begin{align}
\bm{Z}_{ij\alpha k}=|\eta_{\alpha k}|^2\begin{pmatrix}
\kappa_{ik}\kappa^*_{jk}-\kappa'^*_{ik}\kappa'_{jk}&
\kappa_{ik}\kappa'^*_{jk}-\kappa'^*_{ik}\kappa_{jk}\\
\kappa^*_{ik}\kappa'_{jk}-\kappa'_{ik}\kappa^*_{jk}&
\kappa_{ik}^*\kappa_{jk}-\kappa'_{ik}\kappa'^*_{jk}
\end{pmatrix}.
\end{align}
The Green function $\bm{G}(\epsilon_{\alpha k})$ is related to the self-energy function $\bm{\Sigma}(\epsilon_{\alpha k})$, and they are given by
\begin{align}
\bm{G}(\epsilon_{\alpha k})=[\epsilon_{\alpha k}\bm{I}-\bm{\epsilon_S}-\bm{\Sigma}(\epsilon_{\alpha k})]^{-1},\ 
\bm{\Sigma}(\epsilon_{\alpha k})=\sum_{\alpha'}\int\dfrac{d\epsilon}{2\pi}\dfrac{\sqrt{\bm{\mathcal{J}^+}_{\alpha'}(\epsilon)\bm{\mathcal{J}^-}_{\alpha'}(\epsilon)}}{\epsilon_{\alpha k}-\epsilon}.
\end{align}
Then the initial correlation can be expressed as
\begin{align}
[\bm{N_{E_\alpha S}}]_{jk}(t_0)=&\sum_{i}\begin{pmatrix}
\eta_{\alpha k}^*\kappa^*_{\alpha i}& -\eta_{\alpha k}\kappa'_{\alpha i}\\
-\eta_{\alpha k}^*\kappa^{\prime*}_{\alpha i}& \eta_{\alpha k}\kappa_{\alpha i}
\end{pmatrix}\bm{\chi}_{ij}(\epsilon_{\alpha k}),
\end{align}
where
\begin{align}
\label{chi}
\bm{\chi}_{ij}(\epsilon_{\alpha k})=&\bm{G}_{ij}(\epsilon_{\alpha k})\sum_{i'j'}\begin{pmatrix}
(\eta_{\alpha k}\kappa_{\alpha j'}-\eta^*_{\alpha k}\kappa'^*_{\alpha j'})n_0(\epsilon_{\alpha k})&0\\
0&(\eta^*_{\alpha k}\kappa^*_{\alpha j'}-\eta_{\alpha k}\kappa'_{\alpha j'})[1-n_0(\epsilon_{\alpha k})]
\end{pmatrix}\bm{Z}^{\dag-1}_{i'j'\alpha k}\begin{pmatrix}
\eta_{\alpha k}\kappa_{\alpha i'}& \eta^*_{\alpha k}\kappa'^*_{\alpha i'}\\
\eta_{\alpha k}\kappa_{\alpha i'}^{\prime}& \eta^*_{\alpha k}\kappa_{\alpha i'}^*
\end{pmatrix}\notag\\
&+\sum_{i'j'\alpha'k'}\bm{G}_{ij}(\epsilon_{\alpha' k'})\begin{pmatrix}
n_0(\epsilon_{\alpha' k'})&0\\
0&1-n_0(\epsilon_{\alpha' k'})
\end{pmatrix}\dfrac{\bm{G}_{i'j'}^\dag(\epsilon_{\alpha' k'})}{\epsilon_{\alpha' k'}-\epsilon_{\alpha k}-i\delta}\bm{Z}_{i'j'\alpha'k'}^\dag,
\end{align}
and $n_0(\epsilon_{\alpha k})=[e^{(\epsilon_{\alpha k}-\mu_0)/k_BT_0}+1]^{-1}$ is the Fermi-Dirac distribution with the initial chemical potential $\mu_0$ and the initial temperature $T_0$ of the total system.
Then Eq.~(\ref{gbar_NES}) can be simply expressed as
\begin{align}
\bm{\bar{g}}_{\alpha}(\tau_1,\tau_2)=-2i\delta(\tau_2-t_0)\int\dfrac{d\omega}{2\pi}\begin{pmatrix}
e^{-i(\omega+\frac{U_\alpha}{\hbar})(\tau_1-t_0)}&0\\
0&e^{i(\omega+\frac{U_\alpha}{\hbar})(\tau_1-t_0)}
\end{pmatrix}\bm{\mathcal{J}}_\alpha^+(\omega)\bm{\chi}(\omega),
\end{align}
which is the last equation in Eq.~(\ref{integral kernel}). It can be expressed as the combination of particle channel and hole channel $\bm{\bar{g}}_{\alpha}(\tau_1,\tau_2)=\bm{\bar{g}}^p_{\alpha}(\tau_1,\tau_2)+\bm{\bar{g}}^h_{\alpha}(\tau_1,\tau_2)$, where
\begin{subequations}
\label{gbar_ph}
\begin{align}
\bar{\bm{g}}^p_{\alpha}(\tau_1,\tau_2)=&-2i\sum\limits_{j,k}\begin{pmatrix}
\eta_{\alpha k}\kappa_{\alpha j}\\
\eta^*_{\alpha k}\kappa^{\prime*}_{\alpha j}
\end{pmatrix}
\begin{pmatrix}
\langle a_j^\dagger(t_0)b_{\alpha k}(t_0)\rangle & \langle a_j(t_0)b_{\alpha k}(t_0)\rangle\\
\end{pmatrix}
e^{-\frac{i}{\hbar}(\epsilon_{\alpha k}-U_\alpha)(\tau_1-t_0)}\delta(\tau_2-t_0)\\
=&-2i\delta(\tau_2-t_0)\int\dfrac{d\omega}{2\pi}e^{-i(\omega+\frac{U_\alpha}{\hbar})(\tau_1-t_0)}\bm{\mathcal{J}}_\alpha(\omega)\bm{\chi}(\omega)\\
\bar{\bm{g}}^h_{\alpha}(\tau_1,\tau_2)=&2i\sum\limits_{j,k}\begin{pmatrix}
\eta_{\alpha k}\kappa'_{\alpha j}\\
\eta^*_{\alpha k}\kappa^*_{\alpha j}
\end{pmatrix}
\begin{pmatrix}
\langle a_j^\dagger(t_0)b_{\alpha k}^\dagger(t_0)\rangle & \langle a_j(t_0)b_{\alpha k}^\dagger(t_0)\rangle
\end{pmatrix}
e^{\frac{i}{\hbar}(\epsilon_{\alpha k}-U_\alpha)(\tau_1-t_0)}\delta(\tau_2-t_0)\notag\\
=&-2i\delta(\tau_2-t_0)\int\dfrac{d\omega}{2\pi}e^{i(\omega+\frac{U_\alpha}{\hbar})(\tau_1-t_0)}\bm{\mathcal{J}'}_\alpha(-\omega)\bm{\chi}(\omega).
\end{align}
\end{subequations}
The other term of integral kernel $\tilde{\bm{g}}^+$ in Eq.~(\ref{VC}) is given by Eq.~(\ref{integral kernel}). Note that in partition-free scheme, the particle distribution of lead $\alpha$, $n_\alpha(\epsilon_{\alpha k})=\langle b_{\alpha k}^\dag b_{\alpha k}\rangle$, should be calculated from Eq.~(\ref{tot_diagonal})
\begin{align}
\label{PFN}
n_\alpha(\epsilon_{\alpha k})\!=\!&\sum_{ii'jj'}
\!\begin{pmatrix}
\eta_{\alpha k}\kappa_{\alpha i}& \eta^*_{\alpha k}\kappa'^*_{\alpha i}
\end{pmatrix}\bm{Z}_{ij\alpha k}^{-1}\begin{pmatrix}
\lambda_{j\alpha k}^*\lambda_{j'\alpha k}n_0(\epsilon_{\alpha k})&0\\
0&\lambda_{j\alpha k}\lambda^*_{j'\alpha k}[1\!-\!n_0(\epsilon_{\alpha k})]
\end{pmatrix}\bm{Z}_{i'j'\alpha k}^{\dag-1}\begin{pmatrix}
\eta^*_{\alpha k}\kappa_{\alpha i}^*\\
\eta_{\alpha k}\kappa_{\alpha i}^{\prime}
\end{pmatrix}\notag\\
&+\sum_{ii'jj'\alpha'k'}\bm{Z}_{ij\alpha' k'}\dfrac{\bm{G}_{ij}(\epsilon_{\alpha' k'})}{\epsilon_{\alpha' k'}-\epsilon_{\alpha k}+i\delta}\begin{pmatrix}
n_0(\epsilon_{\alpha' k'})&0\\
0&1\!-\!n_0(\epsilon_{\alpha' k'})
\end{pmatrix}\dfrac{\bm{G}_{i'j'}^\dag(\epsilon_{\alpha' k'})}{\epsilon_{\alpha' k'}-\epsilon_{\alpha k}-i\delta}\bm{Z}^\dag_{i'j'\alpha' k'}
\end{align}
where $\lambda_{j\alpha k}=\eta_{\alpha k}\kappa_{\alpha j}-\eta^*_{\alpha k}\kappa'^*_{\alpha j}$.

By the solution of Eq.~(\ref{A28}), the time derivative to the correlation function $\boldsymbol{N_S}(t,t)$ of Eq.~(\ref{A27}) can be expressed as
\begin{align}
\label{A31}
\boldsymbol{\dot{N}_S}(t,t)=-\left[\frac{i}{\hbar}\begin{pmatrix}
\boldsymbol{\varepsilon}&\boldsymbol{0}\\
\boldsymbol{0}&-\boldsymbol{\varepsilon}
\end{pmatrix}+\boldsymbol{\mathcal{K}}(t,t_0)\right]\boldsymbol{N_S}(t,t)-\boldsymbol{\Lambda_C}(t,t_0)
\end{align}
where 
\begin{subequations}
\begin{align}
\label{A_kappa}
&\boldsymbol{\mathcal{K}}(t,t_0)=-\frac{i}{\hbar}\begin{pmatrix}
\boldsymbol{\varepsilon}&\boldsymbol{0}\\
\boldsymbol{0}&-\boldsymbol{\varepsilon}
\end{pmatrix}-\boldsymbol{\dot{U}}(t,t_0)\boldsymbol{U}^{-1}(t,t_0)\\
&\boldsymbol{\Lambda_C}(t,t_0)=\boldsymbol{\dot{U}}(t,t_0)\boldsymbol{U}^{-1}(t,t_0)\boldsymbol{V_C}(t,t)-\boldsymbol{\dot{V_C}}(t,t)
\end{align}
\end{subequations}
Substituting this result into Eq.~(\ref{A27}), we have
\begin{align}
\frac{i}{\hbar}
\sum\limits_{\alpha}\begin{pmatrix}
\boldsymbol{\eta_{\alpha}^*}\boldsymbol{\kappa_{\alpha}^*} & -\boldsymbol{\eta_{\alpha}}\boldsymbol{\kappa_{\alpha}^\prime}\\
\boldsymbol{\eta_{\alpha}^*}\boldsymbol{\kappa_{\alpha}^{\prime*}} & -\boldsymbol{\eta_{\alpha}}\boldsymbol{\kappa_{\alpha}}
\end{pmatrix}
\boldsymbol{N_{E_{\alpha}S}}(t)=\boldsymbol{\mathcal{K}}(t,t_0)\boldsymbol{N_S}(t)+\boldsymbol{\Lambda_C}(t,t_0)
\end{align}
Now, combining all these results together with the aid of the relation
\begin{align}
\begin{pmatrix}
\boldsymbol{a}\rho_S(t)\\
\rho_S(t)(\boldsymbol{a^\dagger})^T
\end{pmatrix}=\begin{pmatrix}
\boldsymbol{\Omega_S}(t)& \boldsymbol{\Pi_S}(t)\\
\boldsymbol{\Pi_S}(t)& \boldsymbol{\Omega_S}(t)
\end{pmatrix}\begin{pmatrix}
\rho_S(t)\boldsymbol{a}\\
(\boldsymbol{a^\dagger})^T\rho_S(t),
\end{pmatrix}
\end{align}
then we can finally solve the collective operator
\begin{align}
\label{A35}
\begin{pmatrix}
\boldsymbol{A}(t)\\
-\boldsymbol{A^\dagger}(t)^T
\end{pmatrix}&=\dfrac{1}{i\hbar}\sum\limits_{\alpha}\begin{pmatrix}
\boldsymbol{\eta_{\alpha}^*}\boldsymbol{\kappa_{\alpha}^*} & -\boldsymbol{\eta_{\alpha}}\boldsymbol{\kappa_{\alpha}^\prime}\\
\boldsymbol{\eta_{\alpha}^*}\boldsymbol{\kappa_{\alpha}^{\prime*}} & -\boldsymbol{\eta_{\alpha}}\boldsymbol{\kappa_{\alpha}}
\end{pmatrix}
\begin{pmatrix}
\boldsymbol{1}+\boldsymbol{\Omega_{E_{\alpha}E_{\alpha}}}(t) & -\boldsymbol{\Pi_{E_{\alpha}E_{\alpha}}}(t)\\
-\boldsymbol{\Pi_{E_{\alpha}E_{\alpha}}^*}(t) & \boldsymbol{1}+\boldsymbol{\Omega^*_{E_{\alpha}E_{\alpha}}}(t)
\end{pmatrix}^{-1}\begin{pmatrix}
\boldsymbol{\Omega_{E_{\alpha}S}}(t) & \boldsymbol{\Pi_{E_{\alpha}S}}(t)\\
-\boldsymbol{\Pi^*_{E_{\alpha}S}}(t) & -\boldsymbol{\Omega^*_{E_{\alpha}S}}(t)
\end{pmatrix}\begin{pmatrix}
\rho_S(t)\boldsymbol{a}\\
\boldsymbol{a^\dagger}\rho_S(t)
\end{pmatrix}\notag\\
&=\dfrac{1}{i\hbar}\sum\limits_{\alpha}\begin{pmatrix}
\boldsymbol{\eta_{\alpha}^*}\boldsymbol{\kappa_{\alpha}^*} & -\boldsymbol{\eta_{\alpha}}\boldsymbol{\kappa_{\alpha}^\prime}\\
\boldsymbol{\eta_{\alpha}^*}\boldsymbol{\kappa_{\alpha}^{\prime*}} & -\boldsymbol{\eta_{\alpha}}\boldsymbol{\kappa_{\alpha}}
\end{pmatrix}
\boldsymbol{N_{E_{\alpha}S}}
\left[\boldsymbol{1}-\begin{pmatrix}
\boldsymbol{\Omega_S}(t)&\boldsymbol{\Pi_S}(t)\\
\boldsymbol{\Pi_S^*}(t)&\boldsymbol{\Omega_S^*}(t)
\end{pmatrix}\right]^{-1}
\begin{pmatrix}
\rho_S(t)\boldsymbol{a}\\
\boldsymbol{a^\dagger}\rho_S(t)
\end{pmatrix}\notag\\
&=\left[\boldsymbol{\mathcal{K}}(t,t_0)\boldsymbol{N_S}(t)+\boldsymbol{\Lambda_C}(t,t_0)\right]
\left[\boldsymbol{1}-\begin{pmatrix}
\boldsymbol{\Omega_S}(t)&\boldsymbol{\Pi_S}(t)\\
\boldsymbol{\Pi_S^*}(t)&\boldsymbol{\Omega_S^*}(t)
\end{pmatrix}\right]^{-1}
\begin{pmatrix}
\rho_S(t)\boldsymbol{a}\\
\boldsymbol{a^\dagger}\rho_S(t)
\end{pmatrix}\notag\\
&=\left\{\boldsymbol{\mathcal{K}}(t,t_0)
\begin{pmatrix}
\boldsymbol{\Omega_{S}}(t) & \boldsymbol{\Pi_{S}}(t)\\
-\boldsymbol{\Pi_{S}^*}(t) & -\boldsymbol{\Omega^*_{S}}(t)
\end{pmatrix}\left[\boldsymbol{1}-\begin{pmatrix}
\boldsymbol{\Omega_{S}}(t) & \boldsymbol{\Pi_{S}}(t)\\
\boldsymbol{\Pi_{S}^*}(t) & \boldsymbol{\Omega^*_{S}}(t)
\end{pmatrix}\right]^{-1}
+\boldsymbol{\Lambda_C}(t,t_0)\right\}\notag\\
&\hspace{5mm}\times\left[\boldsymbol{1}-\begin{pmatrix}
\boldsymbol{\Omega_S}(t)&\boldsymbol{\Pi_S}(t)\\
\boldsymbol{\Pi_S^*}(t)&\boldsymbol{\Omega_S^*}(t)
\end{pmatrix}\right]^{-1}
\begin{pmatrix}
\rho_S(t)\boldsymbol{a}\\
\boldsymbol{a^\dagger}\rho_S(t)
\end{pmatrix}\notag\\
&=-\boldsymbol{\mathcal{K}}(t,t_0)\begin{pmatrix}
\boldsymbol{a}\rho_S(t)\\
\boldsymbol{a^\dagger}\rho_S(t)
\end{pmatrix}-\boldsymbol{\Lambda_C}(t,t_0)\begin{pmatrix}
\rho_S(t)\boldsymbol{a}+\boldsymbol{a}\rho_S(t)\\
\rho_S(t)\boldsymbol{a^\dagger}+\boldsymbol{a^\dagger}\rho_S(t)
\end{pmatrix}.
\end{align}
One can find that it exactly the same as the result of the initially decoupled state, except that $\boldsymbol{\Lambda_D}(t,t_0)$ is replaced by $\boldsymbol{\Lambda_C}(t,t_0)$.
By combining Eq.~(\ref{A2}), Eq.~(\ref{A31}), and Eq.~(\ref{A35}), we can obtain Eq.~(\ref{I_superoperator}), which describes a simple relation between the transient electron transport current and the superoperators in the master equation
\begin{align}
I_T(t)=e\Tr_S[\mathcal{L^+}(t)\rho_S(t)]=-e\Tr_S[\mathcal{L^-}(t)\rho_S(t)].
\end{align}
The resulting transport current is given by
\begin{align}
\label{A_current}
I_{T}(t)&=-\frac{e}{\hbar^2}\sum\limits_{\alpha}{\rm Tr}\Big[\int^{t}_{t_0}d\tau\bm{g}^{-}_{\alpha}(t,\tau)\tilde{\bm{\rho}}(\tau,t)-\int^{t}_{t_0}d\tau\tilde{\bm{g}}^{-}_{\alpha}(t,\tau)\bm{U}^{\dag}(t,\tau)+H.c.\Big],
\end{align}
where
\begin{align}
\label{Arho}
\tilde{\bm{\rho}}&(\tau,t)=\bm{U}(\tau,t_0)\tilde{\bm{\rho}}(t_0,t_0)\bm{U}^{\dag}(t,t_0)+\frac{1}{\hbar^2}\int^{\tau}_{t_0}d\tau_1\int^{t}_{t_0}d\tau_2
\bm{U}(\tau,\tau_1)\left[\tilde{\bm{g}}^{+}(\tau_1,\tau_2)+\bar{\bm{g}}(\tau_1,\tau_2)\right]\bm{U}^{\dag}(t,\tau_2),
\end{align}
as given in Eq.~(\ref{I_alpha}) and Eq.~(\ref{rho_tilde}).

\subsection{Partitioned scheme}
\label{App1}
Now we consider the case that the system and the environment are initially decoupled, and the environment is initially in a thermal state, that is
\begin{align}
\rho_{tot}(t_0)=\rho_S(t_0)\otimes\rho_E(t_0),\ \rho_E(t_0)=\frac{1}{Z_E}\exp\left[-\frac{H_{lead}}{k_BT}\right],
\end{align}
where the initial state of the system $\rho_S(t_0)$ can be arbitrary.
To complete the partial trace in Eq.~(\ref{A3}), we shall use the coherent state path integral method \cite{TZ08,JTZ10,LZ12,ZLX12,Z19,LYH18,YZ20,HYZ20,HZ22,HZ22a,HZ22b}. In the coherent state representation, the matrix element of the collective operator Eq.~(\ref{A3}) can be expressed as
\begin{align}
\label{A5}
&\left\langle\boldsymbol{\xi_t}\left\vert\begin{pmatrix}
\boldsymbol{A}(t)\\
\boldsymbol{A^\dagger}(t)^T
\end{pmatrix}
\right\vert\boldsymbol{\xi^\prime_t}\right\rangle=
\int d\mu(\boldsymbol{\xi_0^*},\boldsymbol{\xi_0})d\mu(\boldsymbol{\xi_0^{\prime*}},\boldsymbol{\xi_0^\prime})\left\langle\boldsymbol{\xi_0}\left\vert
\rho_{S}(t_0)
\right\vert\boldsymbol{\xi_0^\prime}\right\rangle\boldsymbol{\mathcal{K}^A}(\boldsymbol{\xi_t^*},\boldsymbol{\xi_t^\prime},t;\boldsymbol{\xi_0},\boldsymbol{\xi_0^{\prime*}},t_0),
\end{align}
where $d\mu(\boldsymbol{\xi})=\prod_j d\xi_j^* d\xi_j e^{-|\xi_{j}|^2}$.
In the above equation, we define the A-operator associated propagating function $\boldsymbol{\mathcal{K}^A}(\boldsymbol{\xi_t^*},\boldsymbol{\xi_t^\prime},t;\boldsymbol{\xi_0},\boldsymbol{\xi_0^{\prime*}},t_0)$ in a similar way as the propagating function for the reduced density matrix in the coherent state representation \cite{TZ08,JTZ10,LZ12,LYH18,YZ20,HZ22b}
\begin{align}
\label{A6}
&\left\langle\boldsymbol{\xi_t}\left\vert\rho_S(t)
\right\vert\boldsymbol{\xi^\prime_t}\right\rangle=
\int d\mu(\boldsymbol{\xi_0^*},\boldsymbol{\xi_0})d\mu(\boldsymbol{\xi_0^{\prime*}},\boldsymbol{\xi_0^\prime})\left\langle\boldsymbol{\xi_0}\left\vert
\rho_{S}(t_0)
\right\vert\boldsymbol{\xi_0^\prime}\right\rangle\mathcal{K}(\boldsymbol{\xi_t^*},\boldsymbol{\xi_t^\prime},t;\boldsymbol{\xi_0},\boldsymbol{\xi_0^{\prime*}},t_0).
\end{align}
The A-operator associated propagating function $\boldsymbol{\mathcal{K}^A}(\boldsymbol{\xi_t^*},\boldsymbol{\xi_t^\prime},t;\boldsymbol{\xi_0},\boldsymbol{\xi_0^{\prime*}},t_0)$ fully determine the evolution of the collective operator $\boldsymbol{A}(t)$, while the propagating function $\mathcal{K}(\boldsymbol{\xi_t^*},\boldsymbol{\xi_t^\prime},t;\boldsymbol{\xi_0},\boldsymbol{\xi_0^{\prime*}},t_0)$ fully describes the time evolution of the reduced density matrix $\rho_S(t)$, and both of them can be obtained by utilizing the coherent state path integrals
\begin{subequations}
\begin{align}
\label{A7}
&\boldsymbol{\mathcal{K}^A}(\boldsymbol{\xi_t^*},\boldsymbol{\xi_t^\prime},t;\boldsymbol{\xi_0},\boldsymbol{\xi_0^{\prime*}},t_0)=\int_{\boldsymbol{\xi_0},\boldsymbol{\xi_0^{\prime*}}}^{\boldsymbol{\xi_t^*},\boldsymbol{\xi_t^\prime}} D[\boldsymbol{\xi^*},\boldsymbol{\xi},\boldsymbol{\xi^{\prime*}},\boldsymbol{\xi^\prime}]\exp[\frac{i}{\hbar}(S_S[\boldsymbol{\xi^*},\boldsymbol{\xi}]-S_S[\boldsymbol{\xi^{\prime*}},\boldsymbol{\xi^\prime}])]\boldsymbol{\mathcal{F}^A}[\boldsymbol{\xi^*},\boldsymbol{\xi},\boldsymbol{\xi^{\prime*}},\boldsymbol{\xi^\prime}]\\
&\mathcal{K}(\boldsymbol{\xi_t^*},\boldsymbol{\xi_t^\prime},t;\boldsymbol{\xi_0},\boldsymbol{\xi_0^{\prime*}},t_0)=\int_{\boldsymbol{\xi_0},\boldsymbol{\xi_0^{\prime*}}}^{\boldsymbol{\xi_t^*},\boldsymbol{\xi_t^\prime}} D[\boldsymbol{\xi^*},\boldsymbol{\xi},\boldsymbol{\xi^{\prime*}},\boldsymbol{\xi^\prime}]\exp[\frac{i}{\hbar}(S_S[\boldsymbol{\xi^*},\boldsymbol{\xi}]-S_S[\boldsymbol{\xi^{\prime*}},\boldsymbol{\xi^\prime}])]\mathcal{F}[\boldsymbol{\xi^*},\boldsymbol{\xi},\boldsymbol{\xi^{\prime*}},\boldsymbol{\xi^\prime}].
\end{align}
\end{subequations}
The A-operator associated influence functional $\boldsymbol{\mathcal{F}^A}[\boldsymbol{\xi^*},\boldsymbol{\xi},\boldsymbol{\xi^{\prime*}},\boldsymbol{\xi^\prime}]$, after taking the partial trace over the environment states, can be reduced as
\begin{align}
&\boldsymbol{\mathcal{F}^A}[\boldsymbol{\xi^*},\boldsymbol{\xi},\boldsymbol{\xi^{\prime*}},\boldsymbol{\xi^\prime}]
=-\dfrac{1}{\hbar^2}\int_{t_0}^t d\tau\left[\bm{g}^{+}(t,\tau)\begin{pmatrix}
\boldsymbol{\xi}(\tau)\\
\boldsymbol{\xi^*}(\tau)
\end{pmatrix}+\tilde{\bm{g}}^{+}(t,\tau)\begin{pmatrix}
\boldsymbol{\xi}(\tau)+\boldsymbol{\xi'}(\tau)\\
\boldsymbol{\xi^*}(\tau)+\boldsymbol{\xi^{\prime*}}(\tau)
\end{pmatrix}
\right]\mathcal{F}[\boldsymbol{\xi^*},\boldsymbol{\xi},\boldsymbol{\xi^{\prime*}},\boldsymbol{\xi^\prime}],
\end{align}
where the integral kernels $\bm{g}^{+}(t,\tau)$ and $\tilde{\bm{g}}^{+}(t,\tau)$ are the two-time correlation functions given by Eq.~(\ref{integral kernel}), and $\mathcal{F}[\boldsymbol{\xi^*},\boldsymbol{\xi},\boldsymbol{\xi^{\prime*}},\boldsymbol{\xi^\prime}]$ is the influence functional of the reduced density matrix \cite{TZ08,JTZ10,LZ12,LYH18,HZ22b}, which is given by
\begin{align}
\mathcal{F}[\boldsymbol{\xi^*},\boldsymbol{\xi},\boldsymbol{\xi^{\prime*}},\boldsymbol{\xi^\prime}]=\exp\bigg\{-\frac{1}{2\hbar^2}\int_{t_0}^t d\tau &\bigg[\int_{t_0}^\tau d\tau'\begin{pmatrix}
\boldsymbol{\xi^*}(\tau)+\boldsymbol{\xi^{\prime*}}(\tau) & 
\boldsymbol{\xi}(\tau)+\boldsymbol{\xi^{\prime}}(\tau)
\end{pmatrix}\boldsymbol{g^+}(\tau,\tau')\begin{pmatrix}
\boldsymbol{\xi}(\tau')\\
\boldsymbol{\xi^*}(\tau')
\end{pmatrix}\notag\\
&-\int_\tau^t d\tau'\begin{pmatrix}
\boldsymbol{\xi^{\prime*}}(\tau) & 
\boldsymbol{\xi^{\prime}}(\tau)
\end{pmatrix}\boldsymbol{g^+}(\tau,\tau')\begin{pmatrix}
\boldsymbol{\xi}(\tau')+\boldsymbol{\xi'}(\tau')\\
\boldsymbol{\xi^*}(\tau')+\boldsymbol{\xi^{\prime*}}(\tau')
\end{pmatrix}\notag\\
&+\int_{t_0}^t d\tau' \begin{pmatrix}
\boldsymbol{\xi^*}(\tau)+\boldsymbol{\xi^{\prime*}}(\tau) & 
\boldsymbol{\xi}(\tau)+\boldsymbol{\xi^{\prime}}(\tau)
\end{pmatrix}\boldsymbol{\tilde{g}^+}(\tau,\tau')\begin{pmatrix}
\boldsymbol{\xi}(\tau')+\boldsymbol{\xi'}(\tau')\\
\boldsymbol{\xi^*}(\tau')+\boldsymbol{\xi^{\prime*}}(\tau')
\end{pmatrix}
\bigg]\bigg\}
\end{align}
The path integrals in the A-operator associated propagating function of Eq.~(\ref{A7}) can be exactly carried out by using the stationary-path approach \cite{TZ08,JTZ10,LZ12,LYH18,HZ22b}, and the result is
\begin{align}
\label{A10}
\boldsymbol{\mathcal{K}^A}(\boldsymbol{\xi_t^*},\boldsymbol{\xi_t^\prime},t;\boldsymbol{\xi_0},\boldsymbol{\xi_0^{\prime*}},t_0)=\left[\begin{pmatrix}
\boldsymbol{\dot{\xi}}(t)\\
\boldsymbol{\dot{\xi}^*}(t)
\end{pmatrix}+\dfrac{i}{\hbar}\begin{pmatrix}
\boldsymbol{\epsilon}&\boldsymbol{0}\\
\boldsymbol{0}&-\boldsymbol{\epsilon}
\end{pmatrix}\begin{pmatrix}
\boldsymbol{\xi}(t)\\
\boldsymbol{\xi^*}(t)
\end{pmatrix}\right]\mathcal{K}(\boldsymbol{\xi_t^*},\boldsymbol{\xi_t^\prime},t;\boldsymbol{\xi_0},\boldsymbol{\xi_0^{\prime*}},t_0),
\end{align}
Its solution that obtained by the stationary-path approach is given by \cite{TZ08,JTZ10,LZ12,LYH18,HZ22b}
\begin{align}
\label{A11}
\mathcal{K}(\boldsymbol{\xi_t^*},\boldsymbol{\xi_t^\prime},t;\boldsymbol{\xi_0},\boldsymbol{\xi_0^{\prime*}},t_0)=\mathcal{N}(t)\exp\left\{\frac{1}{2}\left[
\boldsymbol{\xi_t^*}\boldsymbol{\xi}(t)+\boldsymbol{\xi^*}(t_0)\boldsymbol{\xi_0}+\boldsymbol{\xi^{\prime*}}(t)\boldsymbol{\xi_t^\prime}+\boldsymbol{\xi}_0^{\prime*}\boldsymbol{\xi'}(t_0)\right]\right\},
\end{align}
where $\mathcal{N}(t)$ is the renormalized constant, and $\boldsymbol{\xi}(t),\boldsymbol{\xi^*}(t_0),\boldsymbol{\xi^{\prime*}}(t)$, and $\boldsymbol{\xi'}(t_0)$ are determined by the stationary path
\begin{subequations}
\begin{align}
\label{A12a}
&\dfrac{d}{d\tau}\begin{pmatrix}
\boldsymbol{\xi}(\tau)+\boldsymbol{\xi'}(\tau)\\
\boldsymbol{\xi^*}(\tau)+\boldsymbol{\xi^{\prime*}}(\tau)
\end{pmatrix}+\dfrac{i}{\hbar}\begin{pmatrix}
\boldsymbol{\epsilon}&\boldsymbol{0}\\
\boldsymbol{0}&-\boldsymbol{\epsilon}
\end{pmatrix}\begin{pmatrix}
\boldsymbol{\xi}(\tau)+\boldsymbol{\xi'}(\tau)\\
\boldsymbol{\xi^*}(\tau)+\boldsymbol{\xi^{\prime*}}(\tau)
\end{pmatrix}-\dfrac{1}{\hbar^2}\int_{\tau}^t d\tau' \boldsymbol{g^+}(\tau,\tau')\begin{pmatrix}
\boldsymbol{\xi}(\tau')+\boldsymbol{\xi'}(\tau')\\
\boldsymbol{\xi^*}(\tau')+\boldsymbol{\xi^{\prime*}}(\tau')
\end{pmatrix}=0\\
\label{A12b}
&\dfrac{d}{d\tau}\begin{pmatrix}
\boldsymbol{\xi}(\tau)\\
\boldsymbol{\xi^*}(\tau)
\end{pmatrix}+\dfrac{i}{\hbar}\begin{pmatrix}
\boldsymbol{\epsilon}&\boldsymbol{0}\\
\boldsymbol{0}&-\boldsymbol{\epsilon}
\end{pmatrix}\begin{pmatrix}
\boldsymbol{\xi}(\tau)\\
\boldsymbol{\xi^*}(\tau)
\end{pmatrix}+\dfrac{1}{\hbar^2}\int_{t_0}^\tau d\tau' \boldsymbol{g^+}(\tau,\tau')\begin{pmatrix}
\boldsymbol{\xi}(\tau')\\
\boldsymbol{\xi^*}(\tau')
\end{pmatrix}=-\dfrac{1}{\hbar^2}\int_{t_0}^t d\tau' \boldsymbol{\tilde{g}^+}(\tau,\tau')\begin{pmatrix}
\boldsymbol{\xi}(\tau')+\boldsymbol{\xi'}(\tau')\\
\boldsymbol{\xi^*}(\tau')+\boldsymbol{\xi^{\prime*}}(\tau')
\end{pmatrix}
\end{align}
\end{subequations}
The above equations can be solved by introducing the following transformation \cite{LYH18,YZ20,HZ22b}
\begin{subequations}
\label{A13}
\begin{align}
&\begin{pmatrix}
\boldsymbol{\xi}(\tau)+\boldsymbol{\xi'}(\tau)\\
\boldsymbol{\xi^*}(\tau)+\boldsymbol{\xi^{\prime*}}(\tau)
\end{pmatrix}=\boldsymbol{U^\dagger}(t,\tau)\begin{pmatrix}
\boldsymbol{\xi}(t)+\boldsymbol{\xi_t'}\\
\boldsymbol{\xi^*_t}+\boldsymbol{\xi^{\prime*}}(t)
\end{pmatrix}\\
&\begin{pmatrix}
\boldsymbol{\xi}(\tau)\\
\boldsymbol{\xi^*}(\tau)
\end{pmatrix}=\boldsymbol{U}(\tau,t_0)\begin{pmatrix}
\boldsymbol{\xi_0}\\
\boldsymbol{\xi^*}(t_0)
\end{pmatrix}+\boldsymbol{V_{D}}(\tau,t)\begin{pmatrix}
\boldsymbol{\xi}(t)+\boldsymbol{\xi_t'}\\
\boldsymbol{\xi^*_t}+\boldsymbol{\xi^{\prime*}}(t)
\end{pmatrix},
\end{align}
\end{subequations}
then Eq.~(\ref{A12a}) is reduced to the differential-integral equation of Eq.~(\ref{U_eq}), and Eq.~(\ref{A12b}) is reduced to
\begin{align}
\dfrac{d}{d\tau}\boldsymbol{V_{D}}(\tau,t)+\dfrac{i}{\hbar}\begin{pmatrix}
\boldsymbol{\epsilon}&\boldsymbol{0}\\
\boldsymbol{0}&-\boldsymbol{\epsilon}
\end{pmatrix}\boldsymbol{V_{D}}(\tau,t)+\dfrac{1}{\hbar^2}\int_{t_0}^\tau d\tau' \boldsymbol{g^+}(\tau,\tau')\boldsymbol{V_{D}}(\tau',t)=\dfrac{1}{\hbar^2}\int_{t_0}^t d\tau' \boldsymbol{\tilde{g}^+}(\tau,\tau')\boldsymbol{U^\dagger}(t,\tau').
\end{align}
One can find clearly that the retarded Green function $\boldsymbol{U}(t,t_0)$ is exactly the same as that of the initial correlated state defined in Eq.~(\ref{A28}), while the noise-induced correlation Green function of the initially correlated state $\boldsymbol{V_D}(t,\tau)$ differs from that of the initial decoupled state $\boldsymbol{V_C}(t,\tau)$ of Eq.~(\ref{VC}) in the integral kernel
\begin{align}
\boldsymbol{V_{D}}(\tau,t)=\dfrac{1}{\hbar^2}\int_{t_0}^\tau d\tau_1\int_{t_0}^t d\tau_2 \boldsymbol{U}(\tau,\tau_1) \boldsymbol{\tilde{g}^+}(\tau_1,\tau_2)\boldsymbol{U^\dagger}(t,\tau_2).
\end{align}
Combining Eq.~(\ref{A6}), Eq.~(\ref{A10}), Eq.~(\ref{A11}), and Eq.~(\ref{A13}), we can obtain the solution of Eq.~(\ref{A5})
\begin{align}
\label{A16}
\left\langle\boldsymbol{\xi_t}\left\vert\begin{pmatrix}
\boldsymbol{A}(t)\\
\boldsymbol{A^\dagger}(t)^T
\end{pmatrix}
\right\vert\boldsymbol{\xi'_t}\right\rangle=\left[\boldsymbol{\mathcal{K}}(t,t_0)\begin{pmatrix}
-\frac{\partial}{\partial\boldsymbol{\xi_t^*}}\\
\boldsymbol{\xi_t^*}
\end{pmatrix}+\boldsymbol{\Lambda_D}(t,t_0)\begin{pmatrix}
-\frac{\partial}{\partial\boldsymbol{\xi^*_t}}-\boldsymbol{\xi'_t}\\
\boldsymbol{\xi^*_t}-\frac{\partial}{\partial\boldsymbol{\xi'_t}}
\end{pmatrix}\right]\left\langle\boldsymbol{\xi_t}\left\vert\rho_S(t)
\right\vert\boldsymbol{\xi'_t}\right\rangle,
\end{align}
where $\boldsymbol{\mathcal{K}}(t,t_0)$ is exactly the same as that in Eq.~(\ref{A_kappa}), while $\boldsymbol{\Lambda_D}(t,t_0)$ is give by
\begin{subequations}
\begin{align}
&\boldsymbol{\Lambda_D}(t,t_0)=\boldsymbol{\dot{U}}(t,t_0)\boldsymbol{U^{-1}}(t,t_0)\boldsymbol{V_D}(t,t)-\boldsymbol{\dot{V}_D}(t,t).
\end{align}
\end{subequations}
With the D algebra of the fermionic creation and annihilation operators $a_i|\xi'_i\rangle=\xi'_i|\xi'_i\rangle$, $a^\dagger_i|\xi'_i\rangle=-\frac{\partial}{\partial\xi'_i}|\xi'_i\rangle$, $\langle\xi_i|a_i^\dagger=\langle\xi_i|\xi_i^*$, and $\langle\xi_i|a_i=\frac{\partial}{\partial\xi_i^*}\langle\xi_i|$, Eq.~(\ref{A16}) becomes
\begin{align}
\label{A18}
\begin{pmatrix}
\boldsymbol{A}(t)\\
-\boldsymbol{A^\dagger}(t)^T
\end{pmatrix}
=-\boldsymbol{\mathcal{K}}(t,t_0)\begin{pmatrix}
\boldsymbol{a}\rho_S(t)\\
\boldsymbol{a^\dagger}\rho_S(t)
\end{pmatrix}-\boldsymbol{\Lambda_D}(t,t_0)\begin{pmatrix}
\rho_S(t)\boldsymbol{a}+\boldsymbol{a}\rho_S(t)\\
\rho_S(t)\boldsymbol{a^\dagger}+\boldsymbol{a^\dagger}\rho_S(t)
\end{pmatrix},
\end{align}

On the other hand, the total transient transport current flowing out of the system to the leads is defined as
\begin{align}
\label{A19}
I_T(t)=-e\dfrac{d}{dt}\sum\limits_{\alpha k}\langle b^\dagger_{\alpha k}(t)b_{\alpha k}(t)\rangle=\dfrac{e}{i\hbar}\sum\limits_{j\alpha k}\left[\eta^*_{\alpha k}\kappa^*_{\alpha j}\langle a^\dagger_{j}(t)b_{\alpha k}(t)\rangle+\eta_{\alpha k}\kappa^\prime_{\alpha j}\langle a^\dagger_{j}(t)b^\dagger_{\alpha k}(t)\rangle\right]
\end{align}
Using the Heisenberg equation of motion, we have
\begin{align}
\label{A20}
&\frac{i}{\hbar}\sum\limits_{\alpha}
\begin{pmatrix}
\boldsymbol{\eta_{\alpha}^*}\boldsymbol{\kappa_{\alpha}^*} & -\boldsymbol{\eta_{\alpha}}\boldsymbol{\kappa_{\alpha}^\prime}\\
\boldsymbol{\eta_{\alpha}^*}\boldsymbol{\kappa_{\alpha}^{\prime*}} & -\boldsymbol{\eta_{\alpha}}\boldsymbol{\kappa_{\alpha}}
\end{pmatrix}
\begin{pmatrix}
\langle \boldsymbol{a^\dagger}(t)\boldsymbol{b_{\alpha}}(t)\rangle & \langle \boldsymbol{a}(t)\boldsymbol{b_{\alpha}}(t)\rangle\\
\langle \boldsymbol{a^\dagger}(t)\boldsymbol{b_{\alpha}^\dagger}(t)\rangle & \langle \boldsymbol{a}(t)\boldsymbol{b_{\alpha}^\dagger}(t)\rangle
\end{pmatrix}\notag\\
&=\dfrac{i}{\hbar}\begin{pmatrix}
\boldsymbol{\varepsilon}&\boldsymbol{0}\\
\boldsymbol{0}&-\boldsymbol{\varepsilon}
\end{pmatrix}\begin{pmatrix}
\langle \boldsymbol{a^\dagger}(t)\boldsymbol{a}(t)\rangle & \langle \boldsymbol{a}(t)\boldsymbol{a}(t)\rangle\\
\langle \boldsymbol{a^\dagger}(t)\boldsymbol{a^\dagger}(t)\rangle & \langle \boldsymbol{a}(t)\boldsymbol{a^\dagger}(t)\rangle
\end{pmatrix}
+\boldsymbol{\dot{U}}(t,t_0)\begin{pmatrix}
\langle \boldsymbol{a^\dagger}(t_0)\boldsymbol{a}(t_0)\rangle & \langle \boldsymbol{a}(t_0)\boldsymbol{a}(t_0)\rangle\\
\langle \boldsymbol{a^\dagger}(t_0)\boldsymbol{a^\dagger}(t_0)\rangle & \langle \boldsymbol{a}(t_0)\boldsymbol{a^\dagger}(t_0)\rangle
\end{pmatrix}\boldsymbol{U^{\dagger}}(t,t_0)+\boldsymbol{\dot{V}_D}(t,t)\notag\\
&=\boldsymbol{\mathcal{K}}(t,t_0)\begin{pmatrix}
\langle \boldsymbol{a^\dagger}(t)\boldsymbol{a}(t)\rangle & \langle \boldsymbol{a}(t)\boldsymbol{a}(t)\rangle\\
\langle \boldsymbol{a^\dagger}(t)\boldsymbol{a^\dagger}(t)\rangle & \langle \boldsymbol{a}(t)\boldsymbol{a^\dagger}(t)\rangle
\end{pmatrix}+\boldsymbol{\Lambda_D}(t,t_0)
\end{align}
Combining Eq.~(\ref{A18}), Eq.~(\ref{A19}), Eq.~(\ref{A20}), and Eq.~(\ref{A2}), we can obtain
\begin{align}
I_T(t)&=e\Tr_S[\mathcal{L^+}(t)\rho_S(t)]=-e\Tr_S[\mathcal{L^-}(t)\rho_S(t)]\notag\\
&=-\frac{e}{\hbar^2}\sum\limits_{\alpha}{\rm Tr}\Big[\int^{t}_{t_0}d\tau\bm{g}^{-}_{\alpha}(t,\tau)\tilde{\bm{\rho}}(\tau,t)-\int^{t}_{t_0}d\tau\tilde{\bm{g}}^{-}_{\alpha}(t,\tau)\bm{U}^{\dag}(t,\tau)+H.c.\Big],
\end{align}
which is exactly the same with the result for initially correlated state of Eq.~(\ref{A_current}), except that the reduced density matrix contains no terms of initial-correlation integral kernel $\boldsymbol{\bar{g}}(\tau_1,\tau_2)$
\begin{align}
\tilde{\bm{\rho}}&(\tau,t)=\bm{U}(\tau,t_0)\tilde{\bm{\rho}}(t_0,t_0)\bm{U}^{\dag}(t,t_0)+\frac{1}{\hbar^2}\int^{\tau}_{t_0}d\tau_1\int^{t}_{t_0}d\tau_2
\bm{U}(\tau,\tau_1)\tilde{\bm{g}}^{+}(\tau_1,\tau_2)\bm{U}^{\dag}(t,\tau_2),
\end{align}
Because that the only difference between the two cases is the presence or absence of $\boldsymbol{\bar{g}}(\tau_1,\tau_2)$ which will vanish after taking the differential of bias $\mu$, the differential conductance of initially decoupled state is exactly the same with that of initially correlated state shown as 
\begin{align}
\label{ADC}
\dfrac{dI_\alpha(t)}{d\mu}=\dfrac{e^2}{\hbar}\sum\limits_{\beta}\Tr\left\{
\begin{array}{l}
\int_{t_0}^t d\tau\int\frac{d\omega}{2\pi}\left[\frac{\partial n_\alpha(\omega)}{\partial \mu}\mathcal{\bm{J}}_{\alpha}(\omega)+\frac{\partial n_\alpha(-\omega)}{\partial \mu}\mathcal{\bm{J}}'_{\alpha}(-\omega)\right]e^{-i\omega(t-\tau)}\boldsymbol{U}^{\dagger}(t,\tau)\\
-\int_{t_0}^t d\tau\int_{t_0}^\tau d\tau_1\int_{t_0}^t d\tau_2\int\frac{d\omega}{2\pi}\int\frac{d\omega^\prime}{2\pi}\left[\mathcal{\bm{J}}_{\alpha}(\omega)-\mathcal{\bm{J}}'_{\alpha}(-\omega)\right]e^{-i\omega(t-\tau)}\boldsymbol{U}(\tau,\tau_1)\\
\times\left[\frac{\partial n_{\beta}(\omega^\prime)}{\partial \mu}\mathcal{\bm{J}}_{\beta}(\omega^\prime)-\frac{\partial n_{\beta}(-\omega^\prime)}{\partial \mu}\mathcal{\bm{J}}'_{\beta}(-\omega^\prime)\right]e^{-i\omega^\prime(\tau_1-\tau_2)}\boldsymbol{U}^{\dagger}(t,\tau_2)
\end{array}
\right\}+H.c.
\end{align}
\end{widetext}

%\bibliographystyle{apsrev4-1}
%\bibliography{Draft_ver02}

\begin{references}
\bibitem{K01} A. Y. Kitaev, \emph{Unpaired Majorana fermions in quantum wires}, Phys. Usp. \textbf{44}, 131 (2001). 

\bibitem{K03} A. Y. Kitaev, \emph{Fault-tolerant quantum computation by anyons}, Ann. Physics \textbf{303}, 2 (2003).

\bibitem{FKL03} M. Freedman, A. Kitaev, M. Larsen, and Z. Wang, \emph{Topological quantum computation}, , Bull. Am. Math. Soc. \textbf{40}, 31 (2003)

\bibitem{NSS08} C. Nayak, S. H. Simon, A. Stern, M. Freedman, and S. Das Sarm, \emph{Non-Abelian anyons and topological quantum computation}, Rev. Mod. Phys. \textbf{80}, 1083 (2008).

\bibitem{AOR11} J. Alicea, Y. Oreg, G. Refael, F. Von Oppen, and M. P. Fisher, \emph{Non-Abelian statistics and topological quantum information processing in 1D wire networks},  Nat. Phys. \textbf{7}, 412 (2011).

\bibitem{VHF15} S. Vijay, T. H. Hsieh, and L. Fu, \emph{Majorana Fermion Surface Code for Universal Quantum Computation} Phys. Rev. X \textbf{5}, 041038 (2015).

\bibitem{PLS16} S. Plugge, L. A. Landau, E. Sela, A. Altland, K. Flensberg, and R. Egger, \emph{Roadmap to Majorana surface codes}, Phys. Rev. B \textbf{94}, 174514 (2016).

\bibitem{AHM16} D. Aasen, M. Hell, R. V. Mishmash, A. Higginbotham, J. Danon, M. Leijnse, T. S. Jespersen, J. A. Folk, C. M. Marcus, K. Flensberg, and J. Alicea, \emph{Milestones Toward Majorana-Based Quantum Computing}, Phys. Rev. X \textbf{6}, 031016 (2016).

\bibitem{LPS16} L. A. Landau, S. Plugge, E. Sela, A. Altland, S. M. Albrecht, and R. Egger, \emph{Towards Realistic Implementations of a Majorana Surface Code}, Phys. Rev. Lett. \textbf{116}, 050501 (2016).

\bibitem{LBK18} R. M. Lutchyn, E. P. Bakkers, L. P. Kouwenhoven, P. Krogstrup, C. M. Marcus, and Y. Oreg, \emph{Majorana zero modes in superconductor–semiconductor heterostructures}, Nat. Rev. Mater. \textbf{3}, 52 (2018).

\bibitem{PSD20} E. Prada, P. San-Jose, M. W. de Moor, A. Geresdi, E. J. Lee, J. Klinovaja, D. Loss, J. Nygard, R. Aguado, and L. P. Kouwenhoven, \emph{From Andreev to Majorana bound states in hybrid superconductor–semiconductor nanowires},  Nat. Rev. Phys. \textbf{2}, 575 (2020).

\bibitem{OV20} Y. Oreg and F. Von Oppen, \emph{Majorana Zero Modes in Networks of Cooper-Pair Boxes: Topologically Ordered States and Topological Quantum Computation}, Annu. Rev. Condens. Matter Phys. \textbf{11}, 397 (2020).

\bibitem{LLN09}  K. T. Law, P. A. Lee, and T. K. Ng,  \emph{Majorana Fermion Induced Resonant Andreev Reflection}, Phys. Rev. Lett. \textbf{103}, 237001 (2009).

\bibitem{F10} K. Flensberg, \emph{Tunneling characteristics of a chain of Majorana bound states}, Phys. Rev. B \textbf{82}, 180516(R) (2010).

\bibitem{DNS16} S. Das Sarma, A. Nag, and J. D. Sau, \emph{How to infer non-Abelian statistics and topological visibility from tunneling conductance properties of realistic Majorana nanowires}, Phys. Rev. B \textbf{94}, 035143 (2016).

\bibitem{MST18} C. Moore, T. D. Stanescu, and S. Tewari, \emph{Two-terminal charge tunneling: Disentangling Majorana zero modes from partially separated Andreev bound states in semiconductor-superconductor heterostructures}, Phys. Rev. B \textbf{97}, 165302 (2018).

\bibitem{ABA19} A. Vuik, B. Nijholt, A. R. Akhmerov, and M. Wimmer, \emph{Reproducing topological properties with quasi-Majorana states}, SciPost Phys. \textbf{7}, 061 (2019).

\bibitem{DHH20} J. Danon, A. B. Hellenes, E. B. Hansen, L. Casparis,
A. P. Higginbotham, and K. Flensberg, \emph{Nonlocal Conductance Spectroscopy of Andreev Bound States: Symmetry Relations and BCS Charges},  Phys. Rev. Lett.
\textbf{124}, 036801 (2020).

\bibitem{MAM20}  G. C. M´enard, G. L. R. Anselmetti, E. A. Martinez, D. Puglia, F. K. Malinowski, J. S. Lee, S. Choi, M. Pendharkar, C. J. Palmstrøm, K. Flensberg, C. M. Marcus, L. Casparis, and A. P. Higginbotham, \emph{Conductance-Matrix Symmetries of a Three-Terminal Hybrid Device},  Phys. Rev. Lett. \textbf{124}, 036802 (2020)

\bibitem{CPS15} J. Cayao, E. Prada, P. San-Jose, and R. Aguado, \emph{SNS junctions in nanowires with spin-orbit coupling: Role of confinement and helicity on the subgap spectrum}, Phys. Rev. B \textbf{91}, 024514 (2015) (2018).

\bibitem{SCP16} P. San-Jose, J. Cayao, E. Prada, and R. Aguado, \emph{Majorana bound states from exceptional points in non-topological superconductors}, Sci. Rep. \textbf{6}, 21427 (2016).

\bibitem{PAS17} E. Prada, R. Aguado, and P. San-Jose, \emph{Measuring Majorana nonlocality and spin structure with a quantum dot},  Phys. Rev. B \textbf{96}, 085418 (2017).

\bibitem{LSS17} C.-X. Liu, J. D. Sau, T. D. Stanescu, and S. Das Sarma, \emph{Andreev bound states versus Majorana bound states in quantum dot-nanowire-superconductor hybrid structures: Trivial versus topological zero-bias conductance peaks}, Phys. Rev. B \textbf{96}, 075161 (2017).

\bibitem{DVP18} M.-T. Deng, S. Vaitiek˙enas, E. Prada, P. San-Jose,
J. Nygard, P. Krogstrup, R. Aguado, and C. M. Marcus, \emph{Nonlocality of Majorana modes in hybrid nanowires}, Phys. Phys. Rev. B \textbf{98}, 085125 (2018).

\bibitem{APP19}  J. Avila, F. Penaranda, E. Prada, P. San-Jose, and
R. Aguado, \emph{Non-hermitian topology as a unifying framework for the Andreev versus Majorana states controversy}, Commun. Phys. \textbf{2}, 133 (2019).

\bibitem{ACB19}  O. A. Awoga, J. Cayao, and A. M. Black-Schaffer, \emph{Supercurrent Detection of Topologically Trivial Zero-Energy States in Nanowire Junctions}, Phys. Rev. Lett. \textbf{123}, 117001 (2019).

\bibitem{CWX12}  Y. Cao, P. Wang, G. Xiong, M. Gong, and X.-Q. Li, \emph{Probing the existence and dynamics of Majorana fermion via transport through a quantum dot}, Phys. Rev. B \textbf{86}, 115311 (2012).

\bibitem{ZR13}  B. Zocher and B. Rosenow, \emph{Modulation of Majorana-Induced Current Cross-Correlations by Quantum Dots}, Phys. Rev. Lett. \textbf{111}, 036802 (2013)

\bibitem{LCL15}  D. E. Liu, M. Cheng, and R. M. Lutchyn, \emph{Probing Majorana physics in quantum-dot shot-noise experiments},  Phys. Rev. B \textbf{91}, 081405(R) (2015).

\bibitem{QFL22}  L. Qin, W. Feng, and X.-Q. Li, \emph{Cross correlation mediated by distant Majorana zero modes with no overlap}, Chin. Phys. B \textbf{31}, 017402 (2022).

\bibitem{S22}  S. Smirnov, \emph{Revealing universal Majorana fractionalization using differential shot noise and conductance in nonequilibrium states controlled by tunneling phases},  Phys. Rev. B \textbf{105}, 205430 (2022).

\bibitem{FK08}  L. Fu and C. L. Kane, \emph{Superconducting Proximity Effect and Majorana Fermions at the Surface of a Topological Insulator}, Phys. Rev. Lett. \textbf{100}, 096407 (2008).

\bibitem{FK09}  L. Fu and C. L. Kane, \emph{Josephson current and noise at a superconductor/quantum-spin-Hall-insulator/superconductor junction}, Phys. Rev. B \textbf{79}, 161408(R) (2009).

\bibitem{A10}  J. Alicea, \emph{Majorana fermions in a tunable semiconductor device}, Phys. Rev. B \textbf{81}, 125318 (2010)

\bibitem{LSD10}  R. M. Lutchyn, J. D. Sau, and S. Das Sarma, \emph{Majorana Fermions and a Topological Phase Transition in Semiconductor-Superconductor Heterostructures}, Phys. Rev. Lett. \textbf{105}, 077001 (2010).

\bibitem{ORv10}  Y. Oreg, G. Refael, and F. von Oppen, \emph{Helical Liquids and Majorana Bound States in Quantum Wires}, Phys. Rev. Lett. \textbf{105}, 177002 (2010).

\bibitem{CZQ11}   S. B. Chung, H.-J. Zhang, X.-L. Qi, and S.-C. Zhang, \emph{Topological superconducting phase and Majorana fermions in half-metal/superconductor heterostructures}, Phys. Rev. B \textbf{84}, 060510(R) (2011)

\bibitem{DB11} M. Duckheim and P. W. Brouwer, \emph{Andreev reflection from noncentrosymmetric superconductors and Majorana bound-state generation in half-metallic ferromagnets}, Phys. Rev. B \textbf{83}, 054513 (2011).

\bibitem{NDB13}  S. Nadj-Perge, I. K. Drozdov, B. A. Bernevig, and
A. Yazdani, \emph{Proposal for realizing Majorana fermions in chains of magnetic atoms on a superconductor},  Phys. Rev. B \textbf{88}, 020407(R) (2013).

\bibitem{PGV13}  F. Pientka, L. I. Glazman, and F. von Oppen, \emph{Topological superconducting phase in helical Shiba chains}, Phys. Rev. B \textbf{88}, 155420 (2013)

\bibitem{ADH11}  A. R. Akhmerov, J. P. Dahlhaus, F. Hassler, M. Wimmer,
and C. W. J. Beenakke, \emph{Quantized Conductance at the Majorana Phase Transition in a Disordered Superconducting Wire}, Phys. Rev. Lett. \textbf{106}, 057001 (2011).

\bibitem{LPL12}  J. Liu, A. C. Potter, K. T. Law, and P. A. Lee, \emph{Zero-Bias Peaks in the Tunneling Conductance of Spin-Orbit-Coupled Superconducting Wires with and without Majorana End-States}, Phys. Rev. Lett. \textbf{109}, 267002 (2012)

\bibitem{SBS15}  F. Setiawan, P. M. R. Brydon, J. D. Sau, and
S. Das Sarma, \emph{Conductance spectroscopy of topological superconductor wire junctions},  Phys. Rev. B \textbf{91}, 214513 (2015).

\bibitem{RVK2018} T. \"{O}. Rosdahl, A. Vuik, M. Kjaergaard, and A. R.
Akhmerov,  \emph{Andreev rectifier: A nonlocal conductance signature of topological phase transitions}, 
Phys. Rev. B \textbf{97}, 045421 (2018).

\bibitem{TZ08} M. W. Y. Tu and W.-M. Zhang, \emph{Non-Markovian decoherence theory for a double-dot charge qubit}, Phys. Rev. B \textbf{78}, 235311 (2008)

\bibitem{JTZ10} J. Jin, M. W.-Y. Tu, W.-M. Zhang, and Y. Yan, \emph{Non-equilibrium quantum theory for nanodevices based on the Feynman–Vernon influence functional},  New J. Phys. \textbf{12}, 083013 (2010)

\bibitem{LZ12} C. U. Lei and W.-M. Zhang, \emph{A quantum photonic dissipative transport theory},  Ann. Phys. \textbf{327}, 1408 (2012)

\bibitem{ZLX12} W.-M. Zhang, P.-Y. Lo, H.-N. Xiong, M.-Y. Tu, and F. Nori, \emph{General Non-Markovian Dynamics of Open Quantum Systems}, Phys. Rev. Lett. \textbf{109}, 170402 (2012)

\bibitem{TZJ12} M.-Y. Tu, W.-M. Zhang, J. Jin, O. Entin-Wohlman, and
A. Aharony, \emph{Transient quantum transport in double-dot Aharonov-Bohm interferometers}, Phys. Rev. B \textbf{86}, 115453 (2012).

\bibitem{TAZ14} M.-Y. Tu, A. Aharony, W.-M. Zhang, and O. Entin-
Wohlman, \emph{Real-time dynamics of spin-dependent transport through a double-quantum-dot Aharonov-Bohm interferometer with spin-orbit interaction},  Phys. Rev. B \textbf{90}, 165422 (2014).

\bibitem{Z19} W.-M. Zhang, \emph{Exact master equation and general non-Markovian dynamics in open quantum systems}, Eur. Phys. J. Spec. Top. \textbf{227}, 1849 (2019).

\bibitem{LYH18} H.-L. Lai, P.-Y. Yang, Y.-W. Huang, and W.-M. Zhang, \emph{Exact master equation and non-Markovian decoherence dynamics of Majorana zero modes under gate-induced charge fluctuations}, Phys. Rev. B \textbf{97}, 054508 (2018)

\bibitem{LZ20} H.-L. Lai and W.-M. Zhang, \emph{Decoherence dynamics of Majorana qubits under braiding operations}, Phys. Rev. B \textbf{101}, 195428 (2020).

\bibitem{YZ20} C.-Z. Yao and W.-M. Zhang, \emph{Probing topological states through the exact non-Markovian decoherence dynamics of a spin coupled to a spin bath in the real-time domain},  Phys. Rev. B \textbf{102}, 035133 (2020).

\bibitem{HYZ20} Y.-W. Huang, P.-Y. Yang, and W.-M. Zhang,  \emph{Quantum theory of dissipative topological systems}, Phys. Rev. B \textbf{102}, 165116 (2020)

\bibitem{XLZ21} F.-L. Xiong, H.-L. Lai, and W.-M. Zhang, \emph{Manipulating Majorana qubit states without braiding}, Phys. Rev. B \textbf{104}, 205417 (2021)

\bibitem{YZ22} C.-Z. Yao and W.-M. Zhang, \emph{The differential conductance tunnel spectroscopy in an analytical solvable two-terminal Majorana device},New J. Phys. \textbf{24}, 073015 (2022).

\bibitem{DRB11} S. Diehl, E. Rico, M. A. Baranov and P. Zoller, \emph{Topology by dissipation in atomic quantum wires}, Nat. Phys. \textbf{7}, 971 (2011).

\bibitem{VRM12} O. Viyuela, A. Rivas, and M. A. M.-Delgado, \emph{Thermal instability of protected end states in a one-dimensional topological insulator}, Phys. Rev. B \textbf{86},  155140 (2012).

\bibitem{CLD12} M. Cheng, R. M. Lutchyn, S. D. Sarma, \emph{Topological protection of Majorana qubits}, Phys. Rev. B \textbf{85}, 165124 (2012).

\bibitem{RVM13} A. Rivas, O. Viyuela, and M. A. M.-Delgado, \emph{Density-matrix Chern insulators: Finite-temperature generalization of topological insulators}, Phys. Rev. B \textbf{88}, 155141 (2013).

\bibitem{BBK13} C.-E. Bardyn, M. A. Baranov, C. V. Kraus, E. Rico, A. İmamoğlu, P. Zoller and S. Dieh, \emph{Topology by dissipation}, New J. Phys. \textbf{15}, 085001 (2013).

\bibitem{MRG14} P. Matthews, P. Ribeiro, and A. M. G.-García, \emph{Dissipation in a Simple Model of a Topological Josephson Junction}, Phys. Rev. Lett. \textbf{112}, 247001 (2014).

\bibitem{GL14} J. R. Colbert and P. A. Lee, \emph{Proposal to measure the quasiparticle
poisoning time of Majorana bound states}, Phys. Rev. B \textbf{89}, 140505(R) (2014).

\bibitem{JL22} J. Jin and X.-Q. Li, \emph{Master equation approach for transport through Majorana zero modes}, New J. Phys. \textbf{24}, 093009 (2022).

\bibitem{Fu10} L. Fu, \emph{Electron Teleportation via Majorana Bound States in a Mesoscopic Superconductor}, Phys. Rev. Lett. \textbf{104}, 056402 (2010).

\bibitem{H1988} F. D. M. Haldane, \emph{Model for a Quantum Hall Effect without Landau Levels: Condensed-Matter Realization
of the "Parity Anomaly"}, Phys. Rev. Lett. \textbf{61}, 2015 (1988).

\bibitem{YLCZ} C.-Z. Yao, H. L. Lai, Y. C. Chang and W.-M. Zhang, \emph{Quantum transport in a superconductor-semiconductor heterostructure} (in preparation).

\bibitem{Yang2017} P. Y. Yang and W. -M. Zhang, \emph{Master equation approach to transient quantum
transport in nanostructures}, Front. Phys. \textbf{12}, 127204 (2017).

\bibitem{QT1993} C. H. Bennett, G. Brassard, C. Cr\'{e}peau, R. Jozsa, A. Peres, and W. K. Wootters, 
\emph{Teleporting an Unknown Quantum State via Dual Classical and Einstein-Podolsky-Rosen Channels}, 
Phys. Rev. Lett. \textbf{70} 1895 (1993).

\bibitem{LLS14} R. López, M. Lee, L. Serra, and J. S. Lim, \emph{Thermoelectrical detection of Majorana states}, Phys. Rev. B \textbf{89}, 205418 (2014).

\bibitem{HSZ21} T.-Y. He, H. Sun and G. Zhou, \emph{Photon-assisted Seebeck effect in a quantum dot coupled to Majorana zero modes}, Front. Phys. \textbf{9}, 28 (2021).

\bibitem{ZWZ22} W.-K. Zou, Q. Wang, H.-K. Zhao, \emph{Aharonov-Bohm oscillations in the Majorana fermion modulated charge and heat transports through a double-quantum-dot interferometer}, Phys. Lett. A \textbf{443}, 128219 (2022).

\bibitem{S23} S. Smirnov, \emph{Majorana differential shot noise and its universal thermoelectric crossover}, Phys. Rev. B \textbf{107}, 155416 (2023).

\bibitem{HZ22} W.-M. Huang and W.-M. Zhang, \emph{Nonperturbative renormalization of quantum thermodynamics from weak to strong couplings}, Phys. Rev. Res. \textbf{4}, 023141 (2022)

\bibitem{HZ22a} W.-M. Huang and W.-M. Zhang, \emph{Strong-coupling quantum thermodynamics far from equilibrium: Non-Markovian transient quantum heat and work}, Phys. Rev. A \textbf{106}, 032607 (2022).

\bibitem{AHZ20} Md. M. Ali, W.-M. Huang and W.-M. Zhang, \emph{Quantum thermodynamics of single particle systems}, Sci. Rep. \textbf{10}, 13500 (2020).

\bibitem{HZ22b} Y.-W. Huang and W.-M. Zhang, \emph{Exact master equation for generalized quantum Brownian motion with momentum-dependent system-environment couplings}, Phys. Rev. Res. \textbf{4}, 033151 (2022)

\bibitem{YLZ15} P.-Y. Yang, C.-Y. Lin and W.-M. Zhang, \emph{Master equation approach to transient quantum transport in nanostructures incorporating initial correlations}, Phys. Rev. B \textbf{92}, 165403 (2015)

\end{references}

\end{document}